\documentclass[a4paper,11pt]{article}
\pdfoutput=1 
\usepackage{jheppub} 
\usepackage{setspace}
\usepackage{graphicx}
\usepackage{wrapfig}
\usepackage{caption}
\usepackage{subcaption}
\usepackage{hyperref}
\usepackage{tikz}
\usepackage{natbib}
\usetikzlibrary{arrows}

\newcommand{\midarrowD}{\tikz \draw[-triangle 90] (0,0) -- +(0,-0.1);}
\newcommand{\midarrowU}{\tikz \draw[-triangle 90] (0,0) -- +(0,0.1);}
\newcommand{\midarrowL}{\tikz \draw[-triangle 90] (0,0) -- +(-0.1,0.0);}
\usepackage{empheq}
\usepackage{empheq}
\definecolor{UI_blue}{RGB}{32, 64, 151}
\definecolor{UI_red}{RGB}{187, 62, 24}
\definecolor{UI_blue2}{RGB}{0, 84, 147}
\definecolor{UI_red2}{RGB}{159, 32, 66}
\definecolor{UI_gray}{RGB}{169, 169, 169}
\definecolor{UI_sepia}{RGB}{112, 66, 20}
\definecolor{UI_bittersweet}{RGB}{254, 111, 94}
\definecolor{UI_emerald}{RGB}{80, 200, 120}
\definecolor{UI_olivegreen}{RGB}{181, 179, 92}
\definecolor{UI_cadetblue}{RGB}{95, 158, 160}
\definecolor{UI_fuchsia}{RGB}{255, 0, 255}
\definecolor{UI_midnightblue}{RGB}{25, 25, 112}
\definecolor{UI_royalblue}{RGB}{0,35, 102}
\definecolor{UI_periwinkle}{RGB}{204, 204, 255}
\definecolor{UI_redorange}{RGB}{255, 83, 73}
\definecolor{UI_brickred}{RGB}{203,65,84}	
\definecolor{UI_forestgreen}{RGB}{34, 139, 34}
\definecolor{UI_tan}{RGB}{210,180,140}	
\definecolor{UI_burlywood}{RGB}{222,184,135}
\definecolor{UI_burlywood}{RGB}{192,64,0}
\definecolor{UI_darkorchid}{RGB}{153,50,204}

\usepackage{url}

\usepackage{mathtools}
\usepackage{physics}
\usepackage{simplewick}
\usepackage[obeyFinal]{todonotes}

\def\beq{\begin{eqnarray}}\def\eeq{\end{eqnarray}}
\def\be{\begin{equation}}\def\ee{\end{equation}}

\def\l{\langle}
\def\r{\rangle}

\def\mes[#1]{d^{3}{#1}}

\def\del{\partial}

\newcommand{\dpk}[1]{\dfrac{\partial}{\partial k_{#1}}}

\newcommand{\eq}[1]{eq.(\ref{#1})}

\def\del{\partial}

\reversemarginpar

\author[a]{Indranil Dey}
\author[a]{Kanhu Kishore Nanda}
\author[a]{Akashdeep Roy}
\author[a]{Sandip P. Trivedi}
\affiliation[a]{\it Department of Theoretical Physics,
	Tata Institute of Fundamental Research,\\  Colaba, Mumbai, India, 400005\\}

\emailAdd{indranil.dey@tifr.res.in}
\emailAdd{kanhu.nanda@tifr.res.in}
\emailAdd{akashdeep.roy@tifr.res.in}
\emailAdd{sandip@theory.tifr.res.in}

\vspace{1cm}

\abstract{ It has been suggested that a $dS_{d+1}$ spacetime of radius $R_{ds}$ has a holographic dual, living at future space-like infinity ${\cal I}^+$, with the bulk wave function being dual to the partition function of the boundary theory, \cite{Malda-NG}. We consider some aspects of this correspondence. For under damped scalars with mass $M^2R_{ds}^2>{d^2\over4}$, 
belonging to the principal series, we show that  for the Bunch Davies vacuum a suitable source in the boundary theory can be identified in terms of the coherent state representation of the wave function. 
We argue that terms in the resulting correlation functions, which are independent of the late time cut-off,  satisfy the Ward identities of a conformal field theory.
We also discuss other ways to identify sources, both in the under damped and the over damped case, where $M^2R_{ds}^2<{d^2\over4}$,  and argue that these too can lead  to correlators satisfying the  Ward identities of a CFT.
  Some comments   on the violation  of reflection positivity, and the cut-off dependent terms, along with   some explicit checks and sample calculations, are also included.  }

\title{Aspects of  dS/CFT Holography}

\preprint{\parbox{3cm}{TIFR/TH/24-13}}

\begin{document}
	\maketitle
	\flushbottom
	\vskip 10pt
	\vskip 20pt
	\newpage
	
	\section{Introduction}
	\label{intro}
	
	de Sitter space holds many mysteries. One question which has attracted considerable attention is whether it has  a dual description in terms of a hologram. For some early references, see \cite{wittends,strominger, Malda-NG, klemm}. The purpose of this paper is to study some aspects  of such a holographic correspondence. 
	
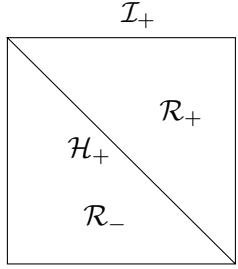
\begin{figure}[h]
		\centering
		\begin{tikzpicture}
			\draw (6,3) -- (6,0) -- (9,0) -- (9,3) -- (6,3) -- node[left]{$\mathcal{H}_+$}(9,0);
			\node[text width=3cm] at (9,3.3) {$\mathcal{I}_+$};
			\node[text width=3cm] at (9.5,2) {$\mathcal{R}_+$};
			\node[text width=3cm] at (8.5,0.6) {$\mathcal{R}_-$};
		\end{tikzpicture}
		\caption{Penrose diagrams for de Sitter space in Poincaré coordinates. }    \label{poinpenrose}
	\end{figure}

	In the Poincaré patch of $d+1$ dimensional de-Sitter space, $dS_{d+1}$, the metric is given by 
	\be
	\label{metpoin}
	ds^2=R_{dS}^2 \left[-{d\eta^2\over \eta^2} + {1\over \eta^2}\sum_{i=1}^ddx^idx_i\right].
	\ee
	Here  $\eta\in(-\infty,0]$ and ${\cal I}_+$, the future space-like boundary, is  obtained by taking  $\eta\rightarrow 0$, see Figure \ref{poinpenrose}. 
	$R_{dS}$ is the radius of $dS$ space which is related to the Hubble constant by 
	\be
	\label{relh}
	H={1\over R_{dS}},
	\ee
	We will often set $R_{dS}=1$. 
	The proposed correspondence which we study, \cite{Malda-NG}, is that the wave function $\Psi$ in $dS$ space, asymptotically, as $\eta\rightarrow 0$, 
	is equal to the partition function $Z_{FT}$ of a dual field theory in the presence of sources.  
 Schematically, 
	\be
	\label{meq}
	\Psi[\phi_i]=Z_{FT}[S_i],
	\ee
	 where $\phi_i$, $i=1,\cdots N$,   denote the values that the various bulk fields take close to 
	${\cal I}_+$, and $S_i$ denote the corresponding sources in the field theory. 	In subsequent sections of the paper we will be much more precise about this correspondence. Let us also mention that we will explore this version of holography below  for fields in the Bunch Davies vacuum. 
	
	We note that the correspondence in eq.(\ref{meq}) is analogous to the $AdS/CFT$ correspondence, for reviews see \cite{magoo, Hubeny-adscft},  the crucial difference  being that it is the bulk wave function, instead of the partition function,  which appears on the LHS in eq.(\ref{meq}). 	 In fact, $dS_{d+1}$ space and Euclidean $AdS_{d+1}$   have the same isometry group, $SO(d+1,1)$,
	and can be related by an analytic continuation.
	 The  metric of Euclidean $AdS_{d+1}$ which is given by 
	 	 \be
	 \label{metads}
	 ds^2=L^2\left[{dz^2\over z^2}+{1\over z^2} \sum_{i=1}^d dx^idx_i\right]
	 \ee
	 with $z\in [0,\infty]$, and $L$ being the radius of $AdS$ space, after the  
	 analytic continuation,
	 \be
	 \label{anac}
	 z\rightarrow (-i)  \eta, \ \ L\rightarrow ( i) R_{\rm dS}
	 \ee
	  becomes the metric of $dS_{d+1}$, eq.(\ref{metpoin}).
	 As a result, some features of the  dS/FT dictionary essentially  follow from the $AdS/CFT $ correspondence, after an   analytic continuation. 	 
	 
	 However, there are  important differences, and some unfamiliar consequences, in   the $dS$ case, and the aim of this paper is to study some of them.   
	 
	  One key issue which we will study is the behaviour of fields with sufficiently high mass in dS space, corresponding to the principal series representations of the $SO(d+1,1)$ group of isometries of $dS_{d+1}$ \cite{Lowe, Zimo-alg}. A free scalar field with a mass $M$  has an action given in 
	  \eq{action} and asymptotically in $dS_{d+1}$, as $\eta\rightarrow 0$, the solutions of its wave equation  are a linear combination of two modes
	  going like 
	 \be
	 \label{scea}
	 \phi\sim (-\eta)^\alpha, \ \ \alpha={{d\over 2} \pm i \sqrt{-{d^2\over 4} +M^2 R_{dS}^2}},
	 \ee
If the mass 
 \be
	 \label{condun}
	 M^2 R_{dS}^2>{d^2\over 4}  
	 \ee
	  we see that the two modes will both continue to oscillate as $\eta\rightarrow 0$. Physically this is because with a sufficiently high mass, the so-called ``Hubble friction term", which arises due to the exponential expansion,   is not enough to cause the field to freeze out. In contrast, when 
\be
\label{condover}
M^2R_{dS}^2 <{d^2\over4}, 
\ee
the behaviour is different and  both  modes  decay towards the boundary with the fall off going like 
$(-\eta)^{{d\over2}\pm \sqrt{{d^2\over4}-M^2R_{dS}^2}}$. The principal series corresponds to the case in eq.(\ref{condun}),  which we will refer to as the underdamped case below. 
We will see that in this case constructing the holographic dictionary,  in particular relating the asymptotic behaviour of the bulk field to a source in the boundary theory,  has some novel features. In contrast,  for the overdamped case where eq.(\ref{condover})  is met, the holographic dictionary  can be constructed more straightforwardly, in close analogy with the AdS case. 

In the underdamped case we  find that it is convenient to consider the wave function in the coherent state basis rather than the basis of eigenstates of the bulk scalar field. A suitable source in the boundary  can then be identified in terms of the eigenvalue  of the coherent state. The corresponding field \footnote{We use the terminology of a boundary field rather than operator since the boundary theory cannot be continued to a field theory in Lorentzian space, at least of a conventional type, see below.} in the boundary which this source couples to   has dimension 
\be
\label{valdelp}
\Delta_+={d\over2}+i\sqrt{M^2R_{dS}^2-{d^2\over4}}.
\ee
The spatial and time reparametrisation invariance conditions on the wave function give rise to constraints  on the correlation functions of the boundary theory.  If we consider the wave function at a late time slice, 
say at  a constant value of $\eta=\eta_1$, the value of  the time coordinate, $\eta_1$,  plays the role of a UV cut-off in the dual theory. We show that the constraints which arise on the cut-off independent terms in correlation functions, due to the invariance of the wave function, take the form of  Ward identities of a conformal field theory. 
The identification of a source in the underdamped case and the resulting Ward identities is one of the main results of the paper.

We also explore other ways of identifying a source for the boundary theory. Working directly in the basis of eigenstates of the bulk scalar, rather that the coherent state representation, we  find that there are  in fact alternate ways of identifying such sources. These are related to the value the bulk field takes  on the hypersurface 
$\eta=\eta_1$ by a spatial non-local transformation. While some of the cut-off dependent terms in the resulting correlation functions are unwieldy\footnote{ E.g. the two point function has a term which is both cut-off, i.e. $\eta_1$  dependent,
and non-local, \eq{posf}.} we show that the cut-off independent terms continue to satisfy the Ward identities of a conformal field theory. 
Most of our analysis in the paper is focused on underdamped fields, but   some of our results, pertaining to the possibility of  identifying sources in an alternate manner,  are  also of interest for  the overdamped case, see section \ref{altrepov}. 

It is worth noting here that underdamped fields were also studied in \cite{Isono}\footnote{We thank the referee for emphasising that we study this paper more thoroughly}, where it was noted that for the free theory, a change in boundary conditions along with an additional boundary term in the action, allowed for a source to be identified with the same conformal dimension, eq.(\ref{valdelp}).
In fact, as we discuss in more detail in section \ref{coherent}, this proposal turns out to be the same as the one we discuss here. The main additional insight we provide, is the connection to  the coherent state representation, with the source in the boundary theory being identified with the coherent state eigenvalue. 
This connection allows us to easily generalise the discussion to include interactions, and it also leads to a transparent proof that the finite terms in the resulting correlation functions of the boundary theory satisfy the Ward identities of conformal invariance, and other symmetries, as discussed in section \ref{WI}. An important point is that the transformation needed to go from the field basis to the coherent state basis does not need to be corrected once interactions are included and as a result one can, in practical terms, calculate correlators using the familiar Witten diagrams and only at the last step carry out the integral transform to go to the coherent state basis, as is discussed further in section \ref{coherent}.

One important difference with the AdS case, as has been noted in \cite{Malda-NG,Brussels, suvrat--Holoinfods} is that the cut-off dependent terms are important for the wave function in dS space. In fact, this dependence gives rise to  the leading contribution in the wave function at late times, and plays an important role in ensuring that the wave function  solves the Wheeler de-Witt equation \cite{nandajt}. This is in contrast to the AdS case where these terms can be removed by adding suitable counter terms.
While we do not analyse them in much detail, it is worth noting that the invariance of the wave function under time and spatial reparametrisations gives rise to important constraints on these cut-off dependent terms  as well, and these  constraints can be obtained in a manner similar to the Ward identities  mentioned above.
In an admittedly  ambitious version of holography, these cut-off dependent terms would also be correctly reproduced by  the boundary theory \footnote{In  referring to the hologram as a field theory above, rather than a conformal field theory,  we are allowing for this possibility.}. 

Finally, we also draw attention to the fact that the dual theory for  dS  violates reflection positivity \cite{Malda-NG}. As a result dimensions of operators and coefficients in correlation functions are often complex numbers, e.g., eq.(\ref{valdelp}). The breaking of reflection positivity means that the dual field theory which is Euclidean cannot 
be continued to a field theory in Lorentzian space, at least  with a hermitian Hamiltonian and states having a  positive norm. This breaking of reflection positivity arises because one is dealing with the hologram at ${\cal I}^+$, rather than the past boundary of de-Sitter on ${\cal I}^-$, see fig \ref{glopenrose}. Another way to put it is that one is constructing the hologram for the expanding rather than the contracting branch of the Hartle-Hawking wave function. 
The hologram at ${\cal I}^-$ is related to the one at ${\cal I}^+$ by a $CP$ transformation, i.e the product of Charge conjugation $C$ and Parity $P$. 
All correlators can  be related in the two holograms through this transformation \footnote{E.g. the dimension in  eq.(\ref{valdelp}) corresponding to the underdamped field 
would take its complex conjugate value in the hologram at ${\cal I}^-$.}. The analytic continuation which takes some $AdS$ correlators to those in $dS$, see \eq{anac} above,
are also correspondingly different for the ${\cal I}^-$ case, resulting in this CP transformation. 

This paper is organised as follows. Section \ref{dSgeo} reviews some essential facts about dS space. Section \ref{setup} calculates the wave function in the Bunch Davies vacuum for free scalars and also discusses some interactions. Section \ref{holo} discusses some holographic aspects and section \ref{coherent} the coherent state representation in the underdamped case.  Ward identities are discussed in section \ref{WI} which also included some discussion of alternate ways of identifying sources  in subsection \ref{addpt}. Additional comments on the violation of reflection positivity, on $dS_3$, etc. are in section \ref{addcmnt}. The appendices contain important supplementary material including some details on computing higher point correlators, the OPE limit, and some concrete checks on the Ward identities. 

We have referred to various important references above and in the following text. Some important additional references are \cite{Balaexplore, MaldaPimen, MataCMB, slowrollTrivedi, Kunduward, Narayan2022, Narayan2023, Q3ddS}.

	\section{de Sitter Geometry}
	\label{dSgeo}
	We begin by reviewing some basic facts about dS space. 
	 $d+1$ dimensional de Sitter space time is described as a $d+1$ dimensional hyperboloid
	\begin{equation}
		\label{embhyp}
		-X_0^2+\sum_{i=1}^{d+1}X_i^2=R_{\rm dS}^2
	\end{equation}
	embedded in $\real^{1,d+1}$ Minkowski space time. We will now set $R_{\rm dS}=1$ and restore the factors when necessary. The global coordinate parametrization of dS$_{d+1}$ is given by 
	\begin{equation}
		\label{globx}
		X_0=\sinh{\tau}~~~~~~~X_i=\cosh{\tau}~\Omega_i~~~(i=1,2,...,d+1)   
	\end{equation}
	where $\Omega_i$ denotes angles on a unit ${\cal S}^{d}$. The metric in these coordinates is 
	\begin{equation}
		\label{globg}
		ds^2=-d\tau^2 + \cosh^2{\tau}~d\Omega_d^2
	\end{equation}
	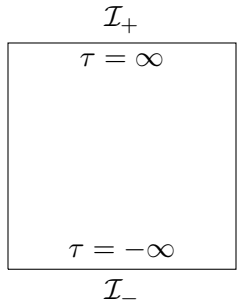
\begin{figure}[h]
		\centering
		\begin{tikzpicture}
			\draw (0,0) -- node[above]{$\tau=-\infty$}node[below]{$\mathcal{I}_-$}(3,0) -- (3,3) -- node[below]{$\tau=\infty$}node[above]{$\mathcal{I}_+$}(0,3) -- (0,0);
		\end{tikzpicture}
		\caption{Penrose diagrams for de Sitter space in Global coordinates}    \label{glopenrose}
	\end{figure}
	These global coordinates cover all of dS space which has two spacelike boundaries namely $\mathcal{I}_+$ and $\mathcal{I}_-$ mentioned in Figure \ref{glopenrose}. 
	Conventionally these boundaries are called future and past spacelike infinity. 
	
	In this paper we will explore a hologram for dS space at future spacelike infinity. For our analysis it will be useful to consider Poincaré coordinates which only cover half of 
	dS space, namely the region ${\mathcal R}_+$. The metric in Poincaré coordinates takes the form
	\begin{equation}
		\label{Poing}
		ds^2=\frac{1}{\eta^2}\left(-d\eta^2+\sum_{i=1}^{d}dx_i^2\right)
	\end{equation}
Poincaré coordinates  are related to global coordinates by the transformation
\begin{equation}
		\label{Poinx}
		X_0=\sinh{t}+\frac{1}{2}e^t\sum_{i=2}^{d+1}x_i^2 ~~~~~~ X_1=\cosh{t}-\frac{1}{2}e^t\sum_{i=2}^{d+1}x_i^2 ~~~~~~ X_i=e^tx_i
	\end{equation}
	with $t\in [-\infty,\infty]$ being related to $\eta$ by 
	\begin{equation}
		\eta=-e^{-t}
	\end{equation}
	It is clear that Poincaré  coordinates only cover the  region 
	\be
	\label{regc}
	X_1\ge X_0
	\ee
	Also in our conventions, note that $\eta$ takes negative values and lies in the range $\eta\in[-\infty,0]$, with ${\mathcal H}_+$ corresponding to  
	$\eta\rightarrow -\infty$, and ${\mathcal I}_+$ corresponding to $\eta\rightarrow 0$. 
		
Note also that in Poincaré coordinates the metric is conformal to $d+1$ dimensional Minkowski space. 

	\section{Basic Setup}
	\label{setup}
	For studying various aspects of the putative hologram at ${\mathcal I}_+$ we will mainly restrict ourselves to bulk scalar fields in this paper. 
	Let $\phi$ be a free real scalar field in the bulk with mass $M$ in $d+1$ dimensional dS space, its action is 
	\begin{equation}
		\label{action}
		S=\int d^{d+1}x\sqrt{-g}\mathcal{L}\left(g^{\mu\nu},\phi,\partial_\mu\phi\right)=\int d^{d+1}x\sqrt{-g}\left[-\frac{1}{2}g^{\mu\nu}\partial_\mu\phi\partial_\nu\phi-\frac{1}{2}M^2\phi^2\right]
	\end{equation}   
The equation of motion for the field is 
	\begin{equation}
		\label{Eq.m}
		\frac{1}{\sqrt{-g}}\partial_\mu(\sqrt{-g}g^{\mu\nu}\partial_\nu \phi)-M^2\phi=0
	\end{equation}
	 Here we study the field in fixed de Sitter background.  
	Inserting the metric given by eq.(\ref{Poing}) in eq.(\ref{Eq.m}), we get 
	\begin{equation}
		\label{EOM}
		\eta^2\partial_\eta^2\phi-(d-1)\eta\partial_\eta\phi-\eta^2\partial^2\phi+M^2\phi=0
	\end{equation}
	where $\partial^2$ denotes Euclidean $d$ dimensional Laplacian. The independent solutions of eq.(\ref{EOM}) are found to be
	\begin{equation}
		\label{sol}
		\phi(\mathbf{x},\eta)=\int \frac{d^d{\bf k}}{(2\pi)^d}\left[\mathbb{C}_1(\mathbf{k})(-\eta)^{d\over2}H^{(1)}_{\nu}(-k\eta)e^{i\mathbf{k}.\mathbf{x}}+\mathbb{C}_2(\mathbf{k})(-\eta)^{d\over2}H^{(2)}_{\nu}(-k\eta)e^{i\mathbf{k}.\mathbf{x}}\right]
	\end{equation}
	where $H^{(1)}_\nu, H^{(2)}_\nu$ denotes Hankel functions with index $\nu$ of the first and second kind respectively,  $\mathbb{C}_1$ and $\mathbb{C}_2$ are constant  coefficients and the index  $\nu$ is given by
	\begin{equation}
		\label{nu}
		\nu=\sqrt{\frac{d^2}{4}-M^2}.
	\end{equation} 
	And 
	\be
	\label{defk}
	k=\sqrt{{\bf k}\cdot {\bf k}}.
	\ee
	denotes the magnitude of ${\bf k}$.\\
	Importantly, for $M^2<\frac{d^2}{4}$, $\nu$ is a real quantity. We refer to this as the overdamped case or the case of light scalars. For $M^2>\frac{d^2}{4}$, $\nu$ is imaginary. We refer to this case as the underdamped case or the case of heavy scalars. $M^2=\frac{d^2}{4}$ is the critically damped case.\\ 
	Eq.(\ref{sol})  is  valid for the overdamped case, $M^2<{d^2\over 4}$. In the underdamped case the corresponding expression can be obtained by analytically continuing 
	\be
	\label{analcont}
	\nu\rightarrow i \mu=i \sqrt{M^2-{d^2\over 4}}
	\ee
	This will also be true for other formulae we discuss  below, the underdamped case can be obtained for these using the continuation eq.(\ref{analcont}) as well.\\ 
	Note that near the horizon ${\mathcal H}_+$ where   $\eta\rightarrow -\infty$, the two Hankel functions behave like 
	\begin{equation}
		\label{horH}
		\lim_{\eta\rightarrow-\infty}H^{(1)}_\nu(-k\eta)= A_h\frac{e^{-ik\eta}}{\sqrt{-k\eta}} \hspace{50pt} \lim_{\eta\rightarrow-\infty}H^{(2)}_\nu(-k\eta)= B_h\frac{e^{ik\eta}}{\sqrt{-k\eta}}
	\end{equation}
	where
	\begin{equation}
		\label{horcoef}
		A_h=\sqrt{2\over\pi}e^{-\frac{i\pi\nu}{2}-\frac{i\pi}{4}} \hspace{50pt} B_h=\sqrt{2\over\pi}e^{\frac{i\pi\nu}{2}+\frac{i\pi}{4}}
	\end{equation}   
	While near ${\mathcal I}_+$, where $\eta\rightarrow 0$,  we have 
	\begin{align}
		\label{bounH2}
		&\lim_{\eta\rightarrow0}H^{(1)}_\nu(-k\eta)=(-k\eta)^{-\nu } \left(-\frac{i 2^{\nu } \Gamma[\nu]}{\pi }\right)+(-k\eta)^{\nu } \left(\frac{2^{-\nu } (1+i \cot (\pi  \nu ))}{\Gamma [\nu +1]}\right)\\ &\lim_{\eta\rightarrow0}H^{(2)}_\nu(-k\eta)=(-k\eta)^{-\nu } \left(\frac{i 2^{\nu } \Gamma[\nu] }{\pi }\right)+(-k\eta)^{\nu } \left(\frac{2^{-\nu } (1-i \cot (\pi  \nu ))}{\Gamma[\nu +1]}\right)\label{bounH1}
	\end{align}
	In obtaining these asymptotic forms we keep track of the fact that in our conventions $\eta$ takes negative values lying in the range $\eta\in [-\infty,0]$. 
	
	\subsection{Field Quantization}
	\label{Qfield&wvf}
	Next we turn to quantising the field. Mode expanding the field $\Phi$ we get 
	\begin{equation}
		\label{phiop}
		\Phi(\mathbf{x},\eta)=\int \frac{d^d{\bf k}}{(2\pi)^d}\left[a_\mathbf{k}\mathbb{C}_{\nu}(-\eta)^{d\over2}H^{(1)}_{\nu}(-k\eta)e^{i\mathbf{k}.\mathbf{x}}
		+a^\dagger_\mathbf{k}\mathbb{C}_{\nu}^\ast(-\eta)^{d\over2}\left(H^{(1)}_{\nu}(-k\eta)\right)^\ast e^{-i\mathbf{k}.\mathbf{x}}\right] 
	\end{equation}
	where $a_\mathbf{k}, a^\dagger_\mathbf{k}$ are the annihilation and creation operators. 
	and 
	\begin{equation}
		\label{defC1}
		\mathbb{C}_\nu=\frac{1}{A_h\sqrt{2}}=\frac{\sqrt{\pi}}{2}e^{\frac{i\pi\nu}{2}+\frac{i\pi}{4}}
	\end{equation}
	The canonical momentum  is defined as
	\begin{equation}
		\label{defPi}
		\Pi(\mathbf{x},\eta)=\sqrt{-g}\frac{\delta \mathcal{L}}{\delta(\partial_\eta\Phi)}=-\sqrt{-g}g^{\eta\eta}\partial_\eta\Phi
	\end{equation}
	which is calculated using eq.(\ref{phiop}) as
	\begin{empheq}{multline}
		\label{piop}
		\Pi(\mathbf{x},\eta)=\frac{1}{(-\eta)^{d-1}}\int \frac{d^d{\bf k}}{(2\pi)^d}\left[a_\mathbf{k}\mathbb{C}_\nu\partial_\eta\left[(-\eta)^{d\over2}H^{(1)}_{\nu}(-k\eta)\right]e^{i\mathbf{k}.\mathbf{x}}\right.\\
		\left. +a^\dagger_\mathbf{k}\mathbb{C}_\nu^\ast\partial_\eta\left[(-\eta)^{d\over2}\left(H^{(1)}_{\nu}(-k\eta)\right)^\ast\right] e^{-i\mathbf{k}.\mathbf{x}}\right]
	\end{empheq}
	Imposing the canonical commutation relations on $\Phi$ and $\Pi$
		\begin{equation}
	\label{[Phi,Pi]}
		\left[\Phi(\mathbf{x},\eta),\Pi(\mathbf{y},\eta) \right]=i\delta^{(d)}(\mathbf{x}-\mathbf{y})
	\end{equation}
	gives rise  to  the commutation relation, 
	\begin{equation}
		\label{[a,a+]}
		[a_{\mathbf{k}},a^\dagger_{\mathbf{k}'}]=(2\pi)^d\delta^d(\mathbf{k}-\mathbf{k}')
	\end{equation}
	This can be conveniently checked  close to the Horizon ${\cal H}_+$.\\ 
	We define the vacuum state to be  one  which is   by the annihilation operators,
	\begin{equation}
		\label{adef}
		a_{\mathbf{k}}\ket{0}=0
	\end{equation}
	This   is called the Bunch Davies vacuum. 
	Multi-particle states can be  obtained by acting with products of the $a^\dagger_\mathbf{k}$ operators on $\ket{0}$.\\ 
	As was mentioned above the mode expansion etc. for the underdamped case can be obtained by the analytic continuation, eq.(\ref{analcont}). It is worth being explicit about some of the  resulting expressions here. 
	The mode expansion, eq.(\ref{phiop}), for underdamped fields,  becomes,
	\be
		\label{phiopu}
		\Phi(\mathbf{x},\eta)=\int \frac{d^d{\bf k}}{(2\pi)^d}\left[a_\mathbf{k}\mathbb{C}_{\mu}(-\eta)^{d\over2}H^{(1)}_{i \mu}(-k\eta)e^{i\mathbf{k}.\mathbf{x}}+a^\dagger_\mathbf{k}\mathbb{C}_{\mu}^\ast(-\eta)^{d\over2}\left(H^{(1)}_{i\mu}(-k\eta)\right)^\ast e^{-i\mathbf{k}.\mathbf{x}}\right] 
	\ee	
with the coefficient 
\be
\label{cmu}
\mathbb{C}_\mu= {{\sqrt \pi}\over 2}e^{-{\pi \mu\over 2}+i {\pi\over 4}}
\ee
and similarly for $\Pi({\bf x},\eta)$. 
	\subsection{Wave Function in Canonical Approach}
	\label{freewvf}
	Here onwards, we  will use the following notation to save clutter,
	\begin{equation}
		\label{defF}
		\mathcal{F}_\nu(k,\eta)=\mathbb{C}_\nu(-\eta)^{d\over2}H^{(1)}_{\nu}(-k\eta)
		\end{equation}
		and denote the boundary value, as $\eta\rightarrow 0$,  of ${\cal F}_{\nu}(k, \eta)$ to be 
		\begin{equation}
		f_\nu(k,\eta)=\mathbb{C}_\nu(-\eta)^{d\over2}\left[(-k\eta)^{-\nu } \left(-\frac{i 2^{\nu } \Gamma[\nu] }{\pi }\right)+(-k\eta)^{\nu } \left(\frac{2^{-\nu } (1+i \cot (\pi  \nu ))}{\Gamma[\nu +1]}\right)\right]
		\label{deff}
	\end{equation}
	We  will also use the notation 
	\begin{equation}
		\label{not1}
		\lim_{\eta\rightarrow0}\partial_\eta\mathcal{F}_\nu(k,\eta)=\dot{f}_\nu(k,\eta).	
	\end{equation} 
	and define 
	\be
	\label{defab}
	\alpha_\nu=\mathbb{C}_\nu\left(-\frac{i 2^{\nu } \Gamma[\nu]}{\pi }\right)\hspace{50pt}\beta_\nu=\mathbb{C}_\nu\left(\frac{2^{-\nu } (1+i \cot (\pi  \nu ))}{\Gamma [\nu +1]}\right)
	\ee
	Using the value for $\mathbb{C}_\nu$ given in eq.(\ref{defC1}) we get that 
	\be
	\label{valtab}
	\alpha_\nu= {2^{\nu-1}\Gamma[\nu]\over \sqrt{\pi}}e^{{i\pi\over2}(\nu-{1\over2})} \hspace{50pt} \beta_\nu = -{2^{-\nu-1}\sqrt{\pi}\over\Gamma[\nu+1] \sin(\pi\nu)}e^{-{i\pi\over2}(\nu+{1\over2})}
	\ee
	The late time value of ${\cal F}_\nu(k,\eta)$ can also then be written as 
	\be
	\label{altf}
	f_\nu(k,\eta) = (-\eta)^{d/2} [ \alpha_\nu (- k\eta)^{-\nu} + \beta_\nu (-k\eta)^\nu]
	\ee
	The field and momentum operator in Fourier space
	\begin{align}
		\label{phiop(k)}
		\Phi(\mathbf{k},\eta)=&\int d^d{\bf x}~\Phi(\mathbf{x},\eta)e^{-i\mathbf{k}.\mathbf{x}}\\
		\Pi(\mathbf{k},\eta)=&\int d^d{\bf x}~\Pi(\mathbf{x},\eta)e^{-i\mathbf{k}.\mathbf{x}}\label{piop(k)}
	\end{align}  
	can be written in terms of $\mathcal{F}_{\nu}(k,\eta)$ as
	\begin{align}
		\label{Phik}
		\Phi(\mathbf{k},\eta)=~&a_\mathbf{k}\mathcal{F}_\nu(k,\eta)+a^\dagger_{-\mathbf{k}}(\mathcal{F}_\nu(k,\eta))^\ast\\
		\Pi(\mathbf{k},\eta)=~&\frac{1}{(-\eta)^{d-1}}\left[a_\mathbf{k}\partial_\eta\mathcal{F}_\nu(k,\eta)+a^\dagger_{-\mathbf{k}}\partial_\eta(\mathcal{F}_\nu(k,\eta))^\ast\right]\label{Pik}
	\end{align}
	The creation and annihilation operators can be represented in terms of $\Phi(\mathbf{k},\eta)$ and $\Pi(\mathbf{k},\eta)$ as
	\begin{align}
		\label{a(k)}
		a_{\textbf{k}}=&\frac{\partial_\eta(\mathcal{F}_\nu(k,\eta))^\ast\Phi(\textbf{k},\eta)-(-\eta)^{d-1}(\mathcal{F}_\nu(k,\eta))^\ast\Pi(\textbf{k},\eta)}{\mathcal{F}_\nu(k,\eta)\partial_\eta(\mathcal{F}_\nu(k,\eta))^\ast-(\mathcal{F}_\nu(k,\eta))^\ast\partial_\eta\mathcal{F}_\nu(k,\eta)}\\ a^\dagger_{-\textbf{k}}=&\frac{\partial_\eta\mathcal{F}_\nu(k,\eta)\Phi(\textbf{k},\eta)-(-\eta)^{d-1}\mathcal{F}_\nu(k,\eta)\Pi(\textbf{k},\eta)}{(\mathcal{F}_\nu(k,\eta))^\ast\partial_\eta\mathcal{F}_\nu(k,\eta)-\mathcal{F}_\nu(k,\eta)\partial_\eta(\mathcal{F}_\nu(k,\eta))^\ast}\label{a+(k)}
	\end{align} 
	Now we construct the ground state wave function in the basis of eigenstates of the field operator. 
	We will work with eigenstates of the operator $\Phi({\bf k},\eta)$, eq.(\ref{Phik}). Note that these operators are time dependent, i.e. dependent on the  coordinate 
	$\eta$.
	The corresponding  eigenstates satisfy the condition,
	\be
	\label{eigphia}
	\Phi({\bf k},\eta)\ket{\phi({\bf k},\eta)}=\phi({-\bf k},\eta) \ket{\phi({\bf k},\eta)}
	\ee
	The corresponding bra $\bra{\phi({\bf k},\eta)}$ then satisfies the condition
	\be
	\label{eigphib}
	\bra{\phi({\bf k},\eta)}\Phi({\bf k}, \eta)=  \phi({\bf k},\eta)\bra{\phi({\bf k},\eta)}
	\ee
	Schematically denoting these eigenstates as   $\ket{\varphi}$ we have that the wave function is given by 
%
	\begin{equation}
		\label{defwvf}
		\psi[\varphi]=\bra{\varphi}\ket{0}
	\end{equation}  
%
%
	The momentum operator  acts on this  wave function as follows:
	\be
	\label{eigpi}
	\Pi(\mathbf{k},\eta)\psi[\varphi]=-i (2\pi)^d\dfrac{\delta}{\delta \varphi(-\textbf{k},\eta)}\psi[\varphi]
	\ee
	Using the fact that the Bunch Davies vacuum is annihilated by   $a_\mathbf{k}$, eq.(\ref{adef}), and the  expression for the annihilation operators in terms of $\Phi(\mathbf{k},\eta)$ and $\Pi(\mathbf{k},\eta)$ given in eq.(\ref{a(k)}), we get a differential equation that $\psi$ should satisfy
	\begin{equation}
		\label{eq.psi}
		\partial_\eta(\mathcal{F}_\nu(k,\eta))^\ast\varphi(\mathbf{k},\eta)\psi[\varphi]+i(-\eta)^{d-1}(\mathcal{F}_\nu(k,\eta))^\ast\frac{\delta\psi[\varphi]}{\delta\varphi(-\mathbf{k},\eta)}=0
	\end{equation}
	The solution to this equation is given by
	\begin{equation}
		\label{psi[phi]}
		\psi[\varphi,\eta]=\mathcal{N}\exp\left[i\int \frac{d^d\mathbf{k}}{(2\pi)^d} ~\varphi(\mathbf{k},\eta)\left(\frac{\partial_\eta(\mathcal{F}_\nu(k,\eta))^\ast}{2(-\eta)^{d-1}(\mathcal{F}_\nu(k,\eta))^\ast}\right)\varphi(-\mathbf{k},\eta)\right]
	\end{equation}
	It is worth drawing attention to the fact  that the wave function of the Bunch Davies vacuum is both  a functional of  $\varphi({\bf k},\eta)$, the eigenvalues of the field operators 
	$\Phi({\bf k},\eta)$, and in addition a  function of time. If we were more explicit  eq.(\ref{psi[phi]}) would  actually have been written as 
	\begin{equation}
		\label{psifa}
		\psi[\varphi({\bf k},\eta), \eta]=\mathcal{N}\exp\left[i\int \frac{d^d\mathbf{k}}{(2\pi)^d} ~\varphi(\mathbf{k},\eta)\left(\frac{\partial_\eta(\mathcal{F}_\nu(k,\eta))^\ast}{2(-\eta)^{d-1}(\mathcal{F}_\nu(k,\eta))^\ast}\right)\varphi(-\mathbf{k},\eta)\right]
	\end{equation}
The explicit time dependence arises because of the time dependence of the dS background.

	The probability functional following from eq.(\ref{psi[phi]}) is given by
	\begin{equation}
		\label{psi^2}
		|\psi[\varphi,\eta]|^2=|\mathcal{N}|^2\exp\left[-\int \frac{d^d\mathbf{k}}{(2\pi)^d} ~\varphi(\mathbf{k},\eta)\left(\frac{1}{2|\mathcal{F}_\nu(k,\eta)|^2}\right)\varphi(-\mathbf{k},\eta)\right]
	\end{equation}
	where we have used the relation
	\begin{equation}
		\label{id1}
		\mathcal{F}_\nu(k,\eta)\partial_\eta(\mathcal{F}_\nu(k,\eta))^\ast-(\mathcal{F}_\nu(k,\eta))^\ast\partial_\eta\mathcal{F}_\nu(k,\eta)=i(-\eta)^{d-1}
	\end{equation}
	This relation follows from the fact that $\mathcal{F}_\nu(k,\eta)$ is defined in eq.(\ref{defF})  and therefore satisfies the equation 
	\begin{equation}
		\label{EOMk}
		\eta^2\partial_\eta^2\mathcal{F}_\nu-(d-1)\eta\partial_\eta\mathcal{F}_\nu+\eta^2k^2\mathcal{F}_\nu+M^2\mathcal{F}_\nu=0
	\end{equation}
	The normalization constant ${\cal N}$ can be fixed by requiring that the total probability given by $\int \mathcal{D}\varphi|\psi[\varphi,\eta]|^2$  is unity. 
	

The bulk correlators can be derived from eq.(\ref{Phik}) and eq.(\ref{Pik}) as
	\begin{align}
		\label{<Phi,Phi>}
		\langle\Phi(\mathbf{k},\eta)\Phi(\mathbf{k}',\eta)\rangle=&(2\pi)^d\mathcal{F}_\nu(k,\eta)(\mathcal{F}_\nu(k,\eta))^\ast\delta^d(\mathbf{k}+\mathbf{k}')\\
		\langle\Pi(\mathbf{k},\eta)\Pi(\mathbf{k}',\eta)\rangle=&\frac{(2\pi)^d}{(-\eta)^{2(d-1)}}\partial_\eta\mathcal{F}_\nu(k,\eta)\partial_\eta(\mathcal{F}_\nu(k,\eta))^\ast\delta^d(\mathbf{k}+\mathbf{k}')\label{<Pi,Pi>}\\
		\langle\Phi(\mathbf{k},\eta)\Pi(\mathbf{k}',\eta)\rangle=&\frac{(2\pi)^d}{(-\eta)^{d-1}}\mathcal{F}_\nu(k,\eta)\partial_\eta(\mathcal{F}_\nu(k,\eta))^\ast\delta^d(\mathbf{k}+\mathbf{k}')\label{<Phi,Pi>}
	\end{align}
	Let us note again that the formulae for the underdamped case can be obtained by analytic continuation, eq.(\ref{analcont}). For clarity, and also to lay down our notation let us discuss some of the resulting expressions. 
	We define in this case the function 
	\be
	\label{defFu}
	{\cal F}_{\mu}(k,\eta)=\mathbb{C}_\mu(-\eta)^{d\over2} H_{i\mu}^{(1)}(-k \eta)
	\ee
	As $\eta\rightarrow 0$ its value  becomes 
	\be
	\label{deffu}
	f_{\mu}(k,\eta)=(-\eta)^{d\over2} [\beta_\mu (-k\eta)^{-i\mu}+\alpha_\mu (-k \eta)^{i \mu}]
	\ee
	where 
	\be
	\label{defabuC}
		\alpha_\mu=\mathbb{C}_\mu\left(\frac{2^{-i\mu } (1+ \coth (\pi  \mu ))}{\Gamma [i\mu +1]}\right)\hspace{50pt}\beta_\mu=\mathbb{C}_\mu\left(-\frac{i 2^{i\mu } \Gamma[i\mu ]}{\pi }\right)
	\ee
	Using \eq{cmu}, we obtain
	\be
	\label{defabu}
	\alpha_\mu=e^{-{\pi \mu\over 2}+i {\pi\over 4}}\left(\frac{{\sqrt \pi}(1+\coth (\pi  \mu ))}{2^{1+i\mu}\Gamma[1+i\mu]}\right)\hspace{50pt}\beta_\mu=-e^{-{\pi \mu\over 2}+i {\pi\over 4}}\left(\frac{i 2^{i\mu-1} \Gamma [i\mu] }{\sqrt{\pi} }\right)
	\ee
		\subsection{Wave Function from Path Integral}
		\label{psipathint}
		
		The wave function   can also be obtained by carrying out a path integral, as has been discussed in \cite{Malda-NG} and \cite{opdic}. 
		The path integral is evaluated as a functional of the  boundary value for the scalar. Here we will be interested in the wave function at late times,  as $\eta\rightarrow 0$. As we will see below, a  suitable boundary condition also needs to be imposed at ${\cal H}_+$. 
		This way of obtaining  the wave function  is related to the calculation of the partition function in AdS space as a functional of the boundary values of fields, after a suitable analytic continuation. Here we will directly describe the calculation in dS space (more correctly with  the time coordinate $\eta$ being given a small imaginary part, as will be explained below). 
		
		For the scalar theory  the wave function at a late instant of time $|\eta_1|\ll 1$ is given by the path integral  
		\be
		\label{efs}
		\psi[\varphi,\eta_1]=\int_{{\cal H}_+}^{\eta_1} D\phi e^{i {\hat S}}
		\ee
		 ${\hat S}$ denotes the action; For a free scalar theory it   is given in eq.(\ref{action}). 
		The limits indicate that  the path integral is to be carried out from ${\cal H}_{+}$, where $\eta\rightarrow -\infty$,  to $\eta_1$. 
		
		In the subsequent discussion, at the risk of causing some confusion perhaps, we will sometimes denote the resulting wave function by 
		\be
		\label{wfe}
		\psi=e^{i S}
		\ee
		At tree level $S$ is the on-shell value of the bulk action ${\hat S}$, obtained after imposing suitable boundary conditions. But at higher orders they will be different. 
		
		In eq.(\ref{efs}) the argument of the wave function, which we are denoting schematically as 
		$\varphi$, equals the value of the field at $\eta=\eta_1$ in the path integral. 
		In turn these are equal to the eigenvalue of the field operator $\Phi(\mathbf{x},\eta_1)$, which we had denoted as $\varphi(\mathbf{x},\eta_1)$ in subsection \ref{freewvf}, see eq.(\ref{eigphia}).  
		The Fourier transform of the field $\varphi({\bf x}, \eta)$ is  
		\be
		\label{ftpf}
		\varphi({\bf k}, \eta)= \int d^d {\bf x} \varphi({\bf x},\eta) e^{-i {\bf k}\cdot {\bf x}}
		\ee
%
		
		The boundary condition in the path integral at ${\cal H}_+$ is that $\phi({\bf k}, \eta)$ vanishes, as $\eta\rightarrow -\infty$, after $\eta$ is given a small  imaginary piece as we explain below.  This will give the correct wave function in the Bunch Davis vacuum for the free theory as we will see below. 
		
		The path integral for the free theory can be simply calculated by evaluating the on shell action for a solution meeting the two boundary conditions mentioned above.  The  on-shell solution for $\phi({\bf k},\eta)$ in general takes the form 
		\be
		\label{gensola}
		\phi({\bf k},\eta)=\phi_1({\bf k}) {\cal F}_\nu({ k}, \eta) + \phi_2({\bf k}) (\mathcal{F}_\nu(k,\eta))^\ast
		\ee
		where ${\cal F}_\nu(k,\eta)$ is defined in eq.(\ref{defF}), and $\phi_1({\bf k}),  \phi_2({\bf k}) $ are independent of $\eta$. 
		
		We now  continue  this solution  to  complex values of $\eta$ by giving $\eta$  an imaginary part, 
		\be
		\label{cone}
		\eta\rightarrow \eta(1-i\epsilon)
		\ee
		where $\epsilon>0, \epsilon \ll 1$ 
		and analyse the resulting   behaviour  near $\eta\rightarrow -\infty$. 
		
		From eq.(\ref{defF}), we see, using eq.(\ref{horH}) that 
		\be
		\label{limfs}
		(\mathcal{F}_\nu(k,\eta))^\ast \sim e^{i k \eta(1-i\epsilon)}
		\ee
		and therefore vanishes as $\eta \rightarrow -\infty$. 
		In contrast ${\cal F}_\nu({k},\eta)\sim e^{-ik\eta (1-i\epsilon)}$ and therefore blows up as $\eta\rightarrow -\infty$.
		The boundary condition at ${\cal H}_+$ we impose is that $\phi({\bf k},\eta)$ must vanish as $\eta\rightarrow -\infty$ after the analytic continuation above. 
		This then  sets $\phi_1({\bf k})=0$ in eq.(\ref{gensola}) leading to  
		\be
		\label{fsola}
		\varphi({\bf k},\eta)=(\mathcal{F}_\nu(k,\eta))^\ast  \phi_2({\bf k})
		\ee

		The on-shell action can then be easily calculated as a boundary term at $\eta=\eta_1$ and gives the wave function
		\be
		\label{wfetaa}
		\psi[\varphi,\eta_1]={\cal N} 
		\exp{i \int {d^d{\bf k}\over (2\pi)^d}\phi_2({\bf k})(\mathcal{F}_\nu(k,\eta_1))^\ast{\partial_\eta(\mathcal{F}_\nu(k,\eta_1))^\ast\over 2 (-\eta_1)^{d-1}} \phi_2(-{\bf k})}
		\ee
		
		We see that this agrees with the wave function  obtained in eq.(\ref{psi[phi]}) after we express $\phi_2(\mathbf{k})$ in terms of $\varphi(\mathbf{k},\eta)$ using eq.(\ref{fsola}) and set $\eta=\eta_1$.

	\subsection{Inclusion of Interactions} 
		\label{interaction1}
		Once interactions are included, the ground state wave function will acquire non-Gaussian terms. 
		
		As an example, consider adding an $\phi^n$ term in the action, which gives  an extra contribution to the action,  
		\begin{equation}
			\label{intterm}
			\delta S_n=\lambda\int \frac{d^{d}{\bf x}d\eta}{(-\eta)^{d+1}}\phi^n(\mathbf{x},\eta)
		\end{equation} 
		
		The correction  at leading order in $\lambda$ in the wave function can be simply calculated by inserting the on-shell solution eq.(\ref{fsola}) discussed in the previous section, into $\delta S$. 
		This gives,
		\begin{equation}
				\label{genpsipert}
				\psi_I[\phi(\mathbf{k},\eta_1),\eta_1]=\exp{\int\frac{d^d\mathbf{k}}{(2\pi)^d}\phi(\mathbf{k},\eta_1)\left(\frac{i}{2(-\eta_1)^{d-1}}\frac{\partial_\eta({\mathcal{F}_\nu}(k,\eta_1))^{\ast}}{(\mathcal{F}_\nu(k,\eta_1))^{\ast}}\right)\phi(-\mathbf{k},\eta_1)}\times e^{i\delta S_n}
			\end{equation}
			where $\delta S_n$ is found to be
			\begin{equation}
				\label{deltasn}
				i\delta S_n=i\lambda\int \prod_{i=1}^{n}\frac{d^d{\mathbf{k}_i}}{(2\pi)^d}(2\pi)^d\delta^{(d)}(\mathbf{k}_1+...+\mathbf{k}_n)\left[I(k_1,k_2,..,k_n,\eta_1)\prod_{i=1}^{n}\frac{\phi(\mathbf{k}_i,\eta_1)}{(\mathcal{F}_\nu(k_i,\eta_1))^\ast}\right]
			\end{equation}
			with
			\begin{equation}
				\label{Iexpression}
				I(k_1,k_2,..,k_n,\eta_1)=\int_{-\infty}^{\eta_1}\frac{d\eta'}{(-\eta')^{d+1}}\prod_{i=1}^{n}(\mathcal{F}_\nu(k_i,\eta'))^{\ast}
			\end{equation}
		Depending on the value of $\nu$ this integral can be  divergent as $\eta_1\rightarrow 0$. 
These divergences are discussed further in section \ref{divergence} and can be understood as arising due to local ``counterterms" or due to operator mixing. 
After subtracting these divergences we get finite value for the  integral which we denote subsequently as $I(k_1, k_2, \cdots, k_n)$. 
		We have denoted the resulting wave function with the suffix $I$ above because it can also be calculated in the operator formalism by doing 
		time dependent perturbation theory in the interaction picture (called the "in-in formalism"). These points are discussed more fully in Appendix \ref{appenboun}.

\begin{figure}[h]
				\centering
				\begin{tikzpicture}
					\draw (0,0)--(8,0);
					\draw (3,-3)--(6,0);
					\draw (3,-3)--(4,0);
					\draw (3,-3)--(2,0);
					\node[text width=3cm] at (4,-3.5) {$(\mathbf{x}',\eta')$};
					\node[text width=3cm] at (3.2,0.3) {$\mathbf{x}_1$};
					\node[text width=3cm] at (5.2,0.3) {$\mathbf{x}_2$};
					\node[text width=3cm] at (7.2,0.3) {$\mathbf{x}_3$};
					\node[text width=3cm] at (9.9,0) {$\eta'=0$};
				\end{tikzpicture}
			\caption{Witten diagram for three point correlator in "in-in formalism".}
			\label{Witten3}
			\end{figure}
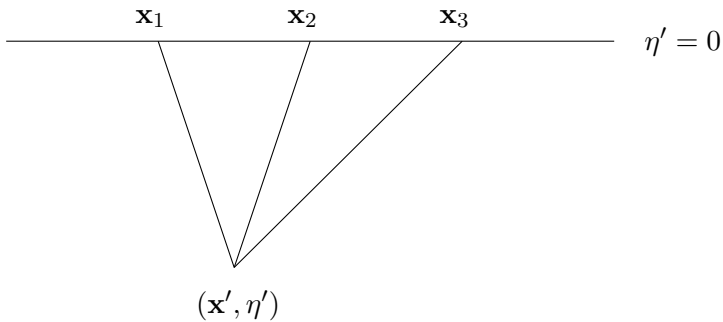

			The cubic case, with  $n=3$ is worth discussing more explicitly. This corresponds to the Witten diagram [see Figure \ref{Witten3}] where the lines indicate the bulk to boundary propagators. We can convert eq.(\ref{Iexpression}) for $n=3$ to an integral over three modified Bessel functions using the transformation properties derived in Appendix \ref{analI}. Up to cut-off dependent terms we get, 
			\begin{equation}
				\label{IexpK}
				I(k_1,k_2,k_3)=\frac{8}{\pi^3}(\mathbb{C}_\nu^\ast)^{3} e^{{i 3 \pi \over 2}\nu^\ast} e^{{i\pi\over 2}(3-\frac{d}{2})}\int_{0}^{\infty}dt~t^{\frac{d}{2}-1}K_{\nu^{\ast}}(k_1t)K_{\nu^{\ast}}(k_2t)K_{\nu^{\ast}}(k_3t)
			\end{equation}
			Since we neglecting the cut-off dependent terms here this integral, in general, is to be taken  as being defined by analytic continuation starting from values of $\nu$ where it converges. After such an analytic continuation, if needed, it agrees with the standard form  for a  three point function in a CFT, as is discussed in \cite{Baut-3ptmom}.

	\section{Holography}
	\label{holo}
	In this paper we will explore the version of holography which relates the late time  wave function of the bulk theory  in the Poincaré patch, for  the Bunch Davies vacuum,  to the generating functional for correlation functions of the dual  CFT.
	We work in Poincaré coordinates, eq.(\ref{Poing}), and late time  corresponds to taking $\eta\rightarrow 0^-$. 
	
	The essential idea of this version of holography, \cite{Malda-NG}, is  similar to that in $AdS/CFT$, with the wave function in the bulk playing the role of the partition function in AdS/CFT.
	To be more precise, in the $dS$  case the late time value  of the bulk fields are  to be appropriately identified with the sources in the dual CFT and the wave function as a functional of these late sources  then is to be equated with the partition function in the CFT in the presence of the same  sources. 
	
	Denoting a generic source by $S$ the wave function at late times can be expanded in a power series in $S$. 
	This is analogous to carrying out a Taylor series expansion of a function. Schematically, the wave function at late times is then given by a sum of terms consisting of   various powers of the  sources $S({\bf x})$ and takes   the form:
	\be
	\label{genfa}
	\log(\Psi[S,\eta_1])=\sum_{n=2}^{\infty}{1\over n!}\int \prod_{j=1}^n (d^d{\bf x}_j S({\bf x}_j)) \l \mathbb{O}({\bf x}_1)\mathbb{O}({\bf x}_2) \cdots \mathbb{O}({\bf x}_n)\r
	\ee
	From the bulk point of view   the functions which appear in this expansion,
	$\l \mathbb{O}({\bf x}_1)\mathbb{O}({\bf x}_2) \cdots \mathbb{O}({\bf x}_n)\r$, are just   coefficients ; the only difference with a conventional Taylor series is that since the wave function is a functional of the  bulk field, and therefore of the sources $S$, the coefficients are now actually coefficient functions, i.e. functions of the spatial locations, ${\bf x}_1,\cdots {\bf x}_n$. The statement of holography is that these coefficient functions are in fact correlation functions of various fields in the dual CFT. And it is   with this  correspondence in mind that we have suggestively denoted the coefficients  as $\l \mathbb{O}({\bf x}_1)\mathbb{O}({\bf x}_2) \cdots \mathbb{O}({\bf x}_n)\r$ above. 
	Before proceeding we note that from eq.(\ref{genfa}) it follows that 
	\be
	\label{genfab}
	{\delta^n \log(\Psi) \over \delta S({\bf x}_1) \delta S({\bf x}_2) ...\delta S({\bf x}_n)}=\l \mathbb{O}({\bf x}_1)\mathbb{O}({\bf x}_2) \cdots \mathbb{O}({\bf x}_n)\r
	\ee

	In general for every bulk field one would obtain a different source  $S_i$ and a correspondingly  different field $\mathbb{O}_i$ in the boundary. 
	Bulk scalars would give rise to scalar fields, the metric to a symmetric two-tensor $T_{ij}$ which is the stress tensor of the CFT, etc. 

	It is worth mentioning, before we proceed, that in 
	 $AdS /CFT$ we usually refer to the  bulk fields  as  being related to sources for operators in the dual CFT. Here we are instead referring to them as sources for fields in the CFT. This is   because, as we will see in section \ref{factorsofi}, and has already been noted in \eq{posspa}, the boundary theory we are dealing with is Euclidean and in general  does not admit a Lorentzian continuation. 
	 Below,  we show how to relate the boundary values of bulk scalar field of various masses  to sources in the CFT.
	And argue  that the  wave function in the Bunch Davies vacuum leads to the correlation functions of the CFT satisfying various  Ward identities including those of conformal invariance, translational invariance and local scale invariance. 
	

		
		
		

We first discuss overdamped case below and  subsequently turn towards heavy fields.
		\subsection{Overdamped Case}
		\label{over}
		In this case the holographic dictionary is quite analogous to the $AdS/CFT$ case as was mentioned above. 
		In particular we will be interested in relating the wave function in the basis of eigenstates of the field operator to the generating functional in the CFT. 
		Following notation of section \ref{Qfield&wvf} we denote the field operator by $\Phi({\bf x},\eta)$, its eigenstates by $\ket{\phi({\bf x},\eta)}$ and their eigenvalues by 
		$\phi({\bf x},\eta)$, so that 
		\be
		\label{evea}
		\Phi({\bf x},\eta)\ket{\phi({\bf x},\eta)}=\phi({\bf x},\eta)\ket{\phi({\bf x},\eta)}
		\ee
		The wave function in this basis of eigenstates is denoted by $\Psi[\phi({\bf x},\eta), \eta]$,  and given by 
		\be
		\label{expa}
		\Psi[\phi({\bf x},\eta), \eta]=\braket{\phi({\bf x},\eta)}{0}
		\ee
		where ${\ket{0}}$ is the Bunch Davies vacuum. 
		We note that besides its dependence on the $\phi({\bf x},\eta)$ the wave function also depends explicitly on time $\eta$. 
		
		The holographic correspondence relates this wave function to the generating function in the CFT once a source is identified in terms of the value the bulk field takes at late times, $\phi({\bf x},\eta)$. We denote the time  at which we are calculating the wave function to be $\eta=\eta_1$ below. Note that $\eta_1\rightarrow 0$.

		Let us first start  with the free field case,  we  will  add interactions in the next subsection. 
		The wave function for the free theory was calculated in general in eq.(\ref{psi[phi]}). 
		We write the late time value of the field $\phi(\mathbf{k},\eta)$ as
		\begin{equation}
			\label{oversource}
			\phi(\mathbf{k},\eta)={f^\ast_\nu(k,\eta) \over f_{\nu}^\ast(k,\eta_1)  }        (-\eta_1)^{{d\over2}-\nu} \hat{\phi}(\mathbf{k})
		\end{equation}
		And identify ${\hat \phi}({\bf k})$ with the source in the boundary theory. 
		
		Note that this is very similar    to what is done in the AdS case. 
		For Euclidean $AdS_{d+1}$, working in Poincaré coordinates, with $AdS$ radius set to unity, 
		\be
		\label{meteuads}
		ds^2={1\over z^2}[dz^2 +(dx^i)^2]
		\ee
		we get that the solution to the wave equation for a scalar of mass $M$ which vanishes at the Poincaré horizon is given by 
		$\phi({\bf k},z)e^{i{\bf k}\cdot {\bf x}}$ where $\phi({\bf k},z)$ is given in terms of the modified Bessel function as 
		\be
		\label{solab}
		\phi({\bf k},z)={z^{d\over2}K_\nu(kz)\over \epsilon^{d\over2} K_\nu(k\epsilon)} \epsilon^{{d\over2}-\nu}{\hat \phi}({\bf k})
		\ee
		with $\nu=\sqrt{{d^2\over4}+M^2}$ and $z=\epsilon$ being the location of the boundary.  ${\hat \phi}({\bf k})$ which appears on the RHS of eq.(\ref{solab}) then acts like  the source
		term for CFT correlation functions. 		
		We should also mention that as a result of the identification in  \footnote{A similar comment applies  for the AdS case, eq.(\ref{solab}).} eq.(\ref{oversource}) the 
		bulk field at time $\eta=\eta_1$ is related to the source ${\hat \phi}$ by the relation in position space
		\be
		\label{relsb}
		\phi({\bf x},\eta_1)=(-\eta_1)^{{d\over2}-\nu} {\hat \phi}({\bf x})
		\ee
		We see that this relation is local in position space along the hypersurface $\eta=\eta_1$.  

		Let us also note before proceeding that from eq.(\ref{bounH2}), eq.(\ref{bounH1})  we see that  the two solutions to the wave equation asymptotically go like 
		$(-\eta)^{{d\over2}-\nu}$, $(-\eta)^{{d\over2}+\nu}$, and 
		since $\nu$ is real and positive the mode falling off as $(-\eta)^{{d\over2}-\nu}$, which is the analogue of the non-normalisable mode in AdS space, will dominate near $\eta\rightarrow 0$ in $f_\nu(k,\eta)$.

		Comparing with eq.(\ref{fsola}) we see that 
		\be
		\label{valso}
		\phi_2({\bf k})={(-\eta_1)^{{d\over2}-\nu} \over f_{\nu}^\ast(k,\eta_1)  } {\hat \phi}({\bf k})
		\ee
		Inserting this in eq.(\ref{psi[phi]})  gives the wave function at late time 
		\begin{equation}
			\label{bounwvf}
			\psi[\hat{\phi},\eta]=\mathcal{N}\exp\left[{i \over 2}\int \frac{d^d\mathbf{k}}{(2\pi)^d} ~
			(-\eta_1)^{1-2\nu}{f_{\nu}^{\ast}(k,\eta)\dot{ f}_{\nu}^{\ast}(k,\eta) \over f_{\nu}^{\ast}(k,\eta_1)^2 } {\hat \phi}(\mathbf{k}){\hat \phi}(-\mathbf{k})\right]
		\end{equation}
       where $f_\nu({ k},\eta)$ is given in eq.(\ref{deff}). We see that at $\eta=\eta_1$, eq.(\ref{bounwvf}) reads
       \be
       \label{bounwvf1}
        \psi[\hat{\phi},\eta_1]=\mathcal{N}\exp\left[{i \over 2}\int \frac{d^d\mathbf{k}}{(2\pi)^d} ~
        (-\eta_1)^{1-2\nu}{\dot{ f}_{\nu}^{\ast}(k,\eta_1) \over f_{\nu}^{\ast}(k,\eta_1) } {\hat \phi}(\mathbf{k}){\hat \phi}(-\mathbf{k})\right]
        \ee
		Putting in the explicit form for $f_\nu({k},\eta)$ for $\eta=\eta_1$ and then expanding about $\eta_1=0$ leads to
		\begin{align}
			\label{bounwvpexp}
			\psi[\hat{\phi},\eta]=\mathcal{N}\exp\left[-\frac{i}{2}\int \frac{d^d\mathbf{k}}{(2\pi)^d} ~\hat{\phi}(\mathbf{k})\left[\left(\frac{d}{2}-\nu\right)(-\eta_1)^{-2\nu}+\frac{\beta^\ast_\nu}{\alpha^\ast_\nu}(2\nu) k^{2\nu}\right]\hat{\phi}(-\mathbf{k})\right]
		\end{align}
		where we have dropped sub-leading terms and the coefficients $\alpha_\nu, \beta_{\nu}$ above are given in \eq{valtab}.
		Note that  in obtaining the last expression we have done an expansion at small $\eta$. 
		
		The first term in the wave function is a contact term. Neglecting  it for now  we have 
		\begin{equation}
			\label{wvfOO}
			\psi[\hat{\phi},\eta_1]=\mathcal{N}\exp\left[\frac{1}{2}\int \frac{d^d\mathbf{k}}{(2\pi)^d} \frac{d^d\mathbf{k'}}{(2\pi)^d}~\hat{\phi}(\mathbf{k})\langle O(\mathbf{k})O(\mathbf{k}')\rangle\hat{\phi}(\mathbf{k}')\right]
		\end{equation}
	 	where the coefficient function 
	 	\begin{equation}
	 		\label{<OkOk'>}
	 		\langle O(\mathbf{k})O(\mathbf{k}')\rangle=-i (2\pi)^d\delta^d(\mathbf{k}+\mathbf{k}')\frac{\beta_\nu^\ast}{\alpha_\nu^\ast}2\nu k^{2\nu}
	 	\end{equation}
		Note that the momentum dependence on the RHS is of the correct form for the position space  field $O({\bf x})$ to have  dimension, 
		\begin{equation}
			\label{overdim}
			\Delta=\frac{d}{2}+\nu
		\end{equation}
		where $\nu$ is  given in eq.(\ref{nu}).
		
%
%
%
		
	Also note that the coefficient in eq.(\ref{<OkOk'>}) is given by, eq.(\ref{valtab}), 
	\be
	\label{ratalbe}
	{\beta_\nu^{\ast}\over \alpha_\nu^{\ast}}=-{\pi 2^{-2\nu} e^{i \pi \nu}\over \nu(\Gamma[\nu])^2\sin{\pi\nu}}
	\ee
	and as  a result eq.(\ref{<OkOk'>}) takes the form
	\begin{equation}
	 		\label{twptb}
	 		\langle O(\mathbf{k})O(\mathbf{k'})\rangle=i (2\pi)^d\delta^d(\mathbf{k}+\mathbf{k}'){\pi 2^{-2\nu+1}e^{i\pi\nu} \over (\Gamma[\nu])^2\sin{\pi\nu}}k^{2\nu}
	\end{equation}
	And the wave function eq.(\ref{wvfOO}) (without the contact term) becomes
	\be
	\label{wfab}
	\psi[\hat{\phi},\eta_1]=\mathcal{N}\exp\left[-{ 2^{-2\nu}\pi  \over \Gamma[\nu]^2} \bigl(1-i \cot(\pi \nu)\bigr) \int \frac{d^d\mathbf{k}}{(2\pi)^d} ~\hat{\phi}(\mathbf{k})\hat{\phi}(\mathbf{-k})k^{2\nu}\right]
	\ee
The negative sign in  the  real part of the exponent leads to an exponentially decaying wave function and thus a normalisable wave function. Of course this is to be expected, but it is worth mentioning a few  details leading to this conclusion. 
Note that  by evaluating eq.(\ref{oversource}) at $\eta=\eta_1$ we get 
\be
\label{reab}
\phi({\bf k},\eta_1)=(-\eta_1)^{{d\over2}-\nu} {\hat \phi}({\bf k})
\ee
The field $\phi({\bf x},\eta)$ is real and thus $\phi({\bf k},\eta_1)^{\ast} =\phi(-{\bf k},\eta_1)$, from here it follows that ${\hat \phi}$ also satisfies the relation 
\be
\label{relhataa}
{\hat \phi}^{\ast}({\bf k})={\hat \phi}(-{\bf k})
\ee
and therefore ${\hat \phi}(-{\bf k}){\hat\phi}({\bf k})=|{\hat \phi}({\bf k})|^2$ is positive definite, leading to the conclusion stated above.

Carrying out a Fourier transform of eq.(\ref{twptb}) gives in position space 
\be
\label{posspa}
\l O({\bf x})O({\bf y})\r=e^{i \pi (2\nu-1)\over 2} 2 \nu {\Gamma[{d\over2}+\nu]\over \pi^{d\over2}\Gamma[\nu]} {1\over |{\bf x}-{\bf y}|^{d+2\nu}}
\ee
The phase factor in front shows that in general the two point function violates reflection positivity. 

 A consequence of reflection positivity, see for example \cite{Refpos, tasilec}, is that 
	\be
	\label{condrefa}
	\l O({\bf x})O({\bf x}^\theta)\r\geq 0
	\ee
	where $ {\bf x}^\theta$ is related to the point ${\bf x}$ by reflection about a coordinate axis, say ${\bf x}^1=0$. 
	This condition is clearly violated by eq.(\ref{posspa}) due to the phase factor in front- we can take the two points to be at $(0, {\bf x}_2, \cdots {\bf x}_n)$ and $(0,-{\bf x}_2,\cdots,,-{\bf x}_n)$ for example. 
	
	The absence of reflection positivity means that in the dS case the correlation functions of the Euclidean field theory cannot be continued to the correlators of a Hermitian operator in a Lorentzian field theory
	
	Note that  the momentum space correlator eq.(\ref{twptb}) is valid for non-integer $\nu$. When $\nu\in {\mathbb Z}$ the leading no-contact term comes from a term going like $k^{2\nu}\log(k)$. The position space correlator continues to be given by eq.(\ref{posspa}) in this case though, as is discussed in appendix \ref{App.analcont}, \eq{dsx2pt}.


	A few more comments are worth making here. 
	The contact term in eq.(\ref{bounwvpexp}) cannot actually be removed in the dS case, as was discussed earlier. 
	It gives rise to a local term in position space which is divergent when $\eta_1\rightarrow 0$, 
	\be
	\label{simwf}
	\psi\sim \exp{-{i \over 2}\left({d\over 2}-\nu\right)(-\eta_1)^{-2\nu}\int d^d{\bf x} {\hat{\phi}({\bf x})}^2}
	\ee
	 and results in a contact term in the two point correlator $\l O({\bf x})O({\bf y})\r$ with an imaginary coefficient. This term, like the finite one discussed above,  is also invariant under scaling.
	 To see this note that  from eq.(\ref{overdim}), and the fact that ${\hat \phi}({\bf x})$ is the source for $O({\bf x})$, it follows that $\hat{\phi}({\bf x})$ has dimension 
	 \be
	 \label{dims}
	 \Delta_-={d\over2}-\nu
	 \ee and transforms under a scaling transformation, ${\bf x}\rightarrow {\bf x}/\lambda$, by ${\hat \phi}({\bf x})\rightarrow \lambda^{\Delta_-} {\hat \phi}(\lambda {\bf x})$, since this would leave $\int d^dx {\hat \phi}({\bf x}) O({\bf x})$ invariant. It then follows that the contact term is also invariant under the scaling transformation once the cut-off $\eta_1$ transforms as $\eta_1\rightarrow {\eta_1\over \lambda}$.
	 

	 Also, and this point will be important when we will consider interactions, for example in the next subsection, we are interested in calculating the wave function in the basis of field eigenstates $\ket{\phi({\bf x},\eta)}$, as was mentioned at the beginning of this section, and in expressing it in terms of a source ${\hat \phi}$. In the free theory this source was defined by the relation eq.(\ref{oversource}) in terms of the bulk field $\phi({\bf k},\eta)$. Once interactions are added we will continue to take source to be defined by the relation eq.(\ref{oversource}), i.e. the definition of the source in terms of $\phi({\bf x},\eta)$ will be uncorrected by the interactions. We will find that with this definition the coefficient functions satisfy the standard CFT Ward identities for fields in a CFT with dimensions $\Delta_+$. 
\subsubsection{Other Representations in the Overdamped Case}\label{altrepov}
Here we consider some other representations of the wave function. For a more extensive discussion see Appendix \ref{app.altov}. 
\paragraph{Momentum Space Representation} In the overdamped case the two solutions at late time both fall off as $(-\eta)^{{d\over2}\pm \nu}$, as discussed above. This is actually reminiscent of the behaviour in AdS space for a field with a negative $M^2$ in the  mass range 
\be
\label{madsaq}
-{d^2\over 4}<M^2<-{d^2\over 4}+1,
\ee see Appendix \ref{altQads}. 
As was discussed in \cite{Balakraus,Breit,wittklev} scalars in the range eq.(\ref{madsaq}) in AdS space  can be quantised in two alternate ways and correspondingly the source can be chosen to correspond to either of the two fall-off. 
The   fall off, $z^{{d\over2}-\nu}$, corresponds to a   source for an operator of dimension $\Delta_+={d\over2}+\nu$, where $z$ is the Poincaré coordinate in AdS, \eq{metads} and $\nu=\sqrt{{d^2\over4}+M^2}$.  While the fall off, 
$z^{{d\over2}-\nu}$, corresponds to the source for an operator of dimension $\Delta_-={d\over2}-\nu$.
The correlation functions in the two cases are related to each other by a Legendre transformation. 

In the dS case, instead of a Legendre transformation it is more natural to consider a Fourier transformation which takes the wave function in the field eigenstate basis to one in the basis of its conjugate momentum  eigenstate, see Appendix \ref{app.altov}. This Fourier transformed wave function will continue to be normalisable. 
Carrying out the Fourier transformation starting with eq.(\ref{psifa}), we get \eq{altovw} restated below
\be
\label{wvmoma}
{\cal W}[\pi({\bf k},\eta_1),\eta_1]=\exp\left[i\int \frac{d^d\mathbf{k}}{(2\pi)^d} ~\pi(\mathbf{k},\eta)\left(\frac{(-\eta)^d(\mathcal{F}_\nu(k,\eta))^\ast}{2\eta \partial_\eta(\mathcal{F}_\nu(k,\eta))^\ast}\right)\pi(-\mathbf{k},\eta)\right]
\ee
If we define a source ${\hat \pi}({\bf k})$  in terms of the asymptotic behaviour of $\pi({\bf k},\eta_1)$ by the relation  
\be
\label{sta}
\pi({\bf k},\eta_1)=(-\eta_1)^{-{d\over2}-\nu} {\hat \pi}({\bf k})
\ee
we get  
\be
\label{momwf}
{\cal W}[{\hat \pi}({\bf k}),\eta_1]=\exp\left[\frac{1}{2}\int \frac{d^d\mathbf{k}}{(2\pi)^d} ~\hat{\pi}(\mathbf{k})\left[{i(-\eta_1)^{-2\nu}\over\left(\frac{d}{2}-\nu\right)}-\frac{\beta^\ast_\nu}{\alpha^\ast_\nu}{(2i\nu)\over\left(\frac{d}{2}-\nu\right)^2} k^{2\nu}\right]\hat{\pi}(-\mathbf{k})\right]
\ee
where ${\beta^{\ast}_{\nu}\over \alpha_{\nu}^{\ast}}$ is given in \eq{ratalbe}. From this we see the two point function for the operator sourced by ${\hat \pi}({\bf k})$  
\be
\label{twomom}
\l \bar{O}({\bf k})\bar{O}(-{\bf k})\r=-\frac{\beta^\ast_\nu}{\alpha^\ast_\nu}{(2i\nu)\over\left(\frac{d}{2}-\nu\right)^2} k^{2\nu}
\ee
Interestingly, this operator is also of dimension 
$\Delta_+$, \eq{overdim}, although the coefficient in front is different from that when  $\hat \phi$ is  the source, eq.(\ref{bounwvpexp}).
Similarly, the  cut-off dependent term in eq.(\ref{momwf}) also has a coefficient different from that in eq.(\ref{bounwvpexp}).

%
%

Also note that had we   directly Fourier transformed the wave function as expressed in terms of ${\hat \phi}({\bf k})$, \eq{bounwvpexp}, we 
would have obtained the same result as in eq.(\ref{twomom}), after a suitable field redefinition.


\paragraph{Coherent State Representation}
We can also consider the wave function in the coherent state representation. In fact this representation will be very useful in the underdamped case we will consider subsequently. For the over damped case this representation is discussed further in Appendix \ref{app.altov}. 
The summary is that one finds that the eigenvalue of the coherent state can also be related suitably to a source in the dual theory and the dual operator then also turns out to have dimension $\Delta_+$, \eq{overdim}. 

\subsubsection{ Interaction in the Overdamped Case}
			\label{holoint}
			
 		Now let us  turn to considering interactions in the overdamped case. Specifically, we consider adding a  $\lambda\phi^n$ interaction to the free theory in the overdamped domain. 
		The resulting correction to the wave function was discussed in eq.(\ref{genpsipert}), eq.(\ref{deltasn}),  at first order in $\lambda$. 
		Inserting the asymptotic value of the field in the form given in eq.(\ref{oversource}) then leads to 
		\begin{equation}
 			\label{overdelSn1}
 			i\delta S_n|=i\lambda\int \prod_{i=1}^{n}\frac{d^d{\mathbf{k}_i}}{(2\pi)^d}(2\pi)^d\delta^{(d)}(\Sigma_{j=1}^n\mathbf{k}_j) \frac{I(k_1,k_2,..,k_n,\eta_1)}{\prod_{i=1}^n f_{\nu}^{\ast}(k_i,\eta_1)}(-\eta_1)^{n(\frac{d}{2}-\nu)}\prod_{i=1}^{n}\hat\phi(\mathbf{k}_i)
 		\end{equation}
		In general the integral on the RHS can be  divergent depending on the value of $\nu$, as is discussed in section \ref{interaction1} and \ref{divergence}. 
		Once the divergent terms are removed the finite part, obtained by retaining the finite contribution in the integral $I(k_1, k_2\cdots, k_n,\eta_1)$ which we denote 
		as $I(k_1,k_2,\cdots k_n)$, see section \ref{interaction1} and the leading behaviour of $f_\nu^\ast (k,\eta_1)$,   takes the form 
 		\begin{equation}
 			\label{overdelS1}
 			i\delta S_n|=i\lambda\int \prod_{i=1}^{n}\frac{d^d{\mathbf{k}_i}}{(2\pi)^d}(2\pi)^d\delta^{(d)}(\Sigma_{j=1}^n\mathbf{k}_j)\frac{\prod_{i=1}^n k_i^{\nu}}{(\alpha_{\nu}^{\ast})^n}I(k_1,..,k_n)\prod_{j=1}^{n}\hat\phi(\mathbf{k}_j) +{\cal O}(\eta_1^{2\nu})
 		\end{equation}
%
 		where $\alpha_{\nu}$ is given in \eq{valtab}. We can rewrite eq.(\ref{overdelS1}) in the following way
		 		\begin{equation}
 			\label{overdelSn2}
 			\log\delta\psi[\hat{\phi}]\equiv i\delta S_n={1\over n!}\int \prod_{i=1}^{n}\frac{d^d{\mathbf{k}_i}}{(2\pi)^d}\langle O(\mathbf{k}_1)...O(\mathbf{k}_n)\rangle\prod_{j=1}^{n}\hat\phi(\mathbf{k}_j)
 		\end{equation}
 		where the $n$ point function 
 		\begin{equation}
 			\label{nptover}
 			\langle O(\mathbf{k}_1)...O(\mathbf{k}_n)\rangle=n!{i\lambda\over( \alpha_{\nu}^{\ast} )^n} (2\pi)^d\delta^{(d)}(\mathbf{k}_1+...+\mathbf{k}_n)\left(\prod_{i=1}^n k_i^{\nu}\right) I(k_1,k_2,\cdots k_n)
 		\end{equation}
 		For $n=3$, this becomes
 		\begin{equation}
 			\label{3ptover}
 			\langle O(\mathbf{k}_1)O(\mathbf{k}_2)O(\mathbf{k}_3)\rangle= i\lambda  {3!\over (\alpha_{\nu}^{\ast})^3} (2\pi)^d\delta^{(d)}(\Sigma_{j=1}^3\mathbf{k}_j)\left(\prod_{i=1}^3 k_i^{\nu}\right) I(k_1,k_2,k_3)
 		\end{equation}
 		\begin{equation}
 			\label{3ptoverprime}
 			\implies \langle O(\mathbf{k}_1)O(\mathbf{k}_2)O(\mathbf{k}_3)\rangle'=i\lambda {3! \over (\alpha_{\nu}^{\ast})^3}\left(\prod_{i=1}^3 k_i^{\nu}\right) I(k_1,k_2,k_3)
 		\end{equation}
 		where prime indicates we are dropping the $(2\pi)^d\delta^{(d)}(\Sigma_{j=1}^3\mathbf{k}_j)$ term. The integral above was discussed in the previous section, see eq.(\ref{IexpK}) and discussion thereafter. It gives the standard three point function for primary fields  of dimension 
		$\Delta=\frac{d}{2}+\nu$,  see reference \cite{Baut-3ptmom}. For a definition of primary fields see Appendix \ref{conft}. 
		
		Note that the result for the correction to the wave function in the $n=3$ case was derived directly in the interaction picture in Appendix \ref{perturbation} and agrees with
		eq.(\ref{3ptoverprime}). In fact the calculation in the appendix can be easily generalised to the case for general $n$ as well and shown to agree with what we have obtained above.  
		
		It is also worth noting that if we calculate the wave function by evaluating the bulk on-shell action after solving the equations of motion, then the solution for the bulk field changes at ${\cal O}(\lambda)$  from its value in the free case. But since this change must vanish at $\eta=\eta_1$, to ensure that the field continues to take a fixed value, $\phi({\bf k},\eta_1)$ at $\eta=\eta_1$, it   does not lead to an additional contribution to the action at this order, 
		 see Appendix \ref{pathpert}.

		\subsection{Underdamped Case}
		\label{under}
		This case presents us with new features and is the main focus of our study. Note that here  $\nu$  given in eq.(\ref{nu}) is purely imaginary. We re-write it as
		\begin{equation}
			\label{undernu}
			\nu=i\sqrt{M^2-\frac{d^2}{4}}\equiv i\mu
		\end{equation}
		where $\mu$ is a real quantity. As a result, eq.(\ref{defF}) becomes in the form \eq{defFu} 
		and the asymptotic behaviour of the solution of $f_\mu(k,\eta)$ given in eq.(\ref{deff}) is a sum over two modes of the form \eq{deffu}
		with both modes going like $(-\eta)^{{d\over2}\pm i \mu}$ are  equally important as $\eta\rightarrow0$, unlike the overdamped case where  one mode decays  faster than the other. This feature is the essential reason why we will need a different map between the bulk field and a source for the CFT in this case.\\~\\ 
%
		We start with the free theory wave function eq.(\ref{psi[phi]}):
		\begin{equation}
			\label{bounwvfunder}
			\psi[\phi,\eta]=\mathcal{N}\exp\left[i\int \frac{d^d\mathbf{k}}{(2\pi)^d} ~\phi(\mathbf{k},\eta)\left(\frac{\partial_\eta\mathcal{F}_\mu^\ast(k,\eta)}{2(-\eta)^{d-1}\mathcal{F}_\mu^\ast(k,\eta)}\right)\phi(-\mathbf{k},\eta)\right]
		\end{equation}
		where $\mathcal{F}_\mu(k,\eta)$ is given in eq.(\ref{defFu}).
		Now suppose we try to identify a source in the boundary theory ${\hat \phi}({\bf k})$, as was done in the overdamped case, eq.(\ref{oversource}) as follows
		\be
		\label{sourcea}
		\phi({\bf k},\eta)={f^{\ast}_\mu(k,\eta)\over f_\mu^{\ast}(k,\eta_1)}(-\eta_1)^{{d\over2}-i\mu} {\hat \phi}({\bf k})
		\ee
		Eq.(\ref{bounwvfunder}) at a late time $\eta_1$ then  takes the form
		\be
		\label{newform}
		\psi[{\hat \phi}({\bf k}),\eta_1]=\mathcal{N}\exp\left[{i\over2}\int {d^d{\bf k} \over (2\pi)^d} ~{\hat \phi}(-{\bf k}) {\hat \phi}({\bf k})  (-\eta_1)^{1-2i\mu} 
		{{\dot f}^{\ast}_\mu(k,\eta_1)\over f^{\ast}_\mu(k,\eta_1)}\right]
		\ee
		similar to eq.(\ref{bounwvf1}).
However, unlike the overdamped case the factor of $f_\mu^{\ast}(k,\eta_1)$ in the denominator of the exponent cannot be expanded in a power series at small $\eta_1$, since as was mentioned above  it is a sum over two modes  
going like $(-\eta_1)^{{d\over2}\pm i \mu}$ and both modes are equally important as $\eta_1\rightarrow 0$. 
		As a result an attempt to relate $\phi({\bf k},\eta)$ to a source for an operator with a fixed anomalous dimension in the CFT, will not work here.\\~\\  
		Instead we will  find that the wave function when expressed in a suitable coherent state basis, rather than   the basis of the field operator, eq.(\ref{evea}),  
		lends itself to an interpretation as a generating function for correlation functions in the CFT.  And the eigenvalue of the coherent state, instead of $\phi({\bf k},\eta)$, the eigenvalue of the  field operator ${\Phi}({\bf k},\eta)$, can be  conveniently  related to a  source in the CFT.\\~\\
			In this subsection and the next  we will first discuss how to compute	the wave function in the field basis in a convenient way. 
		Then in the following section we will discuss how to construct the wave function in the basis of coherent states, starting from $\Psi[\phi({\bf x},\eta_1),\eta_1]$. 
		For computing the wave function $\Psi[\phi({\bf x},\eta_1),\eta_1]$ it is convenient to define a variable $J_+({\bf k},\eta_1)$ in terms of which the field
		\begin{equation}
			\label{subst1}
			\phi(\mathbf{k},\eta_1)=f_\mu^\ast(k,\eta_1) k^{i\mu} J_+({\bf k})
		\end{equation}
		It will be 	convenient to express $\Psi$ in terms of $J_+$.
		In the following section we will find that $J_+$ is in fact the eigenstate of the coherent state.\\ 
Using eq.(\ref{deffu}) we find at late times that this leads to 
\be
\label{defjpb}
 \phi(\mathbf{k},\eta)= (-\eta)^{{d\over2}-i\mu}\alpha_\mu^{\ast}[1+ {\beta_\mu^{\ast} \over \alpha_\mu^\ast }(-\eta k)^{2i \mu} ]  J_+({\bf k})
 \ee 
		The wave function eq.(\ref{bounwvfunder}) now  gives, 
		\be
		\label{wvr2}
		\psi[J_+,\eta]=\mathcal{N}\exp\left[i\int \frac{d^d\mathbf{k}}{(2\pi)^d} ~J_+({\bf k})\left(k^{2i\mu}\frac{\mathcal{F}_\mu^\ast(k,\eta)\partial_\eta\mathcal{F}_\mu^\ast(k,\eta)}{2(-\eta)^{d-1}}\right)J_+(-{\bf k})\right]
				\ee
		Inserting the late time behaviour of $\mathcal{F}_\mu^\ast(k,\eta)\partial_\eta\mathcal{F}_\mu^\ast(k,\eta)$ from  eq.(\ref{deffu}) as,
		\begin{empheq}{multline}
			\label{kernelpsi}
			\dot{f}_\mu^\ast(k,\eta)f_\mu^\ast(k,\eta)=-(-\eta)^{d-1}\left[\alpha_\mu^\ast\left(\frac{d}{2}-i\mu\right)(-k\eta)^{-i\mu}+\beta_\mu^\ast\left(\frac{d}{2}+i\mu\right)(-k\eta)^{i\mu}\right]\\
			\times\left[\alpha_\mu^\ast(-k\eta)^{-i\mu}+\beta_\mu^\ast(-k\eta)^{i\mu}\right]
		\end{empheq}
		where $\alpha_{\mu}$ and $\beta_\mu$ are given in \eq{defabu}.
		We get that the 
		exponent of eq.(\ref{wvr2}) becomes, 
		 \begin{equation}
		 	\begin{split}
		 	&\int {d^d{\bf k}\over (2\pi)^d} J_+(\mathbf{k})\left(k^{2i\mu}\frac{\dot{f}_\mu^\ast(k,\eta)f_\mu^\ast(k,\eta)}{2(-\eta)^{d-1}}\right)J_+(-\mathbf{k})\\&=
			-\int {d^dk\over (2\pi)^d}( \frac{1}{2})\left[(\alpha_\mu^\ast)^2\left(\frac{d}{2}-i\mu\right)(-\eta)^{-2i\mu}J_+(\mathbf{k})J_+(-\mathbf{k})+\right.\\ &
			\left.(\beta_\mu^\ast)^2\left(\frac{d}{2}+i\mu\right)k^{4i\mu}(-\eta)^{2i\mu}J_+(\mathbf{k})J_+(-\mathbf{k})+\alpha_\mu^\ast\beta_\mu^\ast d~ k^{2i\mu}J_+(\mathbf{k})J_+(-\mathbf{k})\right]	
		 	\end{split}
		 	\label{integrand}		 	
		 \end{equation}
		 The first two terms on the RHS depend on the cut-off. The last term is of the form of a CFT two-point function. 
		 Keeping only this last term for now we get
 		\be
 			\label{Z!contact}
			 			\psi[J_+] =  \exp\left[\frac{1}{2}\int\frac{d^d{\bf k}}{(2\pi)^d}\frac{d^d{\bf k}'}{(2\pi)^d}	
				\left[J_+(\mathbf{k}) J_+(\mathbf{k}')
\langle  {\hat O}_+(\mathbf{k}) {\hat O}_+(\mathbf{k}')\rangle \right]\right]
		\ee
 		where
 		\begin{equation}
 			\label{<O+O+>}
 			\langle{\hat O}_+(\mathbf{k}){\hat O}_+(\mathbf{k}')\rangle=-i(2\pi)^d\delta^d(\mathbf{k}+\mathbf{k}') d\alpha_\mu^\ast\beta_\mu^\ast  ~k^{2i\mu}
			\end{equation}
In position space, using \eq{defabu},  this becomes
\be
\label{psotwo}
\l {\hat O}_+({\bf x}){\hat O}_+({\bf y})\r	= {d\over4\mu}e^{-\pi\mu}(1+\coth(\pi\mu)){2^{2i\mu}\Gamma[{d\over2}+i\mu]\over\pi^{d\over2}\Gamma[-i\mu]}{1\over|{\bf x}-{\bf y}|^{d+2i\mu}} 
\ee	 		
		We note that $\langle {\hat  O}_+{\hat  O}_+ \rangle$ is of the correct form to be the  two point function in a CFT  for an operator of dimension $\Delta_+$ 
		with anomalous dimension
		\begin{equation}
			\label{underdim}
			\Delta_+=\frac{d}{2}+ i\mu
		\end{equation} 
		Importantly, $\Delta_+$ is  complex.\\~\\ 
		Turning to the cut-off dependent terms in eq.(\ref{integrand}) we see that the first term on the RHS is local in position space and of the standard form expected from a local counterterm. However the second term going like $\int {d^d{\bf k}\over (2\pi)^d} J_+({\bf k})J_+(-{\bf k}) (-\eta)^{2i\mu} k^{4i \mu}$ is quite ugly- taking the form in position space,
		\be
		\label{posf}
		\sim  (-\eta)^{2i\mu} \int d^d{\bf x} d^d{\bf y} {J_+({\bf x}) J_+({\bf y}) \over |{\bf x}-{\bf y}|^{d+4i\mu}}
		\ee
		showing that it is  both  cut-off dependent and non-local in space. 
		It will turn out that this term will disappear when we go to the coherent state basis!		
%
		  
		  A few comments are in order before we close this subsection. 
		  Note firstly that the identification we have made between the bulk field and the source in eq.(\ref{defjpb}) above is not local along the $\eta_1$ hypersurface, unlike in the overdamped case, eq.(\ref{relsb}).  In fact in position space,   at $\eta=\eta_1$,  eq.(\ref{defjpb}) takes the form
		  \be
		  \label{forrelsba}
		  \phi({\bf x},\eta_1)=(-\eta_1)^{{d\over2}-i\mu} \alpha_\mu^{\ast} J_+({\bf x}) + (-\eta_1)^{{d\over2}+i\mu}\beta_\mu^{\ast} {4\mu e^{\pi\mu}\over d(1+\coth(\pi\mu))} \int d^d{\bf y} \l {\hat O}_+({\bf x}){\hat O}_+({\bf y})\r J_+({\bf y})
		  \ee
		  where $\l {\hat O}_+({\bf x}){\hat O}_+({\bf y})\r$ is given in eq.(\ref{psotwo}), 
		  which shows the non-local relation between $\phi$ and $J_+$ as a function of the position coordinates ${\bf x}, {\bf y}$. 
		   In section \ref{WI} when we discus the Ward identities further we will need to define the  relation in eq.(\ref{defjpb}) more carefully, due to this non-local nature, for a more general hypersurface.\\
		 Second, note  that instead of  eq.(\ref{subst1}) we could have made the following identification
		  \be
		  \label{subst2}
		  \phi({\bf k},\eta_1)=f_\mu^{\ast}(k,\eta_1) J_- ({\bf k}) k^{-i\mu}
		  \ee
		  between the bulk field and a source  $J_-({\bf k})$. 
		  Comparing eq.(\ref{subst1}) and eq.(\ref{subst2}) we see that the relation between $J_+$ and $J_-$ is given by 
		  \be
		  \label{reljpjm}
		  J_+({\bf k})=k^{-2i\mu }J_-({\bf k})
		  \ee
		  which is  non-local in position space. 
		  Expressing the wave function in terms of $J_-$ we get  eq.(\ref{integrand}) with $J_+$ replaced by $J_-$ using eq.(\ref{reljpjm}). 
		  The first term, proportional to $(\alpha_\mu^{\ast})^2$, is now cut-off dependent and non-local in position space, the second term, proportional to $(\beta_\mu^{\ast})^2$, is cut-off dependent and local, and the third term, proportional to $\alpha_\mu^{\ast}\beta_\mu^{\ast}$, corresponds to a CFT two point correlator for an operator of dimension
		  \be
		  \label{dimdelm}
		  \Delta_- ={d\over2}-i \mu
		  \ee
		  So we see that while both  identifications  of the source as $J_+$ or $J_-$ of course have the same physical content as far the wave function is concerned they, interestingly, lead to   different interpretations in terms of a CFT dual. Namely as the source for operators with dimension ${d\over2}\pm i \mu$ respectively.  We will have more to say about some of these issues in the subsequent sections when we discuss the Ward identities etc.

		  Finally, below we will also explore the wave function in the coherent state representation. It will turn out that  this coherent state description is more closely tied to using the identification made in eq.(\ref{defjpb}) and the source in the CFT, in the coherent state representation, 
		  which we will denote by $\rho^{\ast}$, 
		  up to a simple scaling factor,  will be  related to  $J_+$, instead of $J_-$, see section \ref{coherent}. The scaling factor relating $\rho^{\ast}$ to $J_+$ is given in eq.(\ref{Jrelchi}). 
		 We should mention, before proceeding, that in the coherent  state basis the coefficient in the correlation  function will be different from eq.(\ref{<O+O+>}) even after accounting for this scaling factor.

			\subsubsection{Including Interaction}
			\label{interaction2}
			We now show how the correspondence in the underdamped case can be extended in the presence of $\phi^n$ type interactions discussed above. 
		Our discussion will be  in terms of the source $J_+$, a similar analysis also applies in terms of $J_-$ and the results can be obtained using the relation between the $J_+$, $J_-$ given in eq.(\ref{reljpjm}).
			
			The additional contribution to the exponent in the wave functional due to a $\phi^n$  interaction is
			\begin{equation}
				\label{delSunder1}
				i \delta S_n[\phi(\mathbf{k},\eta)]=i \lambda\int \prod_{i=1}^{n}\frac{d^d{\mathbf{k}_i}}{(2\pi)^d}(2\pi)^d\delta^{(d)}(\mathbf{k}_1+...+\mathbf{k}_n)\left[I(k_1,k_2,..,k_n)\prod_{i=1}^{n}\frac{\phi(\mathbf{k}_i,\eta)}{\mathcal{F}_\mu^{\ast}(k_i,\eta)}\right]
			\end{equation}
			as discussed in eq.(\ref{deltasn}). Following eq.(\ref{defjpb}) 
			we now replace  $\phi({\bf k},\eta)$ in terms of $J_+({\bf k})$ leading to  
%
			\begin{equation}
				\label{delSunderJJ1}
				i \delta S_n[J_{+}]=i\lambda\int \prod_{i=1}^{n}\frac{d^d{\mathbf{k}_i}}{(2\pi)^d}(2\pi)^d\delta^{(d)}(\Sigma_j \mathbf{k}_j) I(k_1,k_2,..,k_n)\prod_{a=1}^{n}\biggl\{ J_{+}(\mathbf{k}_a)k_{a}^{i\mu}\biggr\}
			\end{equation}
			
		From eq.(\ref{delSunderJJ1}) we see that the coefficient function  for the $J_+^n$ terms is 
%
			\be
				\label{formco}
				\langle \hat{O}(\mathbf{k}_1)..\hat{O}(\mathbf{k}_n)\rangle= i \lambda n! (2\pi)^d \delta^{(d)}\left(\mathbf{k}_1+..+\mathbf{k}_n\right)\prod_{j=1}^{n}k_j^{i\mu} I(k_1, \cdots k_j, \cdots k_n) 
			\ee
with the interaction part of the wave function being			
\begin{equation}
	\label{udwfpert}
	\log\delta\psi[J_{+}]=i\delta S_n=\frac{1}{n!}\int \prod_{i=1}^{n}\frac{d^d{\mathbf{k}_i}}{(2\pi)^d}\langle \hat{O}(\mathbf{k}_1)..\hat{O}(\mathbf{k}_n)\rangle \prod_{b=1}^{n} J_{+} (\mathbf{k}_b)
\end{equation}
As was mentioned above we will actually identify the source in  the coherent state basis in the following section, but we have written the coefficient function suggestively as an expectation value of fields above, because, after a trivial   rescaling, the source will be turn out to be equal to $J_+$, and the interaction term in the coherent state basis at leading order will actually be the same as eq.(\ref{udwfpert}), up to this rescaling factor.

The integral $I(k_1,\cdots k_n)$ was defined in eq.(\ref{IexpK}). More correctly to compute the wave function at time $\eta_1$ we would need to compute this integral, eq.(\ref{Iexpression}), with the upper limit being $\eta_1$, instead of $0$, 

\be
\label{defin}
I_n(k_1, \cdots k_n)=\int_{-\infty}^{\eta_1} {d\eta'\over (-\eta')^{d+1}}		\prod_{i} {\cal F}_\mu^{\ast}(k_i,\eta')
\ee
However we will see  in the next subsection that when all fields are under damped there is no divergence at late times and accordingly we can take  the upper limit to vanish.

			\subsection{Divergences}
			\label{divergence}
			Here we briefly consider divergences which can arise at late times, when $\eta\rightarrow 0$. 
			We will focus on scalar fields  but similar   conclusions are valid  more generally. 
			The free scalar case has already been discussed above. As an example of interactions we take the $\phi^n$ term discussed above, a similar discussion applies in other cases as well.			
			In the $AdS$ case the divergences can all be removed after holographic renormalisation; once this is established  the divergences are not of much interest in the study of the AdS/CFT correspondence. However in the dS case the divergences should not be removed, and some discussion on  isolating and dealing with them is worthwhile. 

			 We will mostly consider  the three point interaction below, this example  illustrates the  features present more generally.
			  A three point function  arises from the   cubic term in the action,
			 \be
			 \label{cubicint}
			 \delta S_3=\int d^{d+1}{x} \sqrt{-g} \phi_1({\bf x},\eta) \phi_2({\bf x},\eta) \phi_3({\bf x},\eta)
			 \ee
			   We take the three fields to be different in general. 
			   From eq.(\ref{overdelSn1}), eq.(\ref{delSunderJJ1}) we see that in general the contribution to the wave function cubic in the sources takes the form
			   \be
			   \label{genfd}
			   \delta \log(\Psi)\propto \int \left(\prod_{i=1}^{3}{d^d{\bf k}_i\over(2\pi)^d}\right) (2\pi)^d \delta \left(\sum_{i=1}^{3} {\bf k}_i\right) I(k_1, k_2, k_3,\eta_1 )\prod_i 
			   {\phi_i({\bf k}_i,\eta_1) \over {\cal F}_{\nu_i}^{\ast}(k_i,\eta_1)}
			\ee
			   If the $i^{\rm th}$ field is overdamped, $\nu_i$ is related to its mass $M_i$ by eq.(\ref{nu}) and $\phi({\bf k},\eta_1)$ is related to its  source ${\hat \phi}({\bf k})$ by 
			   eq.(\ref{oversource}). If it is under damped $\nu_i=i \mu_i$, with $\mu$ given in eq.(\ref{undernu}) and the field is related to its source $J_+$ by 
			   eq.(\ref{subst1}). 
			   
			    
			   The integral $I(k_1,k_2,k_3)$ is given by, eq.(\ref{Iexpression}),  
			 \be
			 \label{formt}
			I(k_1,k_2,k_3)=\int_{-\infty}^{\eta_1} {d\eta'\over (-\eta')^{d+1}}\prod_{i=1}^3 {\cal F}_{\nu_i}^{\ast}(k_i,\eta')
			\ee
			Here we have put in a late time cut-off at $\eta_1$ and  we are interested in examining what happens as it vanishes. 
			
			
			When all three fields are  over damped the asymptotic behaviour of ${\cal F}_{\nu_i} (k,\eta)$ is given in eq.(\ref{deff}). 
			The  leading divergence is obtained  by keeping the leading behaviour, i.e., the first term on the RHS in eq.(\ref{deff}), for all three fields.
			\be
			\label{ltf}
			f_{\nu}(k,\eta)\simeq \mathbb{C}_\nu\left(-\frac{i 2^{\nu } \Gamma [\nu] }{\pi }\right)(-\eta)^{{d\over2}-\nu}(k)^{-\nu } 
			\ee
			where $\mathbb{C}_\nu$ is given in \eq{defC1}. The behaviour of the integral eq.(\ref{formt}) is divergent if 
			\be
			\label{diva}
			{d\over2}<\sum_i \nu_i.
			\ee
			It is easy to see that this divergence results in a term in the wave function which is local in position space
			\be
			\label{lot}
			\delta \log(\Psi) \sim {1\over (-\eta_1)^{\nu_1+\nu_2+\nu_3-{d\over2}}} \int d^d{\bf x} {\hat \phi}_1({\bf x}) {\hat \phi}_2({\bf x}) {\hat \phi}_3({\bf x})
			\ee
			A sub leading divergence can arise if we take two of the  ${\cal F}_{\nu_i}( k_i,\eta)$ functions in eq.(\ref{formt}) to have their leading behaviour but the third say $i=3$ to 
			be sub leading, i.e. to have the form corresponding to the second term given in eq.(\ref{deff}). Or  if we  keep a sub leading term in one of the ${\cal F}_{\nu_i}^{\ast}(k_i,\eta_1)$ denominator terms  which appear as part of  the 
			$\prod_i{\phi_i({\bf k}_i,\eta)\over {\cal F}_{\nu_i}^{\ast}(k_i,\eta_1)}$ product  in eq.(\ref{genfd}).

			This gives rise to a contribution in  $I(k_1,k_2,k_3)$ which is divergent if  
			\be
			\label{conddab}
			{d\over2}<\nu_1+\nu_2-\nu_3
			\ee
			The resulting  term in the wave function is  
			\be
			\label{wvfra}
			\delta \log( \Psi) \propto {1\over (-\eta_1)^{\nu_1+\nu_2-\nu_3-{d\over2}} } \int d^d{\bf x} d^d{\bf y} {\hat \phi}_1({\bf x}){\hat \phi}_2({\bf x}) {1\over |{\bf x}-{\bf y}|^{d+2\nu_3}}{\hat \phi}_3({\bf y}) 
			\ee
			We see that  this term is  of the form of an additional contribution to the two point function for the operator $O_3$, so that  in effect the divergent term can be absorbed by redefining the source 
			\be
			\label{redfs}
			{\hat \phi}_3({\bf x}) \rightarrow {\hat \phi} _3({\bf x}) + c {1\over (-\eta_1)^{\nu_1+\nu_2-\nu_3-{d\over2}}} {\hat \phi_1}({\bf x}) {\hat \phi}_2({\bf x})
			\ee
			Another way to say this is that the divergence arises due to operator mixing, the operators $O_1$ and $O_2$ mix with $O_3$ when they come close in position space. 
			
			In contrast, when all three fields are underdamped it is easy to see that there is no divergence from the $\eta_1\rightarrow 0$ region in the integral. 
			
			If two fields are overdamped and one, say $i=3$, is  underdamped, then we  get divergent terms if 
			\be
			\label{divca}
			{d\over2}<\nu_1+\nu_2
			\ee
			In this case taking the term in $f_{\nu_3}^{\ast}(k_3,\eta')$ which goes like $(-\eta')^{{d\over2}}(-k_3\eta')^{i \mu}$, i.e. the second  term in eq.(\ref{deffu}),
			we get a contribution analogous to eq.(\ref{wvfra}) above, with 
			\be
			\label{contwf}
			\delta \log(\Psi)\propto {1\over (-\eta_1)^{\nu_1+\nu_2-{d\over2}+i \mu}} \int d^d{\bf x} {\hat \phi}_1({\bf x}){\hat \phi}_2({\bf x}) {1\over |{\bf x}-{\bf y}|^{d+2 i \mu}}J_{+,3}({\bf y})
			\ee
			which is of the form of a contribution to the two point function for the operator ${\hat O}_{+,3}$. 
			On the other hand taking the term in $f_{\nu_3}^{\ast}(k_3,\eta')$ which goes like $(-\eta')^{d\over2}(-k_3\eta')^{-i \mu}$, i.e. the first  term in eq.(\ref{deffu}),
			we get a divergent term which is local in position space, 
			\be
			\label{adddelps}
			\delta \log (\Psi)\propto {1\over (-\eta_1)^{\nu_1+\nu_2-{d\over2}-i \mu}} \int d^d{\bf x} {\hat \phi}_1({\bf x}){\hat \phi}_2({\bf x})J_{+,3}({\bf x})
			\ee
			In both eq.(\ref{contwf}) and eq.(\ref{adddelps}) the dependence on the cut-off $\eta_1$ involves a complex exponent, as is needed from dimensional analysis,
			since the operator ${\hat O}_+$ has a complex anomalous dimension $\Delta_+$, eq.(\ref{underdim}).

			Finally the finite part for the cubic term,   which is independent of $\eta_1$,  has a form which is fixed by conformal invariance. In Appendix \ref{App.Hankel} we discuss how this term  can be calculated by analytic continuation and give the resulting expression for all values of the anomalous dimension in dS space.

			The analysis above can be extended to more general $\phi^n$ interactions in a similar way. 
			The divergences which arise can be understood as corresponding to local counter terms or due to operator mixing. 
			
			
			Before ending  let us come back to contributions in the wave function only involving under damped fields.  Consider a divergence which arises in a Witten diagram with an interaction vertex of the $\phi^n$ type, involving $n$ factors of the under damped bulk field $\phi$. 
			 This results in a contribution going like 
			 \be
			 \label{conde}
			 \delta \log \Psi\sim \int {d\eta' \over (-\eta')^{d+1}} (-\eta')^{n d\over2} (-\eta')^{ \pm i \mu n}
			 \ee
			 we see that the integral converges for $\eta'\rightarrow 0$ as long as $n>2$. 
			 This shows that all contributions to higher point functions, and also to higher loops in the two point function will be finite. 
			 
			 Note that the estimate in eq.(\ref{conde}) is valid when all $n$ lines meeting at the vertex are bulk to boundary propagators, as shown in Figure \ref{Witten3}, or when some lines are bulk to boundary propagators and others are bulk to bulk propagators. 
			 This is true since a bulk to bulk propagator $G({\bf x},\eta; {\bf x}',\eta')$ also behaves like $(-\eta')^{{d\over2}\pm i \mu}$ when $\eta'\rightarrow 0$. 
			
			To conclude, we see that all divergences can be reproduced by the effects of local terms on the late time  boundary, even in the presence of underdamped fields. In the dS case, as has been emphasised in this paper, the divergences are physical and should not be removed. Our discussion above leads to a useful  strategy to incorporate their effects.  One can think of adding two sets of terms to the wave function, one being the local counter terms with the required sign to render the Witten diagrams finite, and the second being the same counter terms but now with the opposite sign. These two additions of course cancel out and do not change the wave function. 
			However, with the first set of local counter terms in place now, of required sign, the Witten diagrams become finite,   are well defined, and can often be calculated easily  by analytic continuation from AdS space, see Appendix \ref{App.analcont}. This leaves the second set of terms with the opposite sign, but these being local are  easy to compute as well. In this way once the locality of the  counter terms has been established,  we see that interactions, including divergences,  can be included in a straightforward manner. 
			
			One final comment. As we will see below, for the underdamped case we will need to go from the field basis to the coherent state basis to obtain the wave function in a form where it can be related to correlation functions of a CFT. However the integral transform involved is the one which takes the wave function in the free theory to the coherent state basis and does not need to be corrected once interactions are included, see section \ref{coherent}. Thus having calculated the wave function in the field basis, including its divergences as mentioned above, one can do the integral transform at the last stage to conveniently obtain the wave function in the required coherent state basis.

%

\section{Coherent State Basis for the Underdamped Fields}
	\label{coherent}
	In this section we discuss  the wave function  of the Bunch Davies vacuum in a coherent state basis for the underdamped fields. 
	As mentioned above the wave function in this basis  will be identified with the generating function for  CFT correlations.
	
	The coherent states of interest are identified as follows. As discussed in section \ref{setup}, the Hankel functions $H^{(1,2)}_{i\mu}$ take the asymptotic form, near the Poincaré horizon, 
	\be
	\label{assf}
	H^{(1)}_{i\mu}(-k\eta)\sim e^{-ik\eta}, \\ H^{(2)}_{i\mu}(-k\eta) \sim e^{+ik\eta}
	\ee
	and therefore behave as positive and negative frequency modes with respect to $\partial_\eta$. 
	Keeping this mind in the mode expansion for $\phi({\bf x},\eta)$, eq.(\ref{phiopu})  the coefficient of the term containing $H^{(1)}_{i\mu}$ was  related to the destruction operator $a_{\bf k}$,
	and $H^{(2)}_{i\mu}$ with the creation operator $a^\dagger_{\bf k}$. However near the future boundary, where $\eta\rightarrow 0$, the behaviour of the Hankel functions is different. The more natural time variable in this region is $t=-\log(-\eta)$, in terms of which the metric 
	\be
	\label{spameta}
	ds^2=-{d\eta^2\over \eta^2} + {1\over \eta^2} \left[\sum_i(dx^i)^2\right]
	\ee
	becomes
	of FRW type
	\be
	\label{metfrw}
	ds^2=-dt^2 +e^{2t} \left[\sum_i(dx^i)^2\right].
	\ee 
	The asymptotic behaviour of $H^{(1)}_{i\mu}$ in this region  in terms of $t$ then takes the form 
	\be
	\label{behavhan}
	\mathbb{C}_\mu H^{(1)}_{i\mu}(-k\eta)= \alpha_\mu k^{i\mu} e^{-i\mu t}   +\beta_\mu k^{-i\mu} e^{i\mu t}
	\ee
	where the coefficient $\mathbb{C}_\mu$ is given in eq.(\ref{cmu}) and $\alpha_\mu, \beta_\mu$ are given in \eq{defabu}. We see that $H^{(1)}_{i\mu}$ and $H^{(2)}_{i\mu}$  contain both positive and negative frequency modes of $\partial_t$. 
	
	This motivates us to   consider a different notion of creation and destruction operators, ${\tilde a}({\bf k}), {\tilde a}^\dagger({\bf k})$, which correspond to modes with positive and negative frequencies with respect to $t$.
	These are  related to $a_{\bf k}, a^\dagger_{\bf k}$
	by  a Bogoliubov transformation. 
	
	After some algebra one gets that 
%
%
	\begin{eqnarray}
	\label{aadag}
	 {\tilde a}({\bf k})& = &{k^{i\mu}\over \sqrt{|\alpha_{\mu}|^2-|\beta_\mu|^2}}[\alpha_\mu a_{\bf k}+ \beta_\mu^{\ast}a^\dagger_{-{\bf k}}]\\
	 {\tilde a}^\dagger({\bf k}) & = & {k^{-i\mu}\over \sqrt{|\alpha_{\mu}|^2-|\beta_\mu|^2}}[\alpha_\mu^{\ast} a^\dagger_{\bf k}+ \beta_\mu a_{-{\bf k}}]
	 \end{eqnarray}
	 From eq.(\ref{[a,a+]}) it follows that these operators also satisfy the commutation relation
	 \be
	 \label{comm}
	 [{\tilde a}({\bf k}), {\tilde a}^\dagger({\bf k}')]=(2\pi)^d\delta^d({\bf k}-{\bf k}')
	 \ee
	 
	 In terms of $\Phi({\bf k},\eta)$ and $\Pi({\bf k},\eta)$, the  momentum modes for the field $\Phi$ and its conjugate momentum $\Pi$,  eq.(\ref{Phik}), eq.(\ref{Pik}), one finds that 	
	\begin{equation}
		\label{rhoop}
		{\tilde a}({\bf k})=
		\left[{1\over(-\eta)^{\Delta_+} \gamma} \right]
\left[\left({d\over2}-i\mu\right)\Phi({\bf k},\eta)+(-\eta)^d\Pi({\bf k},\eta)\right]   
	\end{equation} 
	where 
	\be
	\label{gamma}
	\gamma=-2i\mu\sqrt{|\alpha_\mu|^2-|\beta_\mu|^2}=-i \sqrt{2\mu}
	\ee
	Before proceeding we note that in position space eq.(\ref{rhoop}) becomes
	\be
	\label{relppos}
	{\tilde a}({\bf x})=
		\left[{1\over(-\eta)^{\Delta_+} \gamma} \right]
\left[\left({d\over2}-i\mu\right)\Phi({\bf x},\eta)+(-\eta)^d\Pi({\bf x},\eta)\right]   
\ee
	Now consider a coherent  state which is an eigenstate of ${\tilde a}({\bf k})$ with eigenvalue $\rho({\bf k})$, 
	\begin{equation}
		\label{rhoeigsys}
		{\tilde a}({\bf k})\ket{\rho}=\rho({\bf k})\ket{\rho}
	\end{equation}
	The eigenbasis $\ket{\phi}$ for the field ${\hat \Phi}$ was defined in eq.(\ref{eigphia}). 
	The wave function for $\ket{\rho}$ in this basis 

	\begin{equation}
		\Psi_\rho[\varphi]=\bra{\varphi}\ket{\rho}
	\end{equation}
	can be obtained as follows. 
	We recall, eq.(\ref{eigpi}),  that the momentum operator in $\ket{\phi}$ basis is given by 
	\be
	\label{defmom}
	\Pi({\bf k},\eta)=-i(2\pi)^d {\delta\over\delta\varphi(-{\bf k},\eta)}
	\ee
	Inserting this in eq.(\ref{rhoop}) we get from eq.(\ref{rhoeigsys}) that the wave function must satisfy the equation 
	\begin{equation}
		\label{equPsi}
		\left[\left({d\over2}-i\mu\right)\varphi({\bf k},\eta)+(-\eta)^d\left(-i(2\pi)^d{\delta\over\delta\varphi(-{\bf k},\eta)}\right)\right]\Psi_\rho[\varphi]=\rho({\bf k})(-\eta)^{\Delta_+}\gamma\Psi_\rho[\varphi]
	\end{equation}
	where
The solution to eq.(\ref{equPsi}) (up to an overall normalisation)  is given as
	\begin{equation}
		\label{solPsi}
		\Psi_\rho[\varphi]=\exp\left[\int {d^d{\bf k}\over(2\pi)^d}{i\over(-\eta)^d}\left\{\gamma(-\eta)^{\Delta_+}\rho({\bf k})\varphi(-{\bf k},\eta)-{\Delta_-\over2}\varphi({\bf k,\eta})\varphi(-{\bf k,\eta})\right\}\right]
	\end{equation}
	where $\Delta_-$ is given in eq.(\ref{dimdelm}). 
	
	Before proceeding we note that the solution of eq.(\ref{equPsi}) as given in eq.(\ref{solPsi}) is unique up to an overall normalisation which can depend on $\rho$. Choosing for this normalisation a dependence on $\rho^\ast$ which is local in position space, e.g. of the form 
	\be
	\label{norma}
	N=\exp\left[{\int d^d{\bf x}  (-\eta)^{2i\mu}{\rho^\ast({\bf x})^2}}\right], 
	\ee
	will only change the contact terms in the field theory. A more non-trivial kernel in the exponent in eq.(\ref{norma}) is allowed by eq.(\ref{equPsi}) but will be significantly constrained by the Ward identities which we discuss next.   In the subsequent discussion we will work with the solution given in eq.(\ref{solPsi}).

The wave function for the Bunch Davies vacuum  $\ket{0}$ in the $\ket{\phi}$ basis, $\psi[\phi,\eta]$, was obtained in eq.(\ref{psi[phi]}); it can be used to obtain the wave function for this vacuum in the coherent state basis, $\psi[\rho]$, using the relation,
\begin{equation}
		\label{der1psirho}
		\begin{split}
			\psi[\rho,\eta]=\bra{\rho}\ket{0}=\int {\cal D}\varphi\bra{\rho}\ket{\varphi}\bra{\varphi}\ket{0}=\int {\cal D}\varphi\Psi_\rho^{\ast}[\varphi]\psi[\varphi,\eta] 
		\end{split}
	\end{equation} 

	From eq.(\ref{solPsi}), \eq{psifa}, we get
	\begin{equation}
		\label{der2psirho}
		\begin{split}
			\psi[\rho,\eta]=\int{\cal D}\varphi&\exp\left[\int {d^d{\bf k}\over(2\pi)^d}\left({-i\gamma^{\ast}(-\eta)^{\Delta_-}\over(-\eta)^d}\rho^{\ast}({\bf k})\right)\varphi({\bf k},\eta)\right]\\
			&\times\exp\left[-{1\over2}\int {d^d{\bf k}\over(2\pi)^d}\left(-{i\Delta_+\over(-\eta)^d}-\frac{i\partial_\eta f_\mu(k,\eta)^{\ast}}{(-\eta)^{d-1}f_\mu(k,\eta)^{\ast}}\right)\varphi({\bf k,\eta})\varphi(-{\bf k,\eta})\right]
		\end{split}
	\end{equation} 
	where $f_\mu(k,\eta)$ is late time behaviour of ${\cal F}_{\mu}(k,\eta)$. Using the expression for $f_\mu(k,\eta)$ given in eq.(\ref{deffu}) this can be simplified and gives, 
	\begin{equation}
		\label{der2psirho1}
		\begin{split}
			\psi[\rho,\eta]=\int{\cal D}\varphi&\exp\left[\int {d^d{\bf k}\over(2\pi)^d}\left({-i\gamma^{\ast}(-\eta)^{\Delta_-}\over(-\eta)^d}\rho^{\ast}({\bf k})\right)\varphi({\bf k},\eta)\right]\\
			&\times\exp\left[-{1\over2}\int {d^d{\bf k}\over(2\pi)^d}\left({2\mu\alpha_\mu^{\ast}(-k\eta)^{-i\mu}\over (-\eta)^d[\alpha^{\ast}_\mu(-k\eta)^{-i\mu}+\beta_\mu^{\ast}(-k\eta)^{i\mu}]}\right)\varphi({\bf k,\eta})\varphi(-{\bf k,\eta})\right]
		\end{split}
	\end{equation} 
	
	Before proceeding we note from the form of the coherent state wave function $\Psi_\rho(\phi)$, eq.(\ref{solPsi}), that   the integral transformation involved in taking the wave function in the field basis, $\psi(\phi,\eta)$, to the coherent state basis, $\psi(\rho,\eta)$, eq. (\ref{der1psirho}), is a kind of ``shifted Legendre transformation", involving also a change in the Gaussian term in the wave function. 
	
		The functional integration in eq.(\ref{der2psirho1}) is  a Gaussian and can be carried out easily. We see it results in a non-trivial transformation on $\psi[\varphi,\eta]$. 
	
	Solving for $\rho$ in terms of $\Psi$, using the saddle point condition we get 
	\begin{equation}
		\label{phirelrho}
		\varphi({\bf k},\eta)={1\over \sqrt{2\mu}}\left[(-\eta)^{\Delta_-}+{\beta_\mu^{\ast}\over\alpha_\mu^{\ast}}k^{2i\mu}(-\eta)^{\Delta_+}\right]\rho^{\ast}(-{\bf k})
	\end{equation}
where we have put in the value of  $\gamma$ given in eq.(\ref{gamma}).


The resulting value of the integral (up to a normalisation which we are not keeping track of ) then takes the  form,
\begin{equation}
		\label{psirho1}
		\begin{split}
			\psi[\rho,\eta]=\exp\left[{1\over2}\int {d^d{\bf k}\over(2\pi)^d}\rho^{\ast}({\bf k})\left((-\eta)^{-2i\mu}+{\beta_\mu^{\ast}\over\alpha_\mu^{\ast}}k^{2i\mu}\right)\rho^{\ast}(-{\bf k})\right]
		\end{split}
	\end{equation}
	

	We see that the final form of the wave function in the coherent state basis is in fact quite simple and also suggestive  of an interpretation as a generating functional 
	 in a dual field theory. We identifying $\rho^{\ast}({\bf k})$ to be the source in the boundary theory, and $\psi$  to be the generating functional. And  see that the expression above leads to a two-point function in the CFT which has a contact term, arising from the first  term in eq.(\ref{psirho1}) that  is cut-off dependent, and a non-contact term, arising from the second term in eq.(\ref{psirho1}) that  is of the form of a  two point function for an operator of dimension $\Delta_+$, eq.(\ref{<O+O+>}). In position space, the two point function, using \eq{defabu} can be written as
\begin{equation}
	\label{rho2pt}
	\l O_+({\bf x})O_+({\bf y})\r={i\Gamma[1-i\mu]\Gamma[{d\over2}+i\mu]\over\pi^{1+{d\over2}}(1+\coth(\pi\mu))}{1\over|{\bf x}-{\bf y}|^{d+2i\mu}}
\end{equation}	
Comparing eq.(\ref{phirelrho}) and eq.(\ref{defjpb})  we also see that they are in fact  of the same form, and that $\rho^{\ast}({\bf k})$ and $J_+({\bf k})$ 
 can be identified up to a normalisation factor
	\begin{equation}
		\label{Jrelchi}
		\rho^{\ast}({\bf k})=\sqrt{2\mu}\alpha_\mu^{\ast}J_+(-{\bf k})
	\end{equation} 
	The source $J_+$  first defined  in section \ref{holo}  is then, as promised, found  to be  closely related to the eigenvalue of the coherent 
	state. 
	
	It is worth comparing the expression obtained for $\Psi$  in the coherent state basis with what we had obtained in  eq.(\ref{wvr2}), eq.(\ref{integrand}) 
	 in the field eigenstate basis. 
	The most important difference is that the second term in eq.(\ref{integrand}), which is cut-off dependent and non-local is missing in eq.(\ref{psirho1}) above. As a result the remaining divergent term in eq.(\ref{psirho1}) is of a conventional type, being local in space. 
	Also, the coefficient of the cut-off independent term is different even after we account for the rescaling factor eq.(\ref{Jrelchi}). Using eq.(\ref{Jrelchi})   to express $J_+$ in terms of $\rho^{\ast}$
	we see that this term in eq.(\ref{integrand}) becomes
	\be
	\label{twpaf}
	\delta \log\Psi[\rho]={1\over 2} ({-id\over 2\mu})\int{d^d{\bf k}\over (2\pi)^d} \rho^{\ast}({\bf k})\rho^{\ast}(-{\bf k}) k^{2i\mu} {\beta_\mu^{\ast}\over \alpha_\mu^{\ast}}
	\ee
	Comparing with the second term in eq.(\ref{psirho1}) we see that there is a difference in the coefficient by a factor of ${-i d\over 2\mu}$. This difference will be important when we consider the Ward identities in the next section. Finally, coming back to the remaining divergent term in eq.(\ref{psirho1}) we note that the first term in eq.(\ref{integrand}) is also of  the same form, being cut-off dependent and local,  but differs in its coefficient  even after the rescaling eq.(\ref{Jrelchi}).\\~\\ 
	Our discussion above has been in the free field limit. Going further we can incorporate interactions. Including  a $\phi^n$ interaction which was discussed in section \ref{interaction2}, and working to first order in $\lambda$, one can easily obtain the corresponding interaction term in the coherent state basis, by simply inserting eq.(\ref{Jrelchi})
	in eq.(\ref{delSunderJJ1}) which has the interaction term expressed as a $n^{\rm th}$ order  polynomial in $J_+$. This gives the coherent state wave function
%
%
%
see Appendix \ref{chiinter} 
	\begin{empheq}{multline}
		\label{intwvfchi}
			\psi_I[\rho,\eta]=\psi[\rho,\eta]\exp\left[{i\lambda\over(\sqrt{2\mu}\alpha_\mu^{\ast})^n}\int\left(\prod_{i=1}^{n}{d^d{\bf k}_i\over(2\pi)^d}\right)(2\pi)^d\delta({\bf k}_1+\cdots+{\bf k}_n)\right.\\
			\left.{\left(\prod_{i=1}^{n}k_i^{i\mu}\right)I(k_1,\cdots,k_n)}\left(\prod_{i=1}^{n}\rho^{\ast}({\bf k_i})\right)\right]
	\end{empheq}     
	where  the integral $I(k_1,\cdots,k_n)$ is given in eq.(\ref{defin}). 
	From our earlier discussion in section \ref{interaction2} it follows that this is of the correct form to be the $n$ point correlator for an operator of dimension $\Delta_+$. Explicitly for $n=3$,
	\be
	\label{3ptcoh}
	\l O_+({\bf k}_1)O_+({\bf k}_2)O_+({\bf k}_3)\r= {3!i\lambda\over(\sqrt{2\mu}\alpha_\mu^{\ast})^3}(2\pi)^d\delta({\bf k}_1+{\bf k}_2+{\bf k}_3) {\left(\prod_{i=1}^{3}k_i^{i\mu}\right)I(k_1,k_2,k_3)}
	\ee	
	More generally, going beyond leading order in $\lambda$, we would need to start with  $\psi[\varphi,\eta_1]$, with  the correct coefficient function  up to the required order for the $\phi^n$ term,
	 and then carry out the integral over 
	$\varphi$ in eq.(\ref{der1psirho}). Denoting the radius of $dS_{d+1}$ space as $R_{\rm dS}$, we note that we have been 
	 suppressing an overall factor of ${R_{\rm dS}^{d-1}\over G_N}$,  which appears in front of the action for the scalar field, \eq{action}. Once we keep track of this factor it multiplies 
	 both terms in the exponent in $\Psi_\rho[\phi]$, eq.(\ref{solPsi}), as well as all the terms in the exponent of $\psi[\phi]$, as is discussed in Appendix \ref{chiinter}. 
	 As a result in the limit where ${R_{\rm dS}^{d-1}\over G_N} \rightarrow \infty$, we can evaluate the integral in eq.(\ref{der1psirho}) in the saddle point approximation. 
	  For this purpose it is   best to express $\Psi_\rho[\phi]$, eq.(\ref{solPsi}),  in terms of $J_+$. Using the relation eq.(\ref{defjpb}) it takes the form 
	 \begin{empheq}{multline}
	 \label{formfa}
	 \Psi_\rho[J_+]=\exp\left[\int {d^d{\bf k}\over(2\pi)^d}{i\over(-\eta)^d}\left\{\gamma f_\mu^\ast(k,\eta) k^{i\mu}(-\eta)^{\Delta_+}\rho({\bf k}) J_+(-{\bf k})\right.\right.\\
	 \left.\left.-{\Delta_-\over2}(f_\mu^\ast(k,\eta))^2 k^{2i\mu} J_+({\bf k})J_+(-{\bf k})\right\}\right]
	 \end{empheq}
	 We also then  express $\Psi[\phi]$ in terms of $J_+$ which we have done in previous sections anyway. 
	 In the saddle point approximation we then extremize the resulting full exponent in eq.(\ref{der1psirho}) as a functional of $J_+$ and solve the resulting equation to obtain $J_+$ in terms of $\rho^{\ast}$ 
	  order by order in $\lambda$. Inserting this solution for $J_+$ and evaluating the on-shell value of the  exponent in eq.(\ref{der1psirho}) to the required order 
	  in $\lambda$ then gives the coherent state wave function. More details can be found in Appendix \ref{chiinter}. 
	  
	  In summary the correspondence for underdamped fields is to consider the wave function in the coherent state basis and to 
	  expand it in a Taylor series in $\rho^{\ast}$, i.e.  in an expansion of the  form   eq.(\ref{genfa}) with the source $S({\bf x})$ being $\rho^{\ast}({\bf x})$,
	  \be
	  \label{expwfa}
	  \log(\psi[\rho,\eta])=\sum_{n=1}^\infty {1\over n!}\int \prod_{j=1}^{n} d^d{\bf x}_j \rho^{\ast}({\bf x}_j) \l O_+({\bf x}_1) O_+({\bf x}_2) \cdots O_+({\bf x}_n)\r
	  \ee
	  The coefficient functions then correspond to correlation functions of fields in the dual CFT which have dimension 
	  \be
	  \label{defdeltap}
	  \Delta_+={d\over2}+i \mu
	  \ee
	  We will see below that these coefficient functions indeed satisfy the Ward identities of conformal invariance for fields with dimension $\Delta_+$. 
	  
	  Before ending this section it is important to connect the discussion above to the important reference, \cite{Isono}. 
	  In \cite{Isono} it was proposed that new boundary conditions could be imposed in the underdamped case and this would lead to a satisfactory identification of  a source in the boundary CFT, see eq (2.14), \cite{Isono}. The new boundary condition requires, for a well defined variational principle to yield the equations of motion, that the action is also modified by adding an extra boundary term, see eq(2.16), \cite{Isono}. 
	  A little thought shows that, in fact, this proposal  completely agrees with our conclusions  above. The source $\chi^+_{\bf k}$, eq.(2.14), \cite{Isono}, after a rescaling, is the source $\rho^*({\bf k})$ above. The main new addition  our discussion  provides is the understanding that the resulting transformation  has the simple interpretation of leading to   the  wave function in the familiar coherent state basis, and the resulting source in the boundary theory, $\rho^*$,   is in fact  just the eigenvalue of the coherent state. To make this point clearer we show in appendix \ref{Harmo} that for a harmonic oscillator, the change in boundary conditions, similar to what was done in \cite{Isono}, along with an extra boundary term in the action,  gives rise, in the semiclassical approximation,  to the wave function  being obtained in the coherent state basis. 
	  
	  There are some practical advantages that come with the understanding in terms of the coherent state basis that our discussion above  provides.   
	  First, as discussed above and further in appendix \ref{appenboun}, we show how interactions can be added in a straightforward manner.  Doing so {\it does not} require us to change the integral transform, see eq.(\ref{der1psirho}),  eq.(\ref{der2psirho1}) above,  which takes us from the field basis to the coherent state basis, in the free theory. This means  one can continue to use familiar Witten diagrams to calculate the wave function first in the field basis and only at the last step then carry out the transformation to express the wave function in the coherent state basis. This also means that interactions involving both underdamped and overdamped fields can be calculated and related to the CFT in a straightforward manner. Second, as will be discussed in section \ref{WI}, it allows us to derive the Ward identities which must be satisfied by the coefficient functions of the wave function in the coherent state basis in a transparent manner,  and show that they agree with the Ward identities in a CFT for a field of dimension $\Delta_+$, eq.(\ref{defdeltap}). 
	  Finally, since  the strategy for computing the interaction terms does not have to change much, one can continue to rely on the understanding of how divergences arise in the Witten diagrams  in the underdamped case, or  more generally in cases involving both underdamped and overdamped fields, and isolate these divergences easily, see section \ref{divergence} and Appendix \ref{analI}. 
	  
	  We should also mention the authors of \cite{Isono} provide an interesting interpretation of the wave function in the field basis (as opposed to the coherent state basis). They argue that once the eigenvalue of the coherent state is identified as the  source in a CFT, the asymptotic value of the field,  denoted as $\phi_{\bf k}(\tau)$ in \cite{Isono}, acts like a source in a theory obtained by adding a double trace deformation to the CFT. In this paper, see  section (\ref{cohJ}), we discuss another possibility. Namely,  rather than the asymptotic value of the  field, we consider one of its two different falls-offs,  and argue that either can be identified with a source in a CFT, denoted by $J_+$ or $J_-$, provided the identification is made in an appropriate manner.  We  then show how the ward identities of the CFTs will continue to hold, after such an   identification is carried out, see section (\ref{cohJ}) for further details.

			\section{Ward Identities }
			\label{WI}
			Here we will discuss various  Ward identities which are satisfied by the correlators of the field theory, after an  identification is made  between the asymptotic value of the bulk fields and sources in the field theory, as discussed above. 
			
			Compared to the AdS case there are two important things to keep in mind. First,  we will also include the under damped fields. 			
			Second, as has been argued above, in the dS case one cannot neglect the divergent terms which arise as the late time cut-off, $\eta_1\rightarrow 0$, and one is therefore dealing with a field theory which depends on a cut-off. 
			We will argue that the cut-off dependent terms in the correlation functions are also invariant under conformal transformations, as long as the cut-off, $\eta_1$,   transforms appropriately. 
			 For terms which are independent of the cut-off, 
			  the Ward identities   correspond to the Ward identities of a Conformal Field Theory and will include the constraints that follow from conformal invariance,
			   translational  invariance and local scale invariance, i.e., the vanishing trace of the stress tensor.
			   If additional  global symmetries are present they will include the constraints which follow from charge current conservation. 
			   
			  		Various aspects regarding the Ward identities for overdamped fields have been discussed earlier, see \cite{MaldaPimen, BzowskiWI} and in the context of inflation in \cite{Kunduward,slowrollTrivedi}. Here our  focus is on underdamped fields for which we have proposed an identification of the coherent state eigenvalue as the source in the dual CFT.
					The question we  address in this section is whether this identification  made leads to the correlation functions of the CFT satisfying the required ward identities of conformal invariance, and when additional global symmetries, like a $U(1)$ symmetry, are present the Ward identities for these global symmetries. 
		
	\subsection{Summary of Key Points}
	 Before proceeding we summarise some of the key points of this section here. The issue of meeting the Ward identities in Holography is in fact   somewhat subtle. The reader may recall that in the initial  studies of AdS/CFT,  see \cite{Wittenads},  it was proposed that the  non-normalisable fall-off of the bulk field should correspond to the source in the CFT. Subsequent investigation in the context of the $N=4$ SYM theory, see \cite{freedU1}, though revealed that the Ward identities for the $SU(4)$ $R$ symmetry, in SYM theory,  would then not work, unless the correspondence was altered somewhat and the relation between the asymptotic value of the bulk field $\phi({\bf k},z)$ and the source ${\hat \phi}({\bf k})$ was taken to be  of the more precise form  given in eq.(\ref{solab}). Similarly, if  we attempted to  identify  boundary source in dS, for underdamped fields,  in terms of $J_+({\bf k})$ or $J_-({\bf k})$,   as given in eq.(\ref{defjpb}), eq.(\ref{subst1}), eq.(\ref{subst2}), or in an analogous manner for over damped fields,  the Ward identities would also not hold.		
		
		Here we show that in general, in the presence of   gauge invariant local interactions in the bulk, the Ward identities will hold if the source is taken to be complex conjugation of the coherent state eigenvalue, $\rho^*({\bf k})$, \eq{rhoeigsys}. The key point is the following. The identification made in \cite{freedU1}, eq.(\ref{solab}), leads in position space, in the AdS case, to the following relation  between the bulk field at the boundary $z=\epsilon$ and the source in position space, 
		\be
		\label{cariden}
		\phi({\bf x},\epsilon)=\epsilon^{{d\over 2}-\nu}{\hat \phi}({\bf x})
		\ee
	We see that this relation is local along the boundary, relating the left and right hand sides at the same boundary location ${\bf x}$. Instead if one took  the identification with the source ${\hat \phi}$ to be 
	\be
	\label{naida}
	\phi({\bf k},\epsilon)= \epsilon^{{d\over2}-\nu} 
	{\hat \phi}({\bf k}) + c_\nu \epsilon^{{d\over2}+\nu} k^{2\nu} {\hat \phi}({\bf k}),
	\ee
	,so that only the leading non-normalisable mode was identified  with the source in the boundary theory, then   while the first term on the RHS would be locally related along the boundary to the bulk field, the second term would be related in a non-local manner. It is this non-local aspect that spoils the Ward identities and leads to  the more precise identification in eq.(\ref{solab}) being needed. 
	
	In the underdamped case the source  $\rho^*$ which we have identified is locally related to  bulk data on the  boundary $\eta=\eta_1$. This follows from the fact that  $\rho$ is the eigenvalue of the destruction operator ${\tilde a}$ defined in \eq{rhoeigsys}, and this operator is related to the bulk field $\Phi $ and its conjugate momentum $\Pi$, at the boundary  $\eta=\eta_1$ by, the local  relation, eq.(\ref{relppos}),
	\be
	\label{leads2aa}
	{\tilde a}({\bf x})=\left[{1\over (-\eta_1)^ {\Delta_+}\gamma}\right]
	\left[\left({d\over 2}-i\mu\right)
	\Phi({\bf x},\eta_1) + (-\eta_1)^d \Pi({\bf x},\eta_1)\right]
	\ee
	This local aspect is the key and then leads to all the Ward identities holding as we explain in this section. 
	
	In subsection \ref{cohJ} towards the end  of this section we use these insights and argue that  a source can also be identified in terms of $J_\pm$, such that the finite terms in the correlation functions satisfy the Ward identities, but this requires the second term in eq.(\ref{defjpb}), and similarly the second term that would arise in  eq.(\ref{subst2}),   to be changed to include an appropriate and non-trivial dependence on the metric.


	\subsection{More Details}	
		There are two ways to understand the arguments leading to the Ward identities. 
			From the perspective of Canonical quantisation we note that we are working throughout this paper  in ADM gauge.
			Including perturbations,  the  metric in this gauge is given by 
			  \be
			  \label{metadm}
			  ds^2=-{d\eta^2\over \eta^2} + {1\over \eta^2} ({\hat g}_{ij} dx^i dx^j) 
			  \ee
			 so  that the lapse function  and shift function have been set to $N^2={1\over \eta^2}$ and  $N^i=0$ respectively. 
			  The equations of motion obtained by varying $N$ and $N^i$ then lead to 
			  the conditions of  time and spatial reparametrisation invariance. Any physical state, including the Bunch Davies vacuum,  must satisfy these conditions,
			  which ensure that the wave function is invariant under residual gauge transformations which preserve the ADM gauge. 
			   
			 The resulting conditions then give rise to the Ward identities. Specifically, spatial reparametrisation invariance gives rise to the Ward identities of translational invariance  and time reparametrisation invariance gives rise to the Ward identity of local scale invariance, i.e. 
			vanishing of the   trace of the stress tensor, $T^i_i=0$, up to contact terms. The $SO(1,d+1)$ isometries of $dS_{d+1}$ space are a combination of spatial and time reparametrisation transformations and their Ward identities also then follows from the above invariances. 
			 
			 For additional gauge symmetries, e.g. a $U(1)$ gauge symmetry,   we work in $A_\eta=0$ gauge and the Gauss law constraint on the wave function then  leads, in a similar manner, to the Ward identities of charge conservation. 
			 
			 From the path integral perspective the wave function is  evaluated by carrying out a path integral subject to appropriate boundary conditions. This path integral is invariant under general coordinate transformations, as long as the boundary conditions are also transformed appropriately to account for the coordinate transformations. The resulting conditions on the wave function then give rise to  the Ward identities. The boundary conditions need to be imposed both on the Poincaré horizon and the late time slice. 
						We argued in section \ref{psipathint}, eq.(\ref{cone})  that after a suitable $i \epsilon$ prescription the wave function vanishes at the Poincaré horizon. This will continue to be true after the  coordinate transformations we consider. Therefore  the invariance of the path integral  follows as long as the boundary values of fields are transformed appropriately on the late time slice.
			 
			In practice one often works in the classical limit where the path integral can be evaluated in the saddle point approximation, i.e., by solving the equations of motion subject to the required boundary conditions. The invariance of the wave function then follows from the invariance of the on-shell action under coordinate transformations, once the boundary values are also suitable transformed. The classical limit is justified by taking 
			${R_{\rm dS}^{d-1}\over G_N}\rightarrow \infty$. 
			

			In most of the discussion below  we only consider scalar  fields. The extension to other fields, scalars with higher spin, or fermions,  is straightforward but we do not discuss them here. 
			The one exception is the metric, which we include. This field  is of course special since it is dual to the stress tensor in the boundary.

From the point of view of our present investigation the non-trivial aspect of the analysis is as follows. Consider evaluating the late time wave function on the slice $\eta=\eta_1$, with the bulk scalar taking the value $\phi({\bf x},\eta_1)$ and the metric components, eq.(\ref{metadm}), taking the value ${\hat g}_{ij}({\bf x},\eta_1)$.
Under a  coordinate transformation which takes
\begin{eqnarray}
\label{transgen}
\phi & \rightarrow &  \phi-\delta \phi \\
{\hat g}_{ij} & \rightarrow &  {\hat g}_{ij}-\delta {\hat g}_{ij} \\
\eta_1& \rightarrow & \eta_1-\delta \eta_1
\end{eqnarray}
 the invariance of the wave function $\Psi$ leads to the condition
\be
\label{condinva}
\Psi[\phi-\delta \phi,{\hat g}_{ij}-\delta {\hat g}_{ij}, \eta_1-\delta \eta_1]-\Psi[\phi,{\hat g}_{ij},\eta_1]=0
\ee
To relate  this to constraints on the correlation functions of the CFT one needs to find out how the source one is associating with the bulk field transforms under the coordinate transformation eq.(\ref{transgen}), obtain the invariance condition eq.(\ref{condinva}) in terms of the sources, and then finally argue that the
coefficient functions which arise when the wave function is expanding in terms of the sources  satisfy the identities in a CFT. 

When the relation between the boundary value of the fields and the source is local along the hypersurface $\eta=\eta_1$ as is the case for the overdamped case, or in the AdS/CFT correspondence, obtaining the transformation laws for the source is straight forward and  the subsequent argument that  the invariance of the wave function leads to the required Ward identities is also straightforward. However, if the relation between the bulk fields and the source is non-local, as  happens for the identification of the source we made in the underdamped case in eq.(\ref{subst1}) or eq.(\ref{subst2}), see section \ref{under} for discussion, then obtaining the transformation laws for the sources is more non-trivial and formulating the arguments requires more care. In the case of the coherent state basis the relation between the bulk fields $\phi({\bf x},\eta)$ and $\partial_\eta  \phi({\bf x},\eta)$ and the coherent state eigenvalue, $\rho$, is more straightforward and the discussion leading to the Ward identities  is also then straightforward. 

Accordingly we will first consider the coherent state basis below in section \ref{cohward} and then turn to other source $J_+$ or $J_-$ in our subsection \ref{cohJ} later. 
Additional discussion and some explicit calculations are contained in Appendix \ref{Wardtest}. 

Before proceeding let us note  an important conclusion one arrives at from the analysis of the Ward identities.
As has been discussed above, the dual theory in dS space is Euclidean, and also  breaks reflection positivity. It  therefore cannot be continued  in general to Lorentzian space.  And the boundary fields in the Euclidean theory, to which  the bulk sources couple,  also  cannot then be continued to operators in a Lorentzian theory. Nevertheless, we  find that these fields in the boundary theory transform under conformal transformations like primary operators in a CFT. Moreover, as  follows from the discussion in section \ref{discuss} and also Appendix \ref{OPEint},  the nature of short distance singularities when these fields come together in correlation functions is entirely analogous to that of primary operators in a CFT, allowing us to define an expansion analogous to the operator product expansion, and use the bulk correlators to extract operator product coefficients. Thus, despite the obstruction to a continuation to Lorentzian space, much of the structure of the resulting boundary theory is identical to that of   conventional CFTs which admit such a Lorentzian continuation. 

\subsection{Ward Identities in the Coherent State Representation}
\label{cohward}
We note that the operator ${\tilde a}({\bf x})$ whose eigenstates we will work with in this representation is related to $\Phi({\bf x},\eta)$, $\Pi({\bf x},\eta)$ by the relation
eq.(\ref{relppos}). 
The subsections below give the general arguments leading to the Ward identities. Some explicit calculations are presented in Appendix \ref{Wardtest}. 

\subsubsection{Ward Identities for Spatial Reparametrisations}
\label{WISR}
The spatial transformations which keep us in the ADM gauge, eq.(\ref{metadm}) are given by 
\be
\label{sptr}
x^i \rightarrow x^i +v^i({\bf x})
\ee
where $v^i({\bf x})$ are infinitesimal parameters.
Under it $\Phi({\bf x},\eta)$ and $\Pi({\bf x},\eta)$ both transform as scalars, i.e., their changes are, 
\be
\label{chnga}
\delta \Phi=-v^i\partial_i \Phi, \ \ \delta \Pi({\bf x},\eta)=-v^i \partial_i \Pi({\bf x},\eta)
\ee
As a result we see from eq.(\ref{relppos}) that ${\tilde a}({\bf x})$ also behaves like a scalar and its eigenvalue must also transform like a scalar,
i.e. 
\be
\label{chngrhospat}
\delta \rho({\bf x})=-v^i\partial_i \rho({\bf x})
\ee

The metric transforms in the usual fashion 
\be
\label{mettransspat}
\delta {\hat g}_{ij}=-[\hat{\nabla}_i v_j + \hat{\nabla}_j v_i]
\ee
where $\hat{\nabla}$ denotes covariant derivative compatible with $\hat{g}_{ij}$.
It follows then from eq.(\ref{condinva}) that the condition for invariance of the wave function in the coherent state basis   leads to 
\be
\label{condinaa}
\Psi[\rho+\delta \rho,{\hat g}_{ij}+\delta {\hat g}_{ij}, \eta_1]= \Psi[\rho, {\hat g}_{ij}, \eta_1]
\ee

To proceed we first note that the expansion of the wave function in terms of coefficient functions was schematically given in eq.(\ref{genfa}).
Let us be more precise about this. We will expand the metric ${\hat g}_{ij}$ about its flat space value
\be
\label{exfl}
{\hat g}_{ij}=\delta_{ij}+\gamma_{ij}
\ee
Then the Taylor series expansion of $\Psi$ will be carried out in powers of $\gamma_{ij}$ and of the scalar source $\rho$, in the coherent state basis, or in terms of $J_\pm$ in the field basis, as we will discuss later. 
The coefficient functions will be defined as follows:
\begin{empheq}{multline}
\label{coeffiexp}
\log[\Psi]  =  {1\over 2} \int d^d{\bf x}d^d{\bf y}\sqrt{{\hat g}({\bf x})}\sqrt{{\hat g}({\bf y})} \rho^\ast({\bf x})\rho^\ast({\bf y})\l O_+({\bf x})O_+({\bf y})\r \\
+
{1\over 4} \int d^d{\bf x}d^d{\bf y}d^d{\bf z} \sqrt{{\hat g}({\bf x})}\sqrt{{\hat g}({\bf y})}\sqrt{{\hat g}({\bf z})}
\gamma_{ij}({\bf z})\rho^\ast({\bf x})\rho^\ast({\bf y}) \l T_{ij}({\bf z})O_+({\bf x})O_+({\bf y})\r
+\cdots	
\end{empheq}
From eq.(\ref{exfl}) we see that on starting with the flat metric $\delta_{ij}$ and carrying out a spatial reparametrisation gives,
\be
\label{chnggamma}
\gamma_{ij}=-[\hat{\nabla}_i v_j +\hat{\nabla}_j v_i]
\ee
Also, the invariance condition eq.(\ref{condinaa}) then leads to the condition 
\be
\label{condpsiaa}
\Psi[\rho+\delta \rho, \delta_{ij}+\gamma_{ij}, \eta_1]=\Psi[\rho,\delta_{ij}, \eta_1]
\ee
For the coefficient functions being considered in eq.(\ref{coeffiexp}) this gives 
a  relation between the two point and three point correlators:
\be
\label{condtwothree}
\del_i^z\l T^{ij}({\bf z})O_+({\bf x})O_+({\bf y})\r + \delta({\bf z}-{\bf x})\del_i^x\l O_+({\bf x})O_+({\bf y})\r + \delta({\bf z}-{\bf y})\del_i^y\l O_+({\bf y})O_+({\bf x})\r=0
\ee
as is discussed further in Appendix \ref{appSrep}. 
Note this is the expected relation in a CFT with the operator $O_+({\bf x})$ transforming as a scalar. 
Similar relations can also be obtained relating correlation functions involving $m$ scalar fields and a stress tensor, or multiple stress tensors. 
In  all cases the stress tensor will be conserved, i.e., will meet the condition 
\be
\label{condaa}
\partial_iT^{ij}=0
\ee 
up to contact terms which involves the scalars or additional stress tensors transforming appropriately under spatial reparametrisations, i.e. as scalar operators or a rank two symmetric tensor. 

We also note that the correlation functions will also have cut-off dependent terms in them. The Ward identities above do not mix the cut-off dependent terms with those which are cut-off independent. 
The cut-off independent terms will satisfy the Ward identities of a standard CFT, the cut-off dependent terms will also satisfy the relations obtained from the conditions discussed above. Physically these Ward identities express the fact that the dual field theory  correlators are translational invariant and the stress tensor is therefore conserved, up to  precise contact terms. 

\subsubsection{Ward Identity for Time Reparametrisation}
\label{witra}

The time reparametrisation which preserves ADM gauge asymptotically, as $\eta_1\rightarrow 0$, takes the form,
\begin{eqnarray}
\eta & \rightarrow  & \eta(1+\epsilon({\bf x})) \label{ass1}\\
x^i & \rightarrow & x^i+{1\over 2}\eta^2 \partial_i \epsilon \label{ass2}
\end{eqnarray}
We see that can neglect the change of $x^i$ to leading order- although that argument might need to be more carefully examined when we consider the Ward identities involving cut-off dependent terms. 

To keep our discussion simple we will mostly focus on the cut-off independent terms below. We can then take $x^i$ to be invariant and take $\eta$ to transform as given in eq.(\ref{ass1}).

The scalar field $\Phi({\bf x},\eta)$ will transform as a bulk scalar under this transformation and its change is therefore 
\be
\label{chngaa}
\delta \Phi({\bf x},\eta)=-\epsilon({\bf x})\eta\partial_\eta \Phi({\bf x},\eta)
\ee
From eq.(\ref{defPi}) we see that 
\be
\label{combpi}
(-\eta)^d \Pi({\bf x},\eta)=-\eta\partial_\eta \Phi({\bf x},\eta)
\ee
From this it also follows that 
\be
\label{changab}
\delta ((-\eta)^d \Pi({\bf x},\eta)) = -\epsilon({\bf x}) \eta\partial_\eta ((-\eta)^d \Pi({\bf x},\eta))
\ee
In other words $(-\eta)^d \Pi({\bf x},\eta)$ also behaves like a scalar under the time reparametrisation of interest. 
At first sight this might  seem unexpected. To see the above result simply, note that RHS in eq.(\ref{combpi}) can be written as $\partial_{\log(-\eta)} \Pi$ and $\delta \log(\eta)= \epsilon({\bf x})$ which is independent of $\eta$. 

Let us now write the expression for $a({\bf x}) $ in terms of $\Phi, \Pi$ as follows, eq.(\ref{relppos}):
\be
\label{newexp}
\tilde{a}({\bf x}) (-\eta)^{\Delta_+}= {i\over\sqrt{2\mu}}\left[\left({d\over2}-i\mu\right)\Phi({\bf x},\eta)+(-\eta)^d\Pi({\bf x},\eta)\right]
\ee
Since both terms on the RHS transform as scalars we conclude that the change of the LHS  is given by 
\be
\label{chngbt}
\delta (\tilde{a}({\bf x})(-\eta)^{\Delta_+})=-\epsilon({\bf x}) \eta\partial_\eta (\tilde{a}({\bf x}) (-\eta)^{\Delta_+})
\ee
From this it follows that 
\be
\label{chngff}
\delta \tilde{a}({\bf x})=-\Delta_+ \epsilon({\bf x}) \tilde{a}({\bf x})
\ee
As a result its eigenvalue $\rho$ also transforms as 
\be
\label{tranrr}
\delta \rho({\bf x})=-\Delta_+ \epsilon({\bf x}) \rho({\bf x})
\ee
and its complex conjugate transforms as 
\be
\label{tanrh}
\delta \rho({\bf x})^\ast=-\Delta_- \epsilon({\bf x})\rho({\bf x})^\ast
\ee

It is also easy to see that under the time parametrisation we are considering the metric perturbation $\gamma_{ij}$, eq.(\ref{exfl})  transform as 
\be
\label{metrcompt}
{\gamma}_{ij}=2 \epsilon({\bf x}) \delta_{ij}
\ee
The invariance of the wave function, \eq{psirho1} then takes the form
\be
\label{condwca}
\Psi[\rho+\delta \rho, \delta_{ij}+ \gamma_{ij}, \eta_1(1+\epsilon)]=\Psi[\rho,{\delta}_{ij},\eta_1]
\ee
For cut-off independent terms we can neglect the change in $\eta_1$. 

Applying this condition to the cut-off independent terms in the  two and three point correlators in eq.(\ref{coeffiexp}) we learn that 
\be
\label{condconf}
\l T^{i}_i({\bf z})O_+({\bf x})O_+({\bf y})\r+\Delta_+\delta({\bf z}-{\bf x})\l O_+({\bf x})O_+({\bf y})\rangle+\Delta_+\delta({\bf z}-{\bf y})\l O_+({\bf y})O_+({\bf x})\r=0
\ee
This is the standard Ward identity for local scale invariance in a CFT. Namely, the trace of the stress tensor vanishes up to contact terms which are proportional to the anomalous dimensions of the operators. 

Similar relations will arise when there are more operators with the $T^i_i$ vanishing again up to appropriate contact terms which are proportional to the anomalous dimensions of the operators involved. 
When we consider cut-off dependent terms we will also have to take into account the change in $\eta_1$, which follows from eq.(\ref{ass1}). 

Appendix \ref{twi} has further discussion of how the bilinear and trilinear terms in eq.(\ref{condconf}) are obtained by doing explicit calculations along with some subtleties that arise.

	\subsubsection{Ward Identities for Conformal Invariance}
	Although, as mentioned above, conformal  transformations are  a combination of spatial and time reparametrisations, it is worth considering them separately since they correspond to isometries of the bulk metric and invariance under them is a defining feature of CFTs. 	
	The $SO(1,d+1)$ isometries of $dS_{d+1}$, which correspond to conformal transformations,  are described in Appendix \ref{conft}, see eq.(\ref{dif1})-eq.(\ref{dif4}). 
	
	Consider a general conformal transformation under which 
		\be
	\label{defc}
	x'^i=x^i+v^i, \eta'=\eta(1+\epsilon({\bf x}))
	\ee
	From Appendix \ref{conft} we see that for translations and rotations $\epsilon({\bf x})=0$ and for dilations and special conformal transformations it takes the values
	$\epsilon({\bf x})=\epsilon$ and $\epsilon({\bf x})=2 b_ix^i$ respectively. 
	
	We see that this transformation is in general a combination of a spatial and time reparametrisation of the kind we had discussed above. 
	The coherent state eigenvalue $\rho$  and metric therefore will transform under this transformation as 
	\be
	\label{transfa}
	\delta \rho=-v^i({\bf x}) \partial_i \rho -\epsilon({\bf x}) \Delta_+ \rho
	\ee
	and 
	\be
	\label{cngga}
	\gamma_{ij}= -\hat{\nabla}_i v_j({\bf x})-\hat{\nabla}_jv_i({\bf x}) + 2 \epsilon({\bf x}) \delta_{ij}=0
	\ee
	The fact that the change in the metric vanishes is because conformal transformations are isometries of the dS metric. 
	As a result the coefficient functions under these transformations will transform homogeneously, i.e., a $n$ point scalar correlator will transform into itself and the resulting conditions which arise from invariance of the wave function on these coefficient functions will only involve it  and not other coefficient functions involve  extra factors of the stress tensor, as happened for the spatial and time reparametrisations individually in the previous subsections. 
	For the two -point correlator under a special conformal transformation we get, for example, 
	
	\be
	\label{condcnfin2}
	\l\delta O_+({\bf x}) O_+({\bf y})\r+\l O_+({\bf x}) \delta O_+({\bf y})\r=0
	\ee
	where the change 
	\be
	\label{chngO}
	\delta O_+({\bf x})=b^j[2\Delta x_j+(2x^ix_j-x^2\delta^{i}_{j})\del_i] O_+({\bf x})
	\ee
	Similarly relations can also be obtained for the other conformation transformations as is discussed in Appendix \ref{conft}. 
	$n$ pt correlators will also be invariant under special conformal   with each operator  transforming as given in  eq.(\ref{chngO}).
	Correlations involving the stress tensor will be invariant if the change in $T_{ij}$ given in \eq{Tch} are also included. 
		
		Note that these relations will apply separately to the  cut-off independent  and dependent terms.

			  \subsubsection{Ward Identities for $U(1)$ Gauge Invariance}
			  \label{WIU1}
			  Finally we  consider  the Ward identity for $U(1)$ gauge invariance. 

			  Consider  a scalar charged under a $U(1)$ gauge field with the quadratic term in its action being 
			  \be
			  \label{actsca}
			  S=-\int d^{d+1}x \sqrt{-g}	[g^{\mu\nu}(D_\mu \phi)^\dagger (D_\mu \phi) -M^2\phi^\dagger \phi	 ]
			  \ee
			  where 
			  \be
			  \label{defcov}
			  D_\mu\phi=(\partial_\mu-ieA_\mu ) \phi
			  \ee
			  We work in 
			  \be
			  \label{gaa}
			  A_\eta=0
			  \ee gauge, i.e. set the time like component to vanish. The wave function then depends on $A_i$, the remaining spatial components, and also on $\phi$. 
			  Gauss' law  implies that the wave function is invariant under the residual gauge symmetry, i.e. when the gauge  parameter  is $\eta$ independent and therefore preserve the condition, eq.(\ref{gaa}). 
			  This leads to the constraint  
			  \be
			  \label{invg}
			  \Psi[\phi,A_i]=\Psi[\phi+\delta \phi, A_i+\delta A_i]
			  \ee
			  with 
			  \begin{eqnarray}
			  \delta \phi & =&  i e \chi({\bf x}) \phi \label{condga}\\
			  \delta A_i  & =  & \partial_i \chi({\bf x}) \label{condgb}
			  \end{eqnarray}
			  The  $U(1)$ symmetry is a global symmetry in the boundary theory and the Ward identities of this global symmetry  follow from the invariance of 
			  the wave function. 
			  
			  We will mostly consider the underdamped case, $M^2>{d^2\over4}$, since the analysis for the over damped case is very similar to that in AdS space.

			  In the underdamped case we go to the coherent state basis. To avoid confusion with respect to complex conjugation it is best to first  decompose the complex scalar field into its real and imaginary components,
			  \be
			  \label{dec}
			  \phi({\bf x},\eta)={1\over \sqrt{2}}[\phi_1({\bf x},\eta)+i \phi_2({\bf x},\eta)].
			  \ee
			  and 
			  \be
			  \label{decb}
			  \phi^\dagger({\bf x},\eta)={1\over \sqrt{2}} [\phi_1({\bf x},\eta)-i \phi_2({\bf x},\eta)].
			  \ee

			  From eq.(\ref{condga}) it follows that under a $U(1)$ transformation,
			  \be
			  \label{trga}
			  \delta \phi_1= -e \chi({\bf x}) \phi_2, \ \ \delta \phi_2=e \chi({\bf x}) \phi_1.
			  \ee
			  
			  One can now go the coherent state basis $\rho_1({\bf x}), \rho_2({\bf x})$ for $\phi_1, \phi_2$ respectively, following the discussion in  section \ref{coherent}. 
			  Let $a_1({\bf x}), a_2({\bf x})$ be the corresponding destruction operators for the two fields whose eigenvalues are $\rho_1,\rho_2$ respectively. 
			  From eq.(\ref{relppos}) it then follows that under a $U(1)$ transformation $a_1({\bf x}),a_2({\bf x})$ will transform, analogous to eq.(\ref{trga}), as,
			  \be
			  \label{trgab}
			  \delta a_1({\bf x})= -e \chi({\bf x}) a_2({\bf x}), \ \ \delta a_2({\bf x})=e \chi({\bf x}) a_1({\bf x}).
			  \ee
			  And therefore $\rho_1({\bf x}),\rho_2({\bf x})$ will also transform as 
			  \be
			  \label{delff}
			  \delta \rho_1= - e \chi({\bf x}) \rho_2, \ \ \delta \rho_2=e \chi({\bf x}) \rho_1.
			  \ee
			  
			  The invariance of the wave function, eq.(\ref{invg}) then leads to the condition that 
			  \be
			  \label{condu1i}
			  \Psi[\rho_1+\delta \rho_1, \rho_2+\delta \rho_2, A_1+\delta A_i]=\Psi[\rho_1,\rho_2,A_i]
			  \ee
			  The Ward identities for $U(1)$ invariance arise from this condition. 
			  
			  To examine the consequences in more detail on the wave function let us note, as we can see from eq.(\ref{der1psirho}), eq.(\ref{der2psirho}), that 
			  the coherent state wave function will actually be  expressed in  terms of the complex conjugate of the coherent state eigenvalue, i.e. in terms of 
			  $\rho_1^\ast$, $\rho_2^\ast$. In expanding the wave function in terms of powers of the sources and obtaining constraints that follow from the invariance condition, eq.(\ref{condu1i}), on the correlators it is useful to adopt the following, admittedly somewhat clumsy,  notation. 
			  We define 
			  \begin{eqnarray}
			  \rho^{\ast} & \equiv & {1\over \sqrt{2}}  [\rho_1^\ast+i \rho^\ast_2 ] \label{defcra} \\
			  (\rho^\ast)^\dagger & \equiv & {1\over \sqrt{2}} [\rho_1^\ast-i \rho^\ast_2]\label{defcrb}
			  \end{eqnarray}
			  			  From eq.(\ref{delff}) it follows that under the $U(1)$ gauge transformation 
			  \be
			  \label{delffs}
			  \delta \rho_1^\ast= - e \chi({\bf x}) \rho_2^\ast, \ \ \delta \rho_2^\ast=e \chi({\bf x}) \rho_1^\ast.
			  \ee
			  And this leads to the transformation laws for $\rho^{\ast}, (\rho^{\ast})^\dagger$,
			  \begin{eqnarray}
			  \label{transrho}
			  \delta \rho^{\ast} & = &  i e \chi({\bf x}) \rho^{\ast}\\
			  \delta (\rho^{\ast})^\dagger & = & -i e\chi({\bf x}) (\rho^{\ast})^\dagger
			  \end{eqnarray}
			  showing that $\rho^{\ast}, (\rho^{\ast})^\dagger$  transform as fields with charge $+1$ and $-1$ under the gauge transformation respectively. 
			  
			  Expanding the wave function and keeping quadratic terms in the scalar and a trilinear term which also contains the gauge field we have 
			  \begin{empheq}{multline}
			  \label{expwv}
			  \log(\Psi)  =  {1\over 2} \int d^d{\bf x} d^d{\bf y} \sqrt{{\hat g({\bf x})}}\sqrt{{\hat g({\bf y})}} \rho^{\ast}({\bf x}) (\rho^{\ast}({\bf y}))^\dagger \l O_+({\bf x}) O_+^\dagger({\bf y})\r\\
			   + {1\over 2} \int d^d{\bf x}d^d{\bf y}d^d{\bf z} \sqrt{{\hat g({\bf x})}}\sqrt{{\hat g({\bf y})}}\sqrt{{\hat g({\bf z})}}A_i({\bf z}) \rho^{\ast}({\bf x}) (\rho^{\ast}({\bf y}))^\dagger \l J_i({\bf z}) O_+({\bf x}) O_+^\dagger({\bf y})\r
			  \end{empheq}
			  
			  The invariance of the wave function then leads to the condition 
			  \be
			  \label{condaab}
			  \del_i^{\bf z}\l J_i({\bf z})O_+^\dagger({\bf x})O_+({\bf y})\r-i\delta({\bf z}-{\bf y})\l O_+^\dagger({\bf x})O_+({\bf z})\r+i\delta({\bf z}-{\bf x})\l O_+^\dagger({\bf z})O_+({\bf y})\r=0
			  \ee
			  
			  Similar conditions will arise when we consider higher point correlators. 
			  Appendix \ref{U(1)ward} discusses some more aspects, including explicit calculations, pertaining to these Ward identities. 

			 \subsection{Ward Identities in  the Field Eigenstate Representation:} 
\label{cohJ}			 
			 Instead of the coherent state representation we can work directly with the wave function in terms of the eigenstate of the bulk field $\Phi({\bf x},\eta_1)$, at the late time slice $\eta_1$, . More precisely, as was discussed in section \ref{under} we can express the wave function in terms of the sources $J_+$ or $J_-$, eq.(\ref{defjpb}) and eq.(\ref{subst2}). 
			 
			 Here we will examine the constraints which arise when the wave function is expressed in terms of these sources and argue that the corresponding coefficient functions will satisfy the Ward identities of a CFT. More precisely this will be true for the cut-off independent terms in the coefficient functions. The cut-off dependent terms will also satisfy conditions which arise  from the invariance of the wave function under the various coordinate transformations. 
			 We will mostly discuss the case where the source is taken to be $J_+$, the case with $J_-$ is quite analogous.

			 The non-trivial part of the analysis, as was noted earlier, is the following. On general grounds, as was discussed above, the wave function is invariant under spatial and time reparametrisations which preserve the ADM gauge. 
			 However the sources $J_\pm$ are related to the bulk field $\phi({\bf x},\eta_1)$ in a non-local manner along the hypersurface $\eta=\eta_1$,
			 i.e. through a non-local relation in ${\bf x}$. This makes it somewhat non-trivial to determine how the sources transform under the spatial and time reparametrisations and then find the resulting constraints on the wave function. 
			 
			 In fact, we will need to specify the relation between the sources and the bulk field in a more precise manner to go forward.
			 The physical picture (drawn from AdS/CFT)  to keep in mind is as follows.  In  eq.(\ref{defjpb}) which we reproduce below
			 \be
			 \label{reprjpb}
			 \phi({\bf k},\eta)=\alpha_\mu^{\ast} (-\eta)^{{d\over2}-i\mu} J_+({\bf k}) + \beta_\mu^{\ast} (-\eta)^{{d\over2}+i\mu}k^{2i\mu} J_+({\bf k})
			 \ee
			 the first term on the RHS   specifies the source in terms of    $J_+$, and the second term is the response which arise due to this source. This suggests that  if we take  $(-\eta)^{{d\over2}-i\mu} J_+({\bf x})$ to transform as a scalar, 
			 just as  the bulk scalar $\phi({\bf x},\eta)$,  ensuring  that it takes the same value at the same physical point, before and after the relevant coordinate transformation, 
			 then it should also be possible to define the response term so that it  transforms as a scalar. 
			 
			 For a spatial reparametrisation this is straightforward to do. Let us first write the relation in eq.(\ref{reprjpb}) in position space as 
			 \be
			 \label{relpos}
			 \phi({\bf x},\eta_1)=\alpha_\mu^{\ast} (-\eta)^{{d\over2}-i\mu} J_+({\bf x}) + {2^{2i\mu}\Gamma[{d\over2}+i\mu]\over\pi^{d\over2}\Gamma[-i\mu]} (-\eta)^{d/2+i\mu}\beta_\mu^{\ast} \int d^d{\bf y} {J_+({\bf y})\over |{\bf x}-{\bf y}|^{d+2i \mu}}
			 \ee
			 Similarly for future reference in terms of $J_-$ from eq.(\ref{subst2}) we write
			 \be
			 \label{relposjm}
			 \phi({\bf x},\eta_1)=\beta^{\ast}_\mu (-\eta)^{{d\over2}+i\mu} J_-({\bf x}) + {2^{-2i\mu}\Gamma[{d\over2}-i\mu]\over\pi^{d\over2}\Gamma[i\mu]} (-\eta)^{d/2-i\mu}\alpha_\mu^{\ast}\int d^d{\bf y} {J_-({\bf y})\over |{\bf x}-{\bf y}|^{d-2i \mu}}
			 \ee
			 
			 This relation holds when the metric components ${\hat g}_{ij}=\delta_{ij}$. 
			 For a more non-trivial metric we then define $J_+$ through the  relation 
			 \be
			 \label{relposa}
			 \phi({\bf x},\eta_1)=\alpha_\mu^{\ast} (-\eta)^{{d\over2}-i\mu} J_+({\bf x}) + {2^{2i\mu}\Gamma[{d\over2}+i\mu]\over\pi^{d\over2}\Gamma[-i\mu]} (-\eta)^{{d\over2}+i\mu}\beta_\mu^{\ast} \int d^d{\bf y} \sqrt{{\hat g}({\bf y})}  {J_+({\bf y})\over s({\bf x},{\bf y})^{d+2i \mu}}
			 \ee
			 where $s({\bf x},{\bf y})$ is now the geodesic distance between the points ${\bf x}$ and ${\bf y}$ for the more general metric ${\hat g}_{ij}$. 
			  It is now clear that with this definition if $J_+$ transforms like a scalar under   spatial reparametrisations , the response term will also 
			  be a scalar. That is,  
			  under the transformation $x^i \rightarrow (x^i)'=x^i +v^i({\bf x})$, 
			  if we  take 
			  \be
			  \label{defchJ+}
			  \delta J_+=-v^i \partial_i J_+
			  \ee
			  and this will lead to the LHS in eq.(\ref{relposa}) transforming in the required manner for a scalar,
			  \be
			  \label{lhre}
			  \delta \phi({\bf x},\eta_1)=-v^i \partial_i \phi({\bf x},\eta_1).
			  \ee
			  
			  Note we will take eq.(\ref{relposa}) to be the defining relation for $J_+$ in terms of $\phi({\bf x},\eta_1)$, even for an interacting theory where the   scalar is not a free   field.

			  Next, for dealing with  time reparametrisations, which asymptotically take the form, 
			  \be
			  \label{asfa}
			  \eta\rightarrow \eta'=(1+\epsilon({\bf x}))\eta
			  \ee
			   	we need to take, following the line of thought above, the first term on the RHS of eq.(\ref{relposa}) to be a bulk scalar, i.e. for the change in $(-\eta)^{{d\over2}-i\mu} J_+({\bf x})$ to be 	
			  \be
			  \label{chngtwo}
			  \delta [(-\eta)^{{d\over2}-i\mu} J_+({\bf x})] =-\epsilon({\bf x}) \eta\partial_\eta [(-\eta)^{{d\over2}-i\mu} J_+({\bf x})]=-\Delta_-\epsilon({\bf x})(-\eta)^{{d\over2}-i\mu} J_+({\bf x})
			  \ee
			  This can be accomplished by taking 
			  \be
			  \label{chngthree}
			  \delta J_+({\bf x})=-\epsilon({\bf x}) \Delta_- J_+({\bf x})
			  \ee
			  
			  Now notice  that under a time reparametrisation eq.(\ref{asfa})  the surface $\eta=\eta_1$ will be described as a surface 
			  $\eta'=(1+\epsilon({\bf x}))\eta_1 $, i.e. it will not be a hypersurface along which the $\eta'$ coordinate is constant. We need to therefore define the second term in eq.(\ref{relposa}), the response term, appropriately   for   such hypersurfaces where the $\eta$ coordinate varies. 
			  
			  This can be done as follows. First consider  the  $\eta= {\rm constant}$ surface, with ${\hat g}_{ij}=\delta_{ij}$. From eq.(\ref{metadm}) we see that the $d$ dimensional metric which is induced on this hypersurface is given by $g_{ij}={\delta_{ij} \over \eta^2}$
			 In terms of this induced metric the response term in eq.(\ref{relposa}) then can be written as 
			  \be
			  \label{respth}
			  {2^{2i\mu}\Gamma[{d\over2}+i\mu]\over\pi^{d\over2}\Gamma[-i\mu]}\beta_\mu^\ast \int d^d{\bf y}\sqrt{g({\bf y})} [(-\eta)^{{d\over2}-i\mu} J_+({\bf y})] \left[{(-\eta)^{d+2i\mu}\over |{\bf x}-{\bf y}|^{d+2i\mu}}\right]
			  \ee
			  The first term on the RHS,  $d^d{\bf y} \sqrt{g({\bf y})}$, is the correct measure for  the induced metric while  the last term $(-\eta)^{d+2i\mu}\over |{\bf x}-{\bf y}|^{d+2i\mu}$ equals ${1\over s({\bf x},{\bf y})^{d+2i \mu}}$ where $s(x,y)$ is the geodesic distance between ${\bf x}, {\bf y}$ computed using the induced metric. 
			  The expression in eq.(\ref{respth})  reveals how we can generalise the response term to apply for the more general hypersurfaces  along which $\eta$ also varies. 
			  We simply use  the induced metric in defining the measure for  integration along the hypersurface  and define the distance $s({\bf x},{\bf y})$ also using this metric, so that the response term becomes
			  \be
			  \label{respf}
			 {2^{2i\mu}\Gamma[{d\over2}+i\mu]\over\pi^{d\over2}\Gamma[-i\mu]}\beta_\mu^\ast \int d^d{\bf y} \sqrt{g({\bf y})} [(-\eta({\bf y}))^{{d\over2}-i\mu} J_+({\bf y})] \left[{1\over s({\bf x},{\bf y})^{d+2i\mu}}\right]
			  \ee
			  If the expression $(-\eta)^{{d\over2}-i\mu} J_+$ now transforms as a bulk scalar, as we mentioned above, then we see that the response term, as we have defined it  now, will also behave as a bulk scalar since the measure term, the middle term $(-\eta({\bf y}))^{{d\over2}-i\mu} J_+({\bf y})$, and the distance $s({\bf x},{\bf y})$, will all transform appropriately.			
			
			In summary, putting  together the various points in our discussion of spatial and time reparametrisations  above, we then  define $J_+$ in terms of $\phi({\bf x},\eta)$ in general by the relation 
			\begin{empheq}{multline}
			\label{sumr}
			\phi({\bf x},\eta)=\alpha_\mu^{\ast} (-\eta({\bf x}))^{{d\over2}-i\mu} J_+({\bf x}) \\+ {2^{2i\mu}\Gamma[{d\over2}+i\mu]\over\pi^{d\over2}\Gamma[-i\mu]} \beta_\mu^{\ast} \int d^d{\bf y} \sqrt{g({\bf y})} [(-\eta({\bf y}))^{{d\over2}-i\mu} J_+({\bf y})] \left[{1\over s({\bf x},{\bf y})^{d+2i\mu}}\right]
			\end{empheq}
			This relation is valid on hypersurfaces where $\eta$ can also vary with ${\bf x}$ (at least for small enough variations, e.g. infinitesimal $\epsilon$). 			The integration in the second term in the RHS in eq.(\ref{sumr}) is along the hypersurface, $g_{ij}$ is the induced metric on this hypersurface and $s({\bf x},{\bf y})$ is the geodesic distance using this metric. We take $(-\eta)^{{d\over2}-i\mu}J_+$ to be  scalar under both spatial and time reparametrisations, i.e. to transform under these reparametrisations as given in eq.(\ref{defchJ+}) and eq.(\ref{chngtwo}),eq.(\ref{chngthree}). The second term in the RHS of eq.(\ref{sumr}) will also then be a scalar. This is consistent with the LHS of eq.(\ref{sumr}), i.e. with $\phi({\bf x},\eta)$, also transforming like a scalar, as indeed it should. 
			
			As in the coherent state basis we will expand the metric ${\hat g}_{ij}$ about its flat space value, eq.(\ref{exfl}).
			And now  express the wave function  as a Taylor series in $J_+$ and $\gamma_{ij}$.
		The invariance condition of the wave function under spatial and time reparametrisations then give rise to the condition
			\be
			\label{chnaapsi}
			\Psi[J_+ +\delta J_+, {\delta}_{ij}+\gamma_{ij}, \eta_1(1+\epsilon({\bf x}))]=\Psi[J_+, {\delta}_{ij}, \eta_1]
			\ee
			where the change in $J_+$, eq.(\ref{defchJ+}), eq.(\ref{chngthree}), is  
			\be
			\label{chnjpt}
			\delta J_+=-v^i\partial_i J_+-\epsilon({\bf x})\Delta_- J_+
			\ee
			and  $\gamma_{ij}$ is given by 
			\be
			\label{chngcc}
			\gamma_{ij}=-\hat{\nabla}_i v_j-\hat{\nabla}_jv_i+2 \epsilon({\bf x}) \delta_{ij}
			\ee
			
			For the two and three point correlators, analogous to eq.(\ref{expwv}), we get on expanding the wave function
			\begin{empheq}{multline}
			\label{condtth}
			\log(\Psi) = {1\over 2} \int d^d{\bf x}d^d{\bf y}\sqrt{\hat{g}({\bf x})}\sqrt{{\hat g}({\bf y})}J_+({\bf x})J_+({\bf y}) \l {\hat O}_+({\bf x}) {\hat O}_+({\bf y})\r\\
			+ {1\over 4}\int d^d{\bf x}d^d{\bf y}d^d{\bf z} \sqrt{\hat{g}({\bf x})}\sqrt{\hat{g}({\bf y})}\sqrt{\hat{g}({\bf z})} \gamma_{ij}({\bf z}) J_+({\bf x})J_+({\bf y})\l T_{ij}({\bf z}){\hat O}_+({\bf x}){\hat O}_+({\bf y})\r
			\end{empheq}
			Eq.(\ref{chnaapsi}) then leads for the $\eta_1$ independent terms in the coefficient functions to the conditions 
			\begin{align}
			\label{condaaf}
			&\del_i^z\l T^{ij}({\bf z}){\hat O}_+({\bf x}){\hat O}_+({\bf y})\r + \delta({\bf z}-{\bf x})\del_i^x\l {\hat O}_+({\bf x}){\hat O}_+({\bf y})\r + \delta({\bf z}-{\bf y})\del_i^y\l {\hat O}_+({\bf y}){\hat O}_+({\bf x})\r=0\\
			&\l T^{i}_i({\bf z}){\hat O}_+({\bf x}){\hat O}_+({\bf y})\r+\Delta_+\delta({\bf z}-{\bf x})\l {\hat O}_+({\bf x}){\hat O}_+({\bf y})+\Delta_+\delta({\bf z}-{\bf y})\l {\hat O}_+({\bf y}){\hat O}_+({\bf x})\r=0\label{condaaf2}
			\end{align}
			These agree in their form with eq.(\ref{condtwothree}), \eq{condconf} and are the Ward identities in a CFT for spatial reparametrisations and local scale invariance. 
			It follows from these relations  that the cut-off independent terms will then satisfy the Ward identities of conformal invariance as well
			We will not discuss these in detail except to note that for the cut-off independent terms they take the standard form for 
			and operator of dimension $\Delta_+$, e.g., for the two point function, they are  given in eq.(\ref{condcnfin2}), eq.(\ref{chngO}),  with $O_+$ replaced by 
			${\hat O}_+$.
			
			A few additional  comment before we proceed are in order. First, it is worth being more explicit about   the definition we have given in eq.(\ref{sumr}) for $J_+$ in terms of $\phi({\bf x},\eta)$ and the resulting coefficient functions which arise after the Taylor series expansion of   the wave function. Consider the wave function  as a functional of the field $\phi$ and metric ${\hat g}_{ij}$, eq.(\ref{metadm}),  on a slice specified by giving $\eta({\bf x})$, 
			$\Psi[\phi, {\hat g}_{ij}, \eta({\bf x})]$. For $\eta\rightarrow 0$, to leading order,  the induced metric on the slice is given by 
			
			\be
			\label{defind}
			g_{ij}({\bf x})={{\hat g}_{ij}\over \eta({\bf x})^2}. 
			\ee
			This induced metric is to be  used in  calculating the second term on the RHS of eq.(\ref{sumr}).  
			Although the relation between the two is non-local, $J_+$ can in principle be  obtained now in terms of $\phi$ using eq.(\ref{sumr}) , and the wave function can then  be expressed in terms of 
			$J_+$ to obtain $\Psi[J_+,{\hat g}_{ij}, \eta({\bf x})]$. After expanding in a Taylor series  in $J_+$, and $\gamma_{ij}$, eq.(\ref{exfl}), one then obtains the coefficient functions which will give the correlation functions of the dual theory. In particular, the relation, eq.(\ref{sumr}), although motivated by the behaviour in the free field limit is to be taken as an exact definition of $J_+$ in terms of $\phi$, uncorrected in the presence of additional interactions, e.g. $\phi^n$ interactions, or in the loop expansion (gauge interactions are an exception and will be discussed below). 
			Second, for the cut-off dependent terms also the conditions of invariance under spatial and time reparametrisation  will give rise to  important constraints. Obtaining these constraints require that the cut-off is also transformed suitably. For example consider the term  in the two point function of the form eq.(\ref{posf}). This can be written for a more general hypersurface where $\eta$ varies, and  a more general metric along the hypersurface as 
			 			\be
			\label{posfgen}
			\int d^d{\bf x}d^d{\bf y}\sqrt{ g({\bf x})} \sqrt{ g({\bf y})} ((-\eta({\bf x}))^{{d\over2}-i\mu }J_+({\bf x})) ((-\eta({\bf y}))^{{d\over2}-i\mu} J_+({\bf y})) {1\over s^{d+4i\mu}}
			\ee
			where we have used the notation above with $g_{ij}$ denoting the induced metric etc. 
We see that  this term will indeed be invariant under both spatial and time reparametrisations,  and conformal transformations, once the cut-off is also transformed. 
			Third, a very similarly analysis could have been carried out for the source $J_-$ instead of  $J_+$. The operator sourced by $J_-$, $\hat{O}_-$, has dimension $\Delta_-$ instead of $\Delta_+$ and Ward identities  for the cut-off independent terms then can be obtained from the  $J_+$ case by simply making the replacements $\hat{O}_+\rightarrow {\hat{O}}_-$ and $\Delta_+\rightarrow \Delta_-$. 
			
			Finally, when  a $U(1)$ gauge symmetry is present for a complex scalar field, $\phi$, with action eq.(\ref{actsca}), we need to change the response term to now include a Wilson line 
			\be
			\label{wl}
			W({\bf x}, {\bf y})=
			 \exp\left[{ie \int_{\bf y}^{\bf x}A_i dx^i}\right]
			\ee
			where the path involved for computing the Wilson line is the geodesic from ${\bf x}$ to ${\bf y}$.
			We then define the bulk field in terms of $J_+$ as follows
			\be
			\label{wla}
			\phi({\bf x},\eta_1)=\alpha_\mu^{\ast} (-\eta_1)^{{d\over2}-i\mu}J_+({\bf x}) + {2^{2i\mu}\Gamma[{d\over2}+i\mu]\over\pi^{d\over2}\Gamma[-i\mu]} \beta_\mu^{\ast} (-\eta_1)^{{d\over2}+i\mu} \int d^d{\bf y}  W({\bf x},{\bf y}) 
			{J_+({\bf y}) \over |{\bf x}-{\bf y}|^{d+2i\mu}}
			\ee
			instead of eq.(\ref{relpos}).
					From eq.(\ref{wl}) we see that if $J_+({\bf x})$ transforms under a $U(1)$ gauge transformation as a charge $1$ field, i.e. with 
			\be
			\label{chja}
			\delta J_+({\bf x})=i e \chi({\bf x}) J_+({\bf x})
			\ee
			then the second term in eq.(\ref{wla}) will also transform as a charge $1$ field, which is consistent with the transformation of the bulk scalar eq.(\ref{condga}). 
			
	  Additional affects of a non-trivial metric and varying $\eta$ values along the hypersurface can be included as above, eq.(\ref{sumr}). In this more general case  the Wilson line would be computed along the geodesic defined from the induced metric and the generalisation of eq.(\ref{sumr}) takes the form
	  \be
	  \label{sumra}
	  \phi({\bf x},\eta)=\alpha_\mu^{\ast} (-\eta)^{{d\over2}-i\mu}J_+({\bf x}) + {2^{2i\mu}\Gamma[{d\over2}+i\mu]\over\pi^{d\over2}\Gamma[-i\mu]}\beta_\mu^{\ast} \int d^d{\bf y} \sqrt{g({\bf y})} (-\eta)^{{d\over2}-i\mu} W({\bf x},{\bf y}) {J_+({\bf y})\over s({\bf x},{\bf y})^{d+2i \mu}}
	  \ee
         As  above, starting with the wave function $\Psi[\phi, {\hat g}_{ij}, A_i, \eta({\bf x})]$ we would use the above relation (with $W({\bf x},{\bf y})$ now depending on $A_i$ and on the geodesic determined by ${\hat g}$ and $ \eta({\bf x})$) to obtain $J_+$, and  then find  the wave function as a functional of $J_+$, $\gamma_{ij}$ and $A_i$. Its coefficient functions would be the correlators in the dual theory.
         
	\subsection{Additional Points}	\label{addpt}	
		We have discussed above how a source in the dual theory can be identified with the coherent state eigenvalue. This identification has many attractive feature. The cut-off independent terms in the correlators satisfy the Ward identities of conformal invariance, local scale invariance, and momentum conservation. And the cut-off dependent term in the two point function is a purely local contact term, eq.(\ref{psirho1}). For higher point correlators,  also our analysis in section \ref{holo} suggests that cut-off dependent terms also have a sensible explanation in the dual field theory - they can be understood as arising from operator mixing. In addition, we also found other ways to identify sources, using  $J_+$ or $J_-$, see eq.(\ref{relpos}) and its more accurate version eq.(\ref{sumr}) for $J_+$ and similarly for $J_-$ , eq.(\ref{relposjm}). These turn out to be   source for  boundary fields of dimension $\Delta_+$ or $\Delta_-$ respectively and  are related to the bulk field eigenstates through a non-local transformation in position space.  The  finite parts of the resulting coefficient functions in these cases also   satisfy the Ward identities mentioned above as was discussed above.  However the cut-off dependent terms in these cases are more unwieldy, in particular the two point correlator has an additional term which is both cut-off dependent and non-local. While we did not analyse them in much detail the general arguments we gave lead to the conclusion that the cut-off dependent terms will also be constrained by the conditions of invariance of the wave function under time and spatial reparametrisation invariance, see eq.(\ref{posfgen}) and the related discussion. 

How exactly holography works in the dS context is still poorly understood. In the most optimistic case the holographic dual will be a field theory with a finite UV cut-off, and will reproduce  the full wave function in the bulk, at late time $\eta_1\rightarrow 0$, including the cut-off dependent terms. In particular, it will therefore encode all information  for both modes which have exited the horizon and those which have not done so by the time $\eta_1$. In this case the correlators for modes with $k\abs{\eta_1}\gtrsim {\cal O}(1)$ will depend on $\eta_1$ in a complicated manner, significantly different from correlations in a CFT, and the dual theory would need to reproduce this   full dependence. The somewhat unwieldy contribution to   the two point function,  eq.(\ref{posf}),  will then be a part of this larger set of cut-off dependent terms. 
In the less optimistic case the best one could do would be to have a hologram which is a CFT reproducing the cut-off independent terms. 
The cut-off dependent non-local term in eq.(\ref{posf}) would not be reproduced in such a theory. 
Restricting ourselves to cut-off independent terms there would still be a difference between using the coherent state eigenvalue $\rho^{\ast}$, eq.(\ref{rhoeigsys}) or $J_+$, eq.(\ref{relpos}), eq.(\ref{sumr}), as a source. The ratio of the two point and three point correlators (suitably corrected for different normalisations ) would differ in these two cases, see Appendix \ref{3by2}. 
Thus given a dual description which only yields cut-off independent terms would be still be able to decide which of the two sources, $\rho^{\ast}$, eq.(\ref{rhoeigsys}) or $J_+$,  correctly reproduce correlations in the boundary dual.



\section{Additional Comments}\label{addcmnt}
Here we make some further comments about correlation functions obtained in the   dS case, their continuation from AdS, etc. 

	\subsection{Issues Connected to Factors of $i$ and $-1$ }
	\label{factorsofi}
	It is well known that factors of $i=\sqrt{-1}$, or $-1$ often appear, and confuse or  bedevil attempts    to construct a dS/CFT correspondence.
	Here we discuss some of these issues. 
	
	For the overdamped case, including massless fields,  these factors  can be understood simply from analytically continuing the corresponding expressions in the 
	AdS case\footnote{In the underdamped case it is more complicated, since we work in the coherent state basis section \ref{coherent} or in terms of the sources $J_\pm$, eq.(\ref{subst1})
	eq.(\ref{subst2}).} \cite{Malda-NG}. Let us review how to carry out this continuation in a convenient manner. More details can be found in Appendix \ref{App.analcont}. 
	The metric in Euclidean $AdS_{d+1}$ in Poincaré coordinates is given by 
	\be
	\label{metadsea}
	ds^2={L^2\over z^2}[dz^2+\sum_i(dx^i)^2]
	\ee
	where $L$ is the AdS radius. The coordinate $z$ takes values in the range $z\in [0,\infty]$.
	Continuing 
	\be
	\label{conta}
	z\rightarrow -i \eta(1-i \epsilon)
	\ee
	with $\eta\in (-\infty,0]$, ,and 
	\be
	\label{contab}
	L\rightarrow \pm i R_{\rm dS}
	\ee
	where $R_{\rm dS}$ is the radius of dS space
	 gives the metric for $dS$ in Poincaré  coordinates 
	 	 \be
	 \label{metds}
	 ds^2={R_{\rm dS}^2\over \eta^2} [-d\eta^2+(dx^i)^2].
	 \ee
	 As will be clear  from the discussion below  if $d$ is an even integer the choice of sign in eq.(\ref{contab}) makes a difference.
	 
	 In this paper we have mostly considered the hologram at ${\cal I}^+$, i.e. for the expanding branch. However one could also have considered the contracting branch and constructed a hologram at ${\cal I}^-$. 
	 It turns out that for the expanding branch we need to take 
	 \be
	 \label{contabc}
	 L \rightarrow i R_{\rm dS}
	 \ee
	 And for the contracting branch its complex conjugate, i.e.
	 \be
	 \label{contad}
	 L\rightarrow -i R_{\rm dS}
	 \ee
	 to get agreement between the dS and AdS results. 
	 
	 In the case of the expanding branch after this continuation the action in AdS space $S_{AdS}$ gets related to the action in dS space as follows, 
	 \be\label{actsads}
	 S_{AdS}\rightarrow -i S_{dS}
	 \ee
	 This can be easily checked, for example for a scalar field. 
	 As a result the 
	 partition function $Z_{AdS}$ in AdS space goes over to the wave function $\Psi$ in dS.
	 \be
	 \label{contp}
	 Z_{AdS}=e^{-S_{AdS}} \rightarrow \Psi=e^{iS_{ds}}
	 \ee
	 In carrying out the continuation it is convenient to normalise the action so an overall factor of ${1\over G_N}$ appears in front of it. By dimensional analysis this will lead to a factor of ${L^{d-1}\over G_N}$ appearing, once we input the metric, in the AdS case. Continuing using eq.(\ref{contab}) will then lead to this turning into the  factor 
\be
\label{factora}
{L^{d-1}\over G_N}\rightarrow (i)^{d-1}{R_{\rm dS}^{d-1}\over G_N}
\ee
We also note that the continuation in eq.(\ref{conta}) is required to ensure that the saddle point solution used in evaluating the action in the Euclidean AdS case goes over correctly to the solution which satisfies the Bunch Davies vacuum condition, eq.(\ref{limfs}) in the dS case. 
E.g., for an underdamped scalar the solution in Euclidean AdS is given by 
\be
\label{eucscalar}
\phi\propto z^{d\over2}K_\nu(kz)
\ee
Under the continuation eq.(\ref{conta}) this goes over to ${\cal F}_\nu^\ast$ which vanishes as $\eta\rightarrow -\infty$, as was discussed in eq.(\ref{limfs}). 
For cut-off dependent terms we must also continue the cut-off in AdS space at $z=\epsilon$  to the cut-off $\eta_1$ in dS through the relation, eq.(\ref{conta}),
\be
\label{contcut}
\epsilon=-i \eta_1.
\ee


Before proceeding we remind the reader that the  $n$ point correlator in the hologram to be given by, eq.(\ref{genfab}),  
\be
\label{npt}
\l O({\bf k}_1) O({\bf k}_2) \cdots O({\bf k}_n) \r={\delta^n \log \psi\over \delta {\hat \phi}({\bf k}_1) \delta {\hat \phi}({\bf k}_2) \cdots \delta {\hat \phi}({\bf k}_n)}
\ee
where ${\hat \phi}({\bf k})$ is the source in the boundary theory. 
In the underdamped case ${\hat \phi}$ is replaced by $\rho^\ast$ or  $J_{\pm}$, eq.(\ref{rhoeigsys}),  eq.(\ref{subst1}), eq.(\ref{subst2}). 
Similarly  in AdS space the correlators are defined to be 
\be
\label{nptads}
\l O({\bf k}_1) O({\bf k}_2) \cdots O({\bf k}_n)\r={\delta^n \log Z\over \delta {\phi}_b({\bf k}_1) \delta{\phi}_b({\bf k}_2) \cdots \delta {\phi}_b({\bf k}_n)}
\ee

Before ending this subsection let us comment that the analytic continuation discussed above from AdS to dS agrees with explicit calculations in dS. The resulting answers also meet  the constraints arising from bulk unitarity and the cosmological optical theorem \cite{cotPajer}. In particular we have  explicitly verified that this is true  for  non-derivative $\phi^n$ interactions. 
	\subsection{ Scalars, Stress Tensor and $2d$ CFTs} \label{2dcft}
	
	The two point function for an overdamped scalar was discussed in section \ref{over}, see \eq{posspa}.
	It was noted, \eq{posspa}, that the two point function generically violates reflection positivity. 

The absence of reflection positivity   in the dS case  means that the correlators in the Euclidean field theory  cannot be analytically continued to correlation functions of a  in a conventional Lorentzian field theory with  positive norm states\footnote{However this may be possible to do in a Lorentzian theory with Ghosts, i.e. negative norm states, \cite{Vasiliev,anninoscftex}}. 

It is easy to check that the  result for the two point function of an overdamped scalar  can be  obtained after the analytic continuation from AdS space,  if we use the continuation eq.(\ref{contabc}) in the expanding branch, see Appendix \ref{App.analcont}. 

The two-point function of the stress tensor can be discussed similarly, and can also be obtained by continuing the $AdS$ result given in \cite{ViswaTTads}, see Appendix \ref{App.2ptTT}. 
In dS space  in momentum space    it is given by 
\begin{align}
\label{stressmom}
\langle	T_{ij}({\bf k})T_{kl}(-{\bf k})\rangle &=-  {i}d{\beta_{d\over2}^{\ast}\over8\alpha_{d\over2}^{\ast}}k^d\Bigl[P_{ik}P_{jl}+P_{il}P_{jk}-{2\over d-1}P_{ij}P_{kl}\Bigr]~~~d={\rm odd}\\
\label{stressmomint}
&= - {i}d{\bar{\beta}_{d\over2}^{\ast}\over8\bar{\alpha}_{d\over2}^{\ast}}k^d\log(k)\Bigl[P_{ik}P_{jl}+P_{il}P_{jk}-{2\over d-1}P_{ij}P_{kl}\Bigr]~~~d={\rm even}
\end{align}
where $P_{ij}\equiv \delta_{ij}-{k_ik_j\over k^2}$.
In position space, both \eq{stressmom} and \eq{stressmomint} becomes 
\be
\label{stresspos}
\langle	T_{ij}({\bf x})T_{kl}({\bf y})\rangle=e^{i \pi(d -1) \over 2}  \frac{\Gamma[d+2]}{8(d-1)\pi^{\frac{d}{2}} \Gamma[\frac{d}{2}] r^{2d}} \left(J^i_l (\textbf{r}) J^j_m (\textbf{r}) + J^{i j} (\textbf{r}) J_{l m} (\textbf{r}) - \frac{2}{d} \delta^i_m \delta^j_l \right)
\ee
where,
\begin{equation}
	\textbf{r}= \textbf{x}-\textbf{y}, r= \abs{\textbf{r}},	J^i_j (\textbf{x}) = \delta^i_j - \frac{2 x^i x_j}{x^2}
\end{equation}
 As discussed in Appendix \ref{App.2ptTT} when $d$ is even, $d=2n$,  or when $d$ is odd and given by $d=4m+3$, with $n,m\in \mathbb{Z}$,  this two point correlator violates reflection positivity. 
 


Returning to the scalar case and turning briefly to higher point correlations, we saw in section \ref{holoint} above that the three point correlator for scalars is given by 
\be
	\label{intwf}
	\delta \log(\psi) ={1\over 3!}\int \prod_{i=1}^{3}\frac{d^d{\mathbf{k}_i}}{(2\pi)^d}(2\pi)^d\delta({\bf k}_1+{\bf k}_2+{\bf k}_3)\langle O(\mathbf{k}_1)O(\mathbf{k}_2)O(\mathbf{k}_3)\rangle'\prod_{i=1}^{3}\hat\phi(\mathbf{k}_i)
	\ee
	where $\langle O(\mathbf{k}_1)O(\mathbf{k}_2)O(\mathbf{k}_3)\rangle$ is given as
	\begin{empheq}{multline}
	\label{ooo}
	\langle O(\mathbf{k}_1)O(\mathbf{k}_2)O(\mathbf{k}_3)\rangle'=-\lambda\frac{8\times3!}{2^{3\nu}(\Gamma[\nu])^3}\left[e^{{i\pi\over2}(3+3\nu-\frac{d}{2})}\right]k_1^\nu k_2^\nu k_3^\nu\\
	\times\int_{0}^{\infty}dt~ t^{\frac{d}{2}-1}K_\nu(k_1t)K_\nu(k_2t)K_\nu(k_3t)
	\end{empheq}
	This means  that the  three point function also has a complex coefficient in general.
	 

From \eq{oversource} we also note  that the bulk  scalar being real  implies that the source ${\hat \phi}({\bf x})$ is also real. 
In momentum space this implies that 
${\hat \phi}(\mathbf{k})={\hat \phi}^{\ast}(-\mathbf{k})$. 
The two point and higher point correlators being complex in momentum space therefore implies that the wave function is also complex. 
To understand this  better recall, as was also mentioned in the previous subsection, that we have been  discussing the wave function for the expanding branch of the Hartle- Hawking state above. Its complex conjugate, more correctly CPT conjugate,  $\Psi^{\ast}$, would describe the time reversed contracting branch, and here the  $n$ point correlators would be  complex conjugates of their values  in the expanding branch. The analytic continuation to the contracting branch also has to be done taking into account this complex conjugation, e.g., eq.(\ref{contad}).

	Finally, let us  conclude with  some comments about the $dS_3$ case. 
	By considering global $dS_3$ and evaluating the boundary stress tensor from the behaviour of the extrinsic curvature close to the boundary, one finds,
	see Appendix \ref{App.Central}, that the central  charge is given by 
	\be
	\label{cenc}
	c=i{3R_{\rm dS}\over 2G_N}.
	\ee
	In particular, it is imaginary. 
	
	Using general arguments \cite{Polchinski}, and after   taking into account  the fact that the stress tensor  behaves like a symmetric two index tensor under spatial reparametrisations,
	 one then learns that in  correlation functions the short distance singularities, when two stress tensors, $T\equiv T_{zz}$, come close together, is of the standard OPE form given by 
	\be
	\label{opestr}
	T(z) T(\omega)={c/2\over (z-\omega)^4}+  {2 T(\omega)\over (z-\omega)^2}+ {\partial_\omega T(\omega)\over (z-\omega)}
	\ee
	Note that  $c$ here is given by eq.(\ref{cenc}) and is   imaginary. 
	Similarly for the anti-holomorphic component ${\bar T}({\bar z})=T_{{\bar z}{\bar z}}({\bar z})$, 
	\be
	\label{opan}
	{\bar T}({\bar z}) {\bar T}({\bar \omega})= {c/2\over (\bar{z}-\bar{\omega})^4}+  {2 \bar{T}(\bar{\omega})\over (\bar{z}-\bar{\omega})^2}+ {\partial_{\bar{\omega}} \bar{T}(\bar{\omega})\over (\bar{z}-\bar{\omega})}
	\ee
	
	A scalar field in the bulk correspond to primary a scalar field $O$ in the boundary, see Appendix \ref{conft}. The Ward  identities we have discussed in the previous section 
	tell us that the  leading singularity, when such a field comes close to $T$, is given by 
	\be
	\label{lesi}
	T(z) O(\omega)=\Delta {O(\omega)\over (z-\omega)^2}+{\partial_\omega O(\omega)\over (z-\omega)}
	\ee
	where $\Delta$ is the dimension of the field, 
	and similarly for ${\bar T}$. 
	
	Thus despite the fact that reflection positivity is violated and the central charge is imaginary, key elements of the general structure of CFTs remains intact in field theory duals. 	

		\section{Discussion}
		\label{discuss}
		We are at an early stage  in our study of  holography in de-Sitter space. 
		There are in fact two versions of holography, pertaining to the static patch or  the late time boundary ${\cal I}^+$, Figure \ref{poinpenrose}, which are being studied. The static patch version is discussed in \cite{staticp2,staticp1}. For a somewhat different approach, see \cite{Loga_phiex}. 
		

		Here we have discussed some aspects of the  late time boundary version, which is perhaps of greater interest from the point of inflationary cosmology and  phenomenology. In this case,  holography  relates  the   late time wave function on a hypersurface $\eta=\eta_1$, with $|\eta_1|\ll 1$, to the partition function of a dual field theory, \eq{meq}. The value of the late time coordinate $\eta_1$, more generally the location of the late time hypersurface, is related to the UV cut-off in the field theory, which vanishes as $\eta_1\rightarrow 0$.

		The  $\eta_1$  or  cut-off dependent terms are in fact important in ensuring that the wave function is gauge invariant and solves the Wheeler de-Witt equation, see \cite{suvrat--Holoinfods, nandajt}. A full holographic understanding of the bulk would require that these cut-off dependent terms are also correctly reproduced in the boundary theory. In fact, and perhaps more importantly,  at any finite value of $\eta_1$ there are modes in the bulk which have not exited the horizon, with momenta $k|\eta_1|\gtrsim {\cal O}(1)$. The wave function includes all  information about these modes too and a complete holographic description for the wave function at the late time slice would have to include this information as well. 
		
		Such a complete hologram seems  quite ambitious to obtain since the dependence on  modes which have not exited the horizon  in the
		wave function is a complicated function of both the momentum and the cut-off. In the boundary theory this would correspond to the behaviour of correlation functions on the scale of the cut-off which is  non-universal. The less ambitious  version  of holography would be of a dual theory    which successfully encodes information in the wave function for modes after they  exit the horizon. More precisely, rather than considering all momenta at a fixed and non-vanishing value of $\eta_1$, we take  the $\eta_1\rightarrow 0$ limit keeping the momenta fixed, discard the $\eta_1$ dependent terms and retain  the remaining $\eta_1$ independent terms  in the wave function. These remaining terms  would then be obtained from the hologram. 
		
		There is an important distinction between two descriptions.  Since a translation in time, i.e. $\eta\rightarrow \lambda \eta$, corresponds to a scale transformation $k \rightarrow k/\lambda$ on the boundary, one expects one general grounds 
		that the time translation of the wave function should  map to scale transformation  in the dual theory. The evolution of the bulk wave function as a function of time should be unitary if all modes are  being retained. In a hologram which keeps  all the modes, and complete information about them,   the scale transformation would  also then have to be a unitary transformation. However, if the hologram only retains modes which have exited the horizon, going to larger length scales  by a scale transformation in the boundary theory, i.e. to earlier times in the bulk,
		would lead to a loss of information and to non-unitary evolution. 
		
		As some  preliminary evidence for how scale transformations could act in unitary manner, we note, as was discussed above,   that the boundary theory in general violates   reflection positivity. This  allows for the trace of the stress tensor which generates local scale transformations  to be imaginary, \eq{cftanom}. In fact there are actually  two holograms one could associate with a Hartle Hawking state, one at ${\cal I}^+$ and the other  at ${\cal I}^-$,   corresponding to the boundaries of the expanding  and contracting branches. These boundary descriptions    are related by complex conjugation. For example,  the anomalous dimensions of operators are  related by complex conjugation, as are  the coefficients which appear in the correlations, e.g., the structure constants in the operator product expansion which was discussed in Appendix \ref{OPEint}. 
		
		In this paper we have studied some aspects of the holographic dictionary for underdamped scalar fields which have a mass $M^2>{d^2\over4}$. These fields do not, strictly speaking, freeze out at late times, rather they continue to oscillate but in a universal manner independent of the momenta ${\bf k}$,  with a dependence $e^{\pm i\mu \log(-\eta)}$ where $\mu$ is given in \eq{undernu}. One of our main results is to show   how a source can be identified in the boundary theory  for such fields by working in the coherent state basis. We also  argued that the cut-off independent terms in the  coefficient functions, obtained by expanding the wave function in a Taylor series expansion in this source,  satisfy the Ward identities of a CFT, namely 
		conformal invariance,  local scale invariance and momentum conservation. Some sample calculations for correlators,  showing that one obtains results   expected in a CFT, have also been included.

		We also argued that there are other ways to identify sources in the boundary theory, besides the coherent state eigenvalue. 
		An alternate identification is   possible in terms of variables $J_+, J_-$, which are related to the bulk field $\Phi$ through a transformation which is non-local in position space, \eq{reljpjm}. The cut-off independent terms we argued continue to satisfy the Ward identities of a CFT in this case, although some of the cut-off dependent terms are rather unwieldy, e.g., the resulting two-point correlator has a  cut-off dependent term which is  non-local in position space, \eq{posf}.

	         Our discussion was focused mostly on the under damped case but some aspects also have a bearing on the over damped case. For example it is possible in the over damped case as well as to identify the source  in the boundary theory in ways different from the conventional one and  argue that the cut-off independent terms in the wave function when expanded in terms of these sources continue to  satisfy the Ward identities, see Appendix \ref{app.alts}. 
	         
	         As has been discussed above, the boundary theory which is Euclidean violates reflection positivity and therefore cannot be continued to Lorentzian space, in general. And the fields in the boundary theory to which the bulk sources couple also cannot then be continued to operators in a Lorentzian theory. Nevertheless, we find, as was discussed in section \ref{WI} and Appendices \ref{OPEint} and \ref{conft}, that these boundary fields behave in a manner entirely analogous to operators in more conventional CFTs. In particular, they  transform under conformal transformations like primary operators  and  the nature of short distance singularities when they come together in correlation functions is   the same as  that in an operator product expansion, allowing us to define a  short distance expansion, operator product coefficients,  etc. 
Also, after analytic continuation, as discussed in section \ref{addcmnt} and Appendix \ref{App.analcont}, the operator product coefficients, even in the under damped case where we work  with the coherent state representation, can be obtained by analytical continuation from  AdS space.

	   		We look forward to further progress in this subject, hopefully spurred on by the study of concrete realisations of  de-Sitter space, including \cite{Vasiliev,HarlowAnni,Dasbilocal,anninoscftex}. 
		
		
	\section{Acknowledgements}
	First and foremost, we are deeply grateful to Suvrat Raju for generously sharing his insights and comments with us. The suggestion that the coherent state representation can be used to identify a source in the under damped case was made by him to us, and we thank him for this important insight  and related discussion.	We also acknowledge discussion with  N. Iizuka, J. Maldacena, K. Narayan,  S. Sake, T. Takayanagi,  and members of the TIFR String Theory Group, especially A. Gadde, G. Mandal, S. Minwalla and O. Parrikar.
	SPT acknowledges support from the KITP, Santa Barbara, enabling him to participate  in the workshop, ``What is String Theory".
	We  acknowledge support from Government of India, 
Department of Atomic Energy, under Project  Identification No. RTI 4002 and from the Quantum Space-Time Endowment of the Infosys Science Foundation.   			
Finally, we  thank the people of India for generously supporting research in String Theory.

	
\newpage

\appendix
\section{Wave Function Including $\phi^n$ Interaction}\label{appenboun}
In this section, detailed calculation of wave function under the inclusion of $\phi^n$ interaction through path integral and verification for $n=3$ with that obtained through perturbation technique are presented. It is also discussed how to represent it in coherent basis.  
\subsection{Using Time Dependent Perturbation Technique}
\label{perturbation}
The wave function in free theory is given by eq.(\ref{bounwvf}) near boundary. It is apparent that it should appear as the unperturbed contribution when we include interaction in a perturbative manner. So we denote it by $\psi_0$ and re-write below. 
\begin{equation}
	\label{C1.1}
	\psi_0[\varphi,\eta]=\mathcal{N}\exp\left[\frac{1}{2}\int \frac{d^d\mathbf{k}}{(2\pi)^d} ~\frac{i}{(-\eta)^{d-1}}\frac{\partial_\eta(\mathcal{F}_\nu(k,\eta))^{\ast}}{(\mathcal{F}_\nu(k,\eta))^{\ast}}\varphi(\mathbf{k},\eta)\varphi(-\mathbf{k},\eta)\right]
\end{equation}
where
$\mathcal{F}_\nu(k,\eta)$ is given by \eq{defF}. Including the normal ordered interaction Hamiltonian $:H_I(\eta'):$ in the theory,
\begin{equation}
	\label{C1.3}
	:H_I(\eta):=-\lambda\int d^d{\bf x}'\sqrt{-g}:\Phi^3(\mathbf{x}',\eta):
\end{equation}
The corresponding change in ground state can be calculated using time dependent perturbation technique. In the 1st order of coupling parameter $\lambda$, we get the modified state $\ket{\varphi_I}$ in terms of unperturbed state $\ket{\varphi_0}$ as
\begin{equation}
	\label{C1.4}
	\bra{\varphi_I}=\bra{\varphi_0}\left[1+i\int_{-\infty}^0 d\eta':H_I(\eta'):\right]
\end{equation} 
The modified wave function is thereby given as
\begin{equation}
	\label{C1.5}
	\psi_I=\bra{\varphi_I}\ket{0}
\end{equation}
Using \eq{C1.4}, \eq{C1.3}, \eq{defwvf}, we get
\begin{multline}
	\psi_I
	=\psi_0-i\lambda\int_{-\infty}^{0} d\eta'\int{d^d{\bf k}_1d^d{\bf k}_2d^d{\bf k}_3\over(2\pi)^{3d}}\sqrt{-g}(\mathcal{F}_\nu(k_1,\eta'))^{\ast}(\mathcal{F}_\nu(k_2,\eta'))^{\ast}(\mathcal{F}_\nu(k_3,\eta'))^{\ast}\\ \bra{\varphi_0}a^\dagger_{-\mathbf{k}_1}a^\dagger_{-\mathbf{k}_2}a^\dagger_{-\mathbf{k}_3}\ket{0}(2\pi)^d\delta\left(\sum_{i=1}^3\mathbf{k}_i\right)\label{C1.8}
\end{multline}
To derive the last line, we use \eq{Phik}. Putting the expression for $a^\dagger_{-{\bf k}}$ from \eq{a+(k)} and then using \eq{id1} in \eq{C1.8}, we get
\begin{empheq}{multline}
	\label{C1.10}
	\psi_I=\psi_0-\lambda\int\prod_{i=1}^{3}{d^d{\bf k}_i\over(2\pi)^d}\bra{\varphi_0}\prod_{i=1}^{3}\frac{\partial_\eta\mathcal{F}_{\nu}(k_i,\eta)\Phi(\textbf{k}_i,\eta)-(-\eta)^{d-1}\mathcal{F}_{\nu}(k_i,\eta)\Pi(\textbf{k}_i,\eta)}{(-\eta)^{d-1}}\ket{0} \\ \times (2\pi)^d\delta\left(\sum_{i=1}^3\mathbf{k}_i\right)\int_{-\infty}^{0} d\eta'\sqrt{-g}\prod_{i=1}^{3}(\mathcal{F}_{\nu}(k_i,\eta'))^{\ast}\hspace{50pt}
\end{empheq}
Using \eq{eigphib}, \eq{eigpi}
\begin{empheq}{multline}
	\label{C1.11}
	\psi_I=\psi_0-i\lambda\int\prod_{i=1}^{3}{d^d{\bf k}_i\over(2\pi)^d}(2\pi)^d\delta\left(\sum_{i=1}^3\mathbf{k}_i\right)\prod_{i=1}^{3}\mathcal{F}_{\nu}(k_i,\eta)D[{\bf k}_1,{\bf k}_2,{\bf k}_3]\psi_0\\
	\times\int_{-\infty}^{0} d\eta'\sqrt{-g}\prod_{i=1}^{3}(\mathcal{F}_{\nu}(k_i,\eta'))^{\ast}
\end{empheq}	
where
\begin{equation}
	\label{C1.12}
	D[{\bf k}_1,{\bf k}_2,{\bf k}_3]=\prod_{i=1}^{3}\left[-\frac{i}{(-\eta)^{d-1}}\frac{\partial_\eta(\mathcal{F}_\nu(k_i,\eta))}{\mathcal{F}_\nu(k_i,\eta)}\varphi(\textbf{k}_i,\eta)+\frac{\partial}{\partial\varphi(-\textbf{k}_i,\eta)}\right]
\end{equation}
Carrying out the action of $D[{\bf k}_1,{\bf k}_2,{\bf k}_3]$ on $\psi_0$ given in \eq{C1.1} results
\begin{align}
	\label{C1.13}
	D[{\bf k}_1,{\bf k}_2,{\bf k}_3]\psi_0
	=-\frac{1}{|\mathcal{F}_{\nu}(k_1,\eta)|^2|\mathcal{F}_{\nu}(k_2,\eta)|^2|\mathcal{F}_{\nu}(k_3,\eta)|^2}\varphi(\mathbf{k}_1,\eta)\varphi(\mathbf{k}_2,\eta)\varphi(\mathbf{k}_3,\eta)\psi_0
\end{align}  
To note, we have omitted the terms with single $\varphi(\mathbf{k},\eta)$ since each of such terms carries appropriate $\delta$-function that selects $\varphi(\mathbf{k}=0,\eta)$ among all possible $\mathbf{k}$ values. Hence the wave function is evaluated as
\begin{empheq}{multline}
	\label{C1.14}
	\psi_I=\psi_0\left[1+i\lambda\int\prod_{i=1}^{3}{d^d{\bf k}_i\over(2\pi)^d}(2\pi)^d\delta\left(\sum_{i=1}^3\mathbf{k}_i\right)\prod_{i=1}^{3}\frac{\varphi(\mathbf{k}_i,\eta)}{(\mathcal{F}_{\nu}(k_i,\eta))^{\ast}}\right.\\
	\left.\times\int_{-\infty}^0 d\eta'\frac{1}{(-\eta')^{d+1}}\prod_{i=1}^{3}(\mathcal{F}_{\nu}(k_i,\eta'))^{\ast}\right]
\end{empheq}
where we have put $\sqrt{-g}=\frac{1}{(-\eta)^{d+1}}$ as obvious from the metric \eq{Poing}. The result for wave function including cubic interaction, hence derived using perturbation technique, reproduces eq.(\ref{genpsipert}) at 1st order of coupling strength $\lambda$.

\subsection{Using Path Integral Formalism}
\label{pathpert}
The total action in the $\lambda\phi^n$ theory is the following 
\begin{equation}
	\label{appenS1}
	S=-\frac{1}{2}\int d^d\mathbf{x}d\eta\sqrt{-g}(\nabla_{\mu}\phi\nabla^{\mu}\phi+m^2\phi^2)+\lambda\int d^d\mathbf{x}d\eta\sqrt{-g}\phi^n
\end{equation}
Now,in the free theory,the equation of motion for $\phi$ is
\begin{equation}
	\label{shelleq}
	(\nabla_{\mu}\nabla^{\mu}-m^2)\phi=0
\end{equation}
The on-shell action $S_{onshell}$ is carried out by satisfying \eq{appenS1} with \eq{shelleq} 
\begin{equation}
	\label{appenS2}
	S_{onshell}=-\frac{1}{2}\int_{\partial} d^d\mathbf{x}\sqrt{-g}g^{\eta\eta}\phi\partial_{\eta}\phi+\lambda\int d^d\mathbf{x}d\eta'\sqrt{-g}\phi^n
\end{equation}
Since $\phi$ satisfies the equation of motion for the free theory, we need to consider $\delta\phi=0$  for this onshell calculation up to ${\cal O}(\lambda)$
Therefore,the correction to onshell action
\begin{equation}
	\delta S_n=+\lambda\int d^d\mathbf{x}d\eta'\sqrt{-g}\phi(\mathbf{x},\eta')^n
\end{equation}
In momentum space,this correction becomes,
\begin{equation}
	\delta S_n=\lambda\int \prod_{i=1}^{n}\frac{d^d{\mathbf{k}_i}}{(2\pi)^d}(2\pi)^d\delta^{(d)}(\mathbf{k}_1+...+\mathbf{k}_n)\int d\eta'\sqrt{-g}\prod_{i=1}^n\phi({\bf k}_i,\eta')
\end{equation}
Lets put 
\begin{equation}
	\label{phiphihat}
	\phi({\bf k}_i,\eta')=(\mathcal{F}_{\nu}(k_i,\eta'))^{\ast}\hat\phi(\mathbf{k}_i)
\end{equation}
where $\mathcal{F}_{\nu}(k_i,\eta)$ is the on-shell solution of the free theory given in \eq{defF}. Then, the above equation becomes
\begin{align}
	\delta S_n
	=\lambda\int\prod_{i=1}^{n}\frac{d^d\mathbf{k}}{(2\pi)^d}(2\pi)^d\delta^{(d)}(\mathbf{k}_1+...+\mathbf{k}_n)I(k_1,k_2,\cdots,k_n)\hat\phi(\mathbf{k}_i)
\end{align}
where 
\begin{equation}
	I(k_1,k_2,..,k_n)=\int_{-\infty}^{\eta_1}\frac{d\eta'}{(-\eta')^{d+1}}\prod_{i=1}^{n}(\mathcal{F}_{\nu}(k_i,\eta'))^{\ast}
\end{equation}
where $\epsilon$ is a small positive number. Now, we again substitute $\hat\phi(\mathbf{k}_i)=\frac{\phi(\mathbf{k}_i,\eta)}{(\mathcal{F}_{\nu}(k_i,\eta))^{\ast}}$ using the relation eq.(\ref{phiphihat}) and get
\begin{equation}
	\label{appendeltasn}
	\delta S_n=\lambda\int\prod_{i=1}^{n}\frac{d^d\mathbf{k}}{(2\pi)^d}(2\pi)^d\delta^{(d)}(\mathbf{k}_1+...+\mathbf{k}_n)I(k_1,\cdots,k_n)\prod_{i=1}^n\frac{\phi(\mathbf{k}_i,\eta)}{(\mathcal{F}_{\nu}(k_i,\eta))^{\ast}}
\end{equation}
which is consistent with eq.(\ref{deltasn}). To note,
In the above equation,in principal we are free to take any value of $\eta \in (-\infty,0)$. Now, to get the correction to the wave function near the boundary, we put the series expansion of $(\mathcal{F}_{\nu}(k_i,\eta))^{\ast}$ in the neighbourhood of $\eta=0$ in \eq{appendeltasn}.

\subsection{In Coherent State Basis}
\label{chiinter}
As discussed in section \ref{coherent}, wave function at boundary in terms of field eigenstate, $\psi[\phi]$, can be represented in terms of coherent eigenstate, $\psi[\rho]$ through the transformation given by \eq{der1psirho}. which is a functional integral over $\phi$. Using the relation \eq{defjpb}, we can write this as
\begin{equation}
	\label{der1jrho}
	\begin{split}
		\psi[\rho]=\int {\cal D}J_+\Psi_\rho^{\ast}[J_+]\psi[J_+] 
	\end{split}
\end{equation}  
where $\Psi_\rho[J_+]$ is given in \eq{formfa} and $\psi[J_+]$ is given by
\begin{empheq}{multline}
\label{psijtot}
\psi[J_+]=\exp\left[i\int \frac{d^d\mathbf{k}}{(2\pi)^d} ~Q_1 J_+({\bf k})J_+(-{\bf k})\right.\\
\left.+{i\lambda}\int\left(\prod_{i=1}^{3}{d^d{\bf k}_i\over(2\pi)^d}\right)(2\pi)^d\delta({\bf k}_1+{\bf k}_2+{\bf k}_3)Q_2\left(\prod_{i=1}^{3}J_{+}({\bf k_i})\right)\right]
\end{empheq}
where
\be
\label{Q1Q2}
Q_1=\left(k^{2i\mu}\frac{f_\mu^\ast(k,\eta)\partial_\eta f_\mu^\ast(k,\eta)}{2(-\eta)^{d-1}}\right)~~~~Q_2={\left(\prod_{i=1}^{3}k_i^{i\mu}\right)I(k_1,k_2,k_3)}
\ee
All together, the integrand in \eq{der1jrho} is evaluated as
\begin{empheq}{multline}
	\label{der2jrho}
	\exp\left[\int \frac{d^d\mathbf{k}}{(2\pi)^d}W_1\rho^{\ast}({\bf k}) J_+({\bf k})+i\int \frac{d^d\mathbf{k}}{(2\pi)^d} ~W_2 J_+({\bf k})J_+(-{\bf k})\right.\\
	\left.+{i\lambda}\int\left(\prod_{i=1}^{3}{d^d{\bf k}_i\over(2\pi)^d}\right)(2\pi)^d\delta({\bf k}_1+{\bf k}_2+{\bf k}_3)Q_2\left(\prod_{i=1}^{3}J_{+}({\bf k_i})\right)\right]
\end{empheq}
where
\be
\label{Wcoef}
W_1=\sqrt{2\mu} f_\mu^\ast(k,\eta) k^{i\mu}(-\eta)^{-\Delta_+} ~~~~W_2=\left(Q_1+{\Delta_+(f_\mu^\ast(k,\eta))^2 k^{2i\mu}\over2(-\eta)^d}\right)
\ee
\eq{der2jrho}, as argued in section \ref{coherent}, has a factor $R_{\rm dS}^{d-1}\over G_N$ which in semi-classical limit goes towards $\infty$. This brings us the opportunity to take saddle point approximation to solve the integral. The saddle point can be found by extremizing the integrand \eq{der2jrho} which gives
\begin{equation}
	\label{saddleco}
	W_1\rho^{\ast}({\bf k})
	+2i W_2 J_+(-{\bf k})+{3i\lambda}\int\left(\prod_{i=2}^{3}{d^d{\bf k}_i\over(2\pi)^d}\right)(2\pi)^d\delta({\bf k}+{\bf k}_2+{\bf k}_3)Q_2\left(\prod_{i=2}^{3}J_{+}({\bf k_i})\right)=0
\end{equation}
Above relation is exact and can be easily generalised for $\phi^n$ interaction. In order to get $J_+$ in terms of $\rho^{\ast}$, we have to invert the above relation order by order in $\lambda$. The 0-th order saddle is found as
\be
\label{0thsad}
J_+(-{\bf k})=i{W_1\over2W_2}\rho^{\ast}({\bf k})
\ee
which reproduces \eq{Jrelchi} using \eq{Wcoef}. 
The ${\cal O}(\lambda)$ correction, $\delta J_+$ to the saddle value can be found from \eq{saddleco} as
\be
\label{saddle1}
\delta J_+(-{\bf k})={3\lambda}\int\left(\prod_{i=2}^{3}{d^d{\bf k}_i\over(2\pi)^d}\right)(2\pi)^d\delta({\bf k}+{\bf k}_2+{\bf k}_3){Q_2W_1^2\over8W_2^3}\left(\prod_{i=2}^{3}\rho^{\ast}({\bf k}_i)\right)
\ee
where we have used \eq{0thsad} in the 3rd term in LHS of \eq{saddleco}. Hence the ${\cal O}(\lambda)$ contribution to the integral \eq{der1jrho} is given by its first term as
\begin{align}
	\int \frac{d^d\mathbf{k}}{(2\pi)^d}W_1\rho^{\ast}({\bf k}) \delta J_+({\bf k})
	= {3\lambda}\int\left(\prod_{i=1}^{3}{d^d{\bf k}_i\over(2\pi)^d}\right)(2\pi)^d\delta(-{\bf k}_1+{\bf k}_2+{\bf k}_3){Q_2W_1^3\over8W_2^3}\left(\prod_{i=1}^{3}\rho^{\ast}({\bf k_i})\right)\label{term1}
\end{align}
The fact that the ${\cal O}(\lambda)$ contribution from 2nd term given by
\begin{align}
	&i\int \frac{d^d\mathbf{k}}{(2\pi)^d} ~W_2 [J_+({\bf k})\delta J_+(-{\bf k})+J_+(-{\bf k})\delta J_+({\bf k})]\\
	=& -{3\lambda}\int\left(\prod_{i=1}^{3}{d^d{\bf k}_i\over(2\pi)^d}\right)(2\pi)^d\delta(-{\bf k}_1+{\bf k}_2+{\bf k}_3){Q_2W_1^3\over8W_2^3}\left(\prod_{i=1}^{3}\rho^{\ast}({\bf k_i})\right)
\end{align}
exactly cancels \eq{term1} ensures that the derivation can be easily generalised to $\phi^n$ interaction. Hence ${\cal O}(\lambda)$ contribution comes only from the last term in \eq{der2jrho} obtained using \eq{Q1Q2} and \eq{Wcoef} as
\begin{empheq}{multline}
\label{delpsirho}
\delta\psi[\rho]=\exp\left[{i\lambda\over (2\mu)^{3\over2}(\alpha_\mu^{\ast})^3}\int\left(\prod_{i=1}^{3}{d^d{\bf k}_i\over(2\pi)^d}\right)(2\pi)^d\delta({\bf k}_1+{\bf k}_2+{\bf k}_3)\right.\\
\left.\times{\left(\prod_{i=1}^{3}k_i^{i\mu}\right)I(k_1,k_2,k_3)}\left(\prod_{i=1}^{3}\rho^{\ast}({\bf k}_i)\right)\right]
\end{empheq}
which matches with the ${\cal O}(\lambda)$ correction for $\phi^3$ interaction in \eq{intwvfchi} . Note that to calculate the contribution at ${\cal O}(\lambda^n)$, we should be given the saddle value up to ${\cal O}(\lambda^{n-1})$. Using that in \eq{saddleco}, we get $J_+$ in terms of $\rho^{\ast}$ correct up to ${\cal O}(\lambda^n)$. Now we put the relation in \eq{der2jrho} to get the coherent state wave function correct up to ${\cal O}(\lambda^n)$.

\section{Analytic Continuation from AdS to dS }\label{App.analcont}

Here we briefly study scalar field in AdS and subsequently discuss how the results in overdamped case in dS can be obtained through analytic continuation.\\
\\
\textbf{Scalar Field in AdS}\\
\\
The action for a massive scalar field in Euclidean AdS is written as follows
\be
\label{scactb}
S={1 \over 2G_N}\int d^{d+1}x\sqrt{g} \left(g^{\mu\nu}\partial_{\mu}\phi \partial_{\nu}\phi+M^2\phi^2\right)
\ee 
where the metric is given in \eq{metadsea}. The dependence on $L$ can be determined by dimensional analysis and we will continue it  using eq.(\ref{contab}). 
Also it is convenient to $G_N=1$ where $G_N$ is Newton's constant of gravitation.
Normalised on-shell solution for $\phi$ in momentum space is given by
\begin{equation}
	\phi({\bf k},z) =  \frac{z^{\frac{d}{2}} K_{\nu} (kz) }{\epsilon^{\frac{d}{2}} K_{\nu} (k\epsilon)} \epsilon^{{d\over2}-\nu} \phi_b({\bf k}) \label{phisol}
\end{equation}
where $$\nu=\sqrt{{d^2\over4}+M^2 L^2}$$
and $\epsilon$ is the cut-off value of z close to the AdS boundary. On-shell action is given in momentum space as
\be
\label{onsads}
S_{\partial} = -\frac{L^{d-1}}{2} \int {d^d\mathbf{k}\over(2\pi)^d} \frac{1}{z^{d-1}} \phi({\bf k},z) \partial_z \phi(-{\bf k},z) |_{z=\epsilon}
\ee
%
The partition function in EAdS up to ${\cal O}(\epsilon^{2\nu})$ is given by
\be
\label{wfactads}
Z_{AdS}=\exp[-S_\del]=\exp[\frac{L^{d-1}}{2}\int \frac{d^d\mathbf{k}}{(2\pi)^d} \left\{\left(\frac{d}{2} - \nu\right) \epsilon^{-2 \nu} + k^{2\nu} \frac{a_{\nu}}{b_{\nu}} 2 \nu \right\} \phi_b({\bf k}) \phi_b (-{\bf k})]
\ee
where $a_\nu$ and $b_\nu$ given by
\be
\label{coefads}
a_\nu=2^{-\nu -1} \Gamma[-\nu]    \hspace{50pt}   b_\nu=2^{\nu -1} \Gamma [\nu]
\ee
are coefficients of leading modes of $K_\nu(kz)$ near boundary given by
\be
\label{Kboun}
K_{\nu}(kz)=a_\nu(kz)^{\nu}+b_\nu(kz)^{-\nu}
\ee
Using the definition \eq{nptads}, the 2pt. correlator is found to be
\begin{equation}
	\label{adstwopt}
	\langle O(\mathbf{k}) O(-\mathbf{k}) \rangle_{AdS} = -  \frac{\pi}{2^{2\nu-1} \Gamma[\nu]^2 \sin(\pi \nu)} k^{2\nu} L^{d-1}
\end{equation}
where we used \eq{coefads} and discarded the local contribution that would arise under position space representation of $Z_{AdS}$. \\
Similarly the partition function for integer $\nu$ is given by
\begin{multline}
	\label{partadsp2}
	\log Z_{AdS}=\frac{L^{d-1}}{2}\int \frac{d^d\mathbf{k}}{(2\pi)^d} \left[\left(\frac{d}{2}-\nu\right)\epsilon^{-2\nu}+\left\{\frac{\tilde{b}_0}{\tilde{a}_0}(1+2\nu \log(\epsilon))+2\nu\frac{\tilde{c}_0}{\tilde{a}_0}\right\}k^{2\nu}\right.\\
	\left.+2\nu\frac{\tilde{b}_{0}}{\tilde{a}_{0}} k^{2\nu}\log(k) \right] \phi_b({\bf k}) \phi_b (-{\bf k})
\end{multline}
where ($\gamma_E$ being the Euler number)
\begin{align}
	\tilde{a}_0=2^{\nu-1}\Gamma[\nu] \hspace{20pt}
	\tilde{b}_0=\frac{(-1)^{\nu-1}}{2^{\nu}\Gamma[\nu+1]}\hspace{20pt}
	\tilde{c}_0=\frac{(-1)^{\nu+1}}{2^{\nu}\Gamma[\nu+1]}\left(\gamma_E-\frac{1}{2}\sum_{m=1}^{d/2}\frac{1}{m}-\log(2)\right)
\end{align}
with the two point correlator in momentum space for integer $\nu$
\begin{align}
	\langle O(\mathbf{k})O(\mathbf{-k})\rangle_{AdS}=-\frac{(-1)^{\nu} }{2^{2\nu-2}\Gamma[\nu]^2}L^{d-1}k^{2\nu}\log(k)\label{okokint}
\end{align}	
In position space, both \eq{adstwopt} and \eq{okokint} takes the form
\begin{equation}
	\label{oxoyads}
	\langle O(\mathbf{x})O(\mathbf{y})\rangle_{AdS}= \frac{2\nu }{\pi^{d\over2}}\frac{\Gamma[\frac{d}{2}+\nu]}{\Gamma[\nu]}{L^{d-1}\over|{\bf x}-{\bf y}|^{d+2\nu}}
\end{equation}
\\
\textbf{Analytic Continuation}\\
\\
Now we can go to dS from EAdS through analytic continuation. The primary rules of the continuation are given by
\begin{align}
	L \rightarrow iR_{\rm dS}, \, z \rightarrow -i \eta \label{ancnt1}
\end{align}
According to \eq{ancnt1}, the cut-off is also continued as
\be
\label{cutcont}
\epsilon=-i\eta_1
\ee
Now, for the Euclidean AdS, we specify the boundary condition to be
\begin{equation}
	\phi(\mathbf{k},z) = \epsilon^{{d\over2}-\nu} \phi_b (\mathbf{k})
\end{equation}
while at the boundary of dS, we have
\begin{equation}
	\phi(\mathbf{k},\eta) = (-\eta_1) ^{{d\over2}-\nu} \hat{\phi}(\mathbf{k})
\end{equation}
Now continuing the cut-off $\epsilon$ to $\eta_1$ from EAdS to dS given by \eq{cutcont}
and  $L$ as given by eq.(\ref{ancnt1}), then  gives the rule of continuation for the sources from AdS to dS as follows
\be
\phi_b ({\bf k}) = i^{-{d\over2}+\nu} \hat{\phi}({\bf k}) \label{ancnts}
\ee
and one can check using eq.(\ref{valtab}) and eq.(\ref{coefads}) that 
\be
\label{coeftwo}
i^{2\nu} \frac{a_{\nu}}{b_{\nu}}={\beta_\nu^{\ast}\over\alpha_\nu^{\ast}}=-{\pi 2^{-2\nu} e^{i \pi \nu}\over \nu(\Gamma[\nu])^2\sin{\pi\nu}}
\ee
where the last equality comes from \eq{ratalbe}.
Note that in the context of this continuation, the source in dS is not the same as that in EAdS but differs by a phase factor. Using the rules eq.(\ref{ancnt1}), eq.(\ref{ancnts}) in eq.(\ref{wfactads}), we get for non-integer $\nu$
\be
\label{wfadsds}
\psi_{AdS\rightarrow dS}=\exp[R_{\rm dS}^{d-1}\int \frac{d^d\mathbf{k}}{(2\pi)^d} \left\{-\frac{i}{2}\left(\frac{d}{2} - \nu\right) (-\eta_1)^{-2 \nu} + {i{\pi 2^{-2\nu} e^{i \pi \nu}\over (\Gamma[\nu])^2\sin{\pi\nu}}}k^{2\nu}  \right\} \hat{\phi}({\bf k}) \hat{\phi} (-{\bf k})]
\ee
which is exactly dS wave function eq.(\ref{bounwvpexp}) in overdamped case taking $R_{\rm dS}=1$, ignoring normalising factor. For integer $\nu$, the continuation results
\begin{multline}
	\label{partdsp2}
	\log \psi_{AdS\rightarrow dS}= -\frac{R_{\rm dS}^{d-1}}{2}\int \frac{d^d\mathbf{k}}{(2\pi)^d}\hat{\phi}(\mathbf{k})\hat{\phi}(-\mathbf{k})\Biggl[i\left(\frac{d}{2}-\nu\right)(-\eta_1)^{-2\nu} \\
	+i\left\{(1+2\nu\log(-\eta_1))\frac{\bar{\beta}_{\nu}^{\ast}}{\bar{\alpha}_{\nu}^{\ast}}+2\nu \frac{\bar{C}_{\nu}^{\ast}}{\bar{\alpha}_{\nu}^{\ast}}\right\}k^{2\nu}+2i \nu \frac{\bar{\beta}_{\nu}^{\ast}}{\bar{\alpha}_{\nu}^{\ast}}k^{2\nu}\log(k)\Biggr]
\end{multline}
where
\begin{align}
	\label{appalp}
	\bar{\alpha}_{\nu}^{\ast}&=i\frac{2^{\nu}\Gamma[\nu]}{\pi} \\
	\label{appbet}
	\bar{\beta}_{\nu}^{\ast}&=-\frac{i}{2^{\nu-1}\pi\Gamma[\nu+1]}\\
	\label{appCint}
	\bar{C}_{\nu}^{\ast}&=-\frac{i}{2^{\nu-1}\pi\Gamma[1+\nu]}\left(\gamma+\frac{i\pi}{2}-\frac{1}{2}\sum_{m=1}^{\nu}\frac{1}{m}-\log 2\right)
\end{align}
 Notice that had the analytic continuation in eq.\eqref{ancnt1} been different, and instead given by  $L \rightarrow -i R_{dS}$, as discussed in \cite{Malda-NG}, then eq.\eqref{wfadsds} and eq.\eqref{partdsp2} would not have agreed with the correct dS answers, when $d$ is even.
The two point correlator can now be extracted by the definition \eq{npt}
and is given by
\begin{equation}
	\label{okoknintds}	
	\langle O(\mathbf{k})O(\mathbf{-k})\rangle= 
	\begin{cases}
		[-1](1-i\cot(\pi\nu)) {\pi 2^{-2\nu+1} \over (\Gamma[\nu])^2}R_{\rm dS}^{d-1}k^{2\nu} & {\rm non-integer ~\nu}\\
		\frac{i}{2^{2\nu-2}\Gamma[\nu]^2}R_{\rm dS}^{d-1}k^{2\nu}\log(k) & {\rm integer~\nu}
	\end{cases}	
\end{equation}
where we have discarded the pieces which gives local contribution while converting in position space. The result \eq{okoknintds} for non-integer case agrees with \eq{twptb}. Similarly, the two point function in position space is obtained by the Fourier transform of eq.(\ref{okoknintds}) as 
\begin{align}
	\label{dsx2pt}
	\langle O(\mathbf{x})O(\mathbf{y})\rangle =e^{{i\pi\over2}(2\nu-1)} \frac{2\nu \Gamma[{d\over2}+\nu]}{\pi^{d\over2}\Gamma[\nu]}\frac{R_{\rm dS}^{d-1}}{|\mathbf{x}-\mathbf{y}|^{d+2\nu}}
\end{align}
irrespective of $\nu$ being integer or non-integer which agrees with \eq{posspa}.\\
In dS, the acceptable wave function in momentum space has to be normalisable. The normalisability issue is satisfied in case of non-integral $\nu$ since the real part of the correlator, eq.(\ref{okoknintds}) non-integral case, is positive. However for integral $\nu$, the issue is more subtle since the two point correlator, eq.(\ref{okoknintds}) integral case, is pure imaginary for integer $\nu$, as mentioned in section \ref{factorsofi}. But the dS wave function, eq.(\ref{partdsp2}) is still normalizable because of the following contribution from the contact term 
\begin{align}
	&-\frac{i}{2}R_{dS}^{d-1}\left\{(1+2\nu\log(-\eta_1))\frac{\bar{\beta}_{\nu}^{\ast}}{\bar{\alpha}_{\nu}^{\ast}}+2\nu \frac{\bar{C}_{\nu}^{\ast}}{\bar{\alpha}_{\nu}^{\ast}}\right\}k^{2\nu}\hat{\phi}(\mathbf{k})\hat{\phi}(-\mathbf{k}) \nonumber\\
	&=\frac{i k^{2\nu}}{2^{2\nu}\nu\Gamma[\nu]^2}R_{dS}^{d-1}\left\{1+2\nu\left(\gamma+\frac{i\pi}{2}-\frac{1}{2}\sum_{m=1}^{\nu}\frac{1}{m}-\log[2]+\log[-\eta_1]\right)\right\}\hat{\phi}(\mathbf{k})\hat{\phi}(-\mathbf{k}) \nonumber\\
	&=-\frac{\pi R_{dS}^{d-1}}{2^{2\nu}\Gamma[\nu]^2}k^{2\nu}\hat{\phi}(\mathbf{k})\hat{\phi}(-\mathbf{k})+{\rm (pure~imaginary~term)}
	\label{dampint}
\end{align}
Therefore, the 2nd term inside the curly bracket in \eq{dampint} provides a damping contribution to the wave functional given by 
\begin{equation}
	\label{contactdamp}
	\exp[-\frac{R_{\rm dS}^{d-1}}{2}\int \frac{d^d\mathbf{k}}{(2\pi)^d}\frac{\pi}{2^{2\nu-1}\Gamma[\nu]^2}k^{2\nu}\hat{\phi}({\bf k})\hat{\phi}(-{\bf k})]
\end{equation}

Before we end the discussion of the analytic continuation from AdS to dS we would like to point out that the analytic continuation used in eq.\eqref{ancnt1} is different from \cite{Malda-NG}, see also \cite{Juan--gravity} while agreeing with \cite{opdic}. The difference is primarily in the continuation of $L$ to $R_{dS}$ and arises when $d$ is even. However this has important consequences as explained earlier in this section. In the following we will give one further check of the correctness of the analytic continuation by evaluating the gravitational on-shell action in both AdS and dS space. 
Consider the Euclidean $AdS_4$ action 
\begin{equation}
	S = -\int d^4x (\textbf{R}+\frac{6}{L^2}) -2 \int d\sigma \sqrt{\gamma} K
\end{equation}
evaluated in the case of ball like AdS 
\begin{equation}
	ds^2 = L^2(d\tau^2 + \sinh[2](\tau) d\Omega_3^2)
\end{equation}
for $\tau \in [0,\tau_0]$. We get,
\begin{align}
	\log(Z)& = \int d^4x (\textbf{R}+\frac{6}{L^2}) +2 \int d\sigma \sqrt{\gamma} K 
	= -V_{S^3} \int d\tau L^4 \sinh[3](\tau)  \frac{6}{L^2}+2V_{S^3} L^2 \sinh[3](\tau_0) 3 \coth(\tau_0) \nonumber \\
	& = L^2 V_{S^3}  \left(\cosh(3\tau_0)+3\cosh(\tau_0)-4\right)	\label{ads4}
\end{align}
The $\tau_0$ independent term in the above action agrees with eq.(5.1) in \cite{Juan--gravity}. An analogous calculation for $AdS_5$ yields,
\begin{equation}
	\log(Z) = L^3 V_{S^4} \frac{3}{4}  \left(\sinh(4\tau_0)-4\tau_0\right)	\label{ads5}
\end{equation}
Using the analytic continuation suggested in \cite{Juan--gravity}, $\tau_0 = \theta_0 + \frac{i \pi}{2}, L= -i R_{dS}$, we get for $dS_4$
\begin{equation}
	\log(\Psi_{HH})= - R_{dS}^2 V_{S^3}  \left(-i\sinh(3\theta_0)+3i\sinh(\theta_0)-4\right)	= R_{dS}^2 V_{S^3} (4 + 4 i \sinh[3](\theta_0)) \label{ds4check}
\end{equation} 
and for $dS_5$,
\begin{equation}
	\log(\Psi_{HH}) = i R_{dS}^3 V_{S^4} \frac{3}{4}(\sinh(4\theta_0)-4 \theta_0) + R_{dS}^3 V_{S^4} \frac{3 \pi}{2} \label{ds5check}
\end{equation}
To check the analytic continuation let us now calculate the expanding branch Hartle Hawking (HH$^+$) wavefunction for both $dS_4$ and $dS_5$. The global metric for $dS_{d+1}$ is given by,
\begin{equation}
	ds^2 = R_{dS}^2(-d\theta^2 + \cosh[2](\theta) d\Omega_{d}^2) \label{dsmet}
\end{equation}
The HH$^+$ wavefunction consists of a real part which is the volume of a hemisphere and an imaginary part obtained by calculating the Lorentzian gravitational action for the above metric. As an example consider $dS_4$. The Lorentzian action is then given by,
\begin{equation}
	i S^L_{dS} = i \int d^4x (\textbf{R}-\frac{6}{R_{dS}^2}) -2 i \int d\sigma \sqrt{\gamma} K
\end{equation}
Using the metric eq.\eqref{dsmet} with $\theta\in[0,\theta_0]$ we obtain,
\begin{equation}
	i S^L_{dS} = - 4 i  R_{dS}^2 V_{S^3} \sinh[3](\theta_0)
\end{equation}
Then the wavefunction is given by,
\begin{equation}
	\log(\Psi_{HH}) = 3 R_{dS}^2 V_{S^4} -  4 i  R_{dS}^2 V_{S^3} \sinh[3](\theta_0) = R_{dS}^2 V_{S^3} (4 - 4 i \sinh[3](\theta_0))  \label{ds4}
\end{equation}
In the case of for $dS_5$ one similarly gets,
\begin{equation}
	\log(\Psi_{HH}) =  -i R_{dS}^3 V_{S^4} \frac{3}{4}(\sinh(4\theta_0)-4 \theta_0) + R_{dS}^3 V_{S^4} \frac{3 \pi}{2} \label{ds5}
\end{equation}
Comparing eq.\eqref{ds4check} and eq.\eqref{ds5check} with eq.\eqref{ds4} and eq.\eqref{ds5} we see that the hemisphere part which is the real part agrees while the imaginary part disagrees.
However if we take 
\begin{equation}
	\tau_0 = \theta_0 - \frac{i \pi}{2}, L= i R_{dS} \label{anacon}
\end{equation}
then one can check that the analytic continuation of eq.\eqref{ads4} and eq.\eqref{ads5} will agree with eq.\eqref{ds4} and eq.\eqref{ds5} respectively. The analytic continuation eq.\eqref{anacon} agrees with eq.(4.6) of \cite{opdic}.

The above analysis shows that the analytic continuations eq.\eqref{ancnt1} and eq.\eqref{anacon} give  the correct continuations from AdS to dS for all values of $d$.

\subsection{Analytic Continuation from Overdamped to Underdamped}
In the preceding discussion we saw that the results obtained for the overdamped scalar fields can be understood through an analytic continuation of the corresponding results in the case of AdS. In this subsection we will show that the results obtained for the underdamped scalar fields can similarly be obtained by doing an analytic continuation of the corresponding results in the overdamped case. 

Consider the two point function obtained for the overdamped scalar fields, \eq{posspa}, reproduced below.
\be
\label{posspa2}
\l O({\bf x})O({\bf y})\r=e^{i \pi (2\nu-1)\over 2} 2 \nu {\Gamma[{d\over2}+\nu]\over \pi^{d\over2}\Gamma[\nu]} {1\over |{\bf x}-{\bf y}|^{d+2\nu}}
\ee
Similarly consider the two point function obtained for the underdamped scalar fields in the coherent state basis,\eq{rho2pt}, reproduced below.
\begin{equation}
	\l O_+({\bf x})O_+({\bf y})\r={i\Gamma[1-i\mu]\Gamma[{d\over2}+i\mu]\over\pi^{1+{d\over2}}(1+\coth(\pi\mu))}{1\over|{\bf x}-{\bf y}|^{d+2i\mu}}
\end{equation}	
Simplifying the above expression we get,
\begin{equation}
	\label{rho2pt2}
	\l O_+({\bf x})O_+({\bf y})\r={e^{-\pi \mu}\Gamma[{d\over2}+i\mu]\over\pi^{{d\over2}} \Gamma[i \mu]}{1\over|{\bf x}-{\bf y}|^{d+2i\mu}}
\end{equation}	
The rule for analytic continuation in going from overdamped to underdamped will be,
\begin{equation}
	\nu \rightarrow i \mu. \label{anconrel}
\end{equation}
Then \eq{posspa2} becomes,
\be
\label{posspa3}
\l O({\bf x})O({\bf y})\r|_{\nu\rightarrow i\mu}=e^{i \pi (2 i \mu -1)\over 2} 2 i \mu {\Gamma[{d\over2}+i \mu]\over \pi^{d\over2}\Gamma[i \mu]} {1\over |{\bf x}-{\bf y}|^{d+2i\mu}}
\ee
Comparing \eq{rho2pt} with \eq{posspa3} we get,
\begin{equation}
	\frac{\langle O_+({\bf x}) O_+({\bf y})\rangle}{\langle O({\bf x})O({\bf y})\rangle|_{\nu \rightarrow i \mu}} = \frac{1}{2 \mu} \label{ratio2pt}
\end{equation}
The factor $\frac{1}{2 \mu} = \frac{1}{(\sqrt{2\mu})^2}$ can be thought of as the ratio of normalizations of $\rho$ and $\hat{\phi}$. To verify this statement let us now compare three point functions of both underdamped and overdamped fields. The three point function for overdamped field, given by \eq{3ptover} in k-space, can be expressed in position space \footnote{Fourier transform of three point function considering fields with different masses is explicitly performed in Appendix \ref{threeancnt}.} as
\begin{equation}
	\label{3ptxov}
	\l O({\bf x}_1)O({\bf x}_2)O({\bf x}_3)\r=-{3\lambda\over \pi^d}
	{\Gamma\left[{d\over4}+{3\nu\over2}\right]\Gamma\left[{\Delta\over2}\right]^3\over\Gamma[\nu]^3}
	\times\\{\exp[{i\pi\over2}(3+3\nu-{d\over2})]\over |{\bf x}_1-{\bf x}_2|^{\Delta}|{\bf x}_2-{\bf x}_3|^{\Delta}|{\bf x}_3-{\bf x}_1|^{\Delta}}	
\end{equation}
where $\Delta$ is given in \eq{overdim}.
Three point function for underdamped field, given by \eq{3ptcoh} in k-space, can be expressed in position space as
\begin{multline}
	\label{3ptxun}
	\l O_+({\bf x}_1)O_+({\bf x}_2)O_+({\bf x}_3)\r= -{3\lambda\over (\sqrt{2\mu})^3\pi^d}
	{\Gamma\left[{d\over4}+{3i\mu\over2}\right]\Gamma\left[{\Delta_+\over2}\right]^3\over\Gamma[i\mu]^3}
	\times\\{\exp[{i\pi\over2}(3+3i\mu-{d\over2})]\over |{\bf x}_1-{\bf x}_2|^{\Delta_+}|{\bf x}_2-{\bf x}_3|^{\Delta_+}|{\bf x}_3-{\bf x}_1|^{\Delta_+}}
\end{multline}
where $\Delta_+$ is given in \eq{underdim}.
Doing the analytic continuation of \eq{3ptxov} using \eq{anconrel} we get,
\begin{multline}
	\label{ovun3}
	\l O({\bf x}_1)O({\bf x}_2)O({\bf x}_3)\r|_{\nu\rightarrow i\mu}= -{3\lambda\over\pi^d}
{\Gamma\left[{d\over4}+{3i\mu\over2}\right]\Gamma\left[{\Delta_+\over2}\right]^3\over\Gamma[i\mu]^3}
\times\\{\exp[{i\pi\over2}(3+3i\mu-{d\over2})]\over |{\bf x}_1-{\bf x}_2|^{\Delta_+}|{\bf x}_2-{\bf x}_3|^{\Delta_+}|{\bf x}_3-{\bf x}_1|^{\Delta_+}}
\end{multline}
Comparing \eq{ovun3} with \eq{3ptxun} we get,
\begin{equation}
	\frac{\langle O_+({\bf x}_1) O_+({\bf x}_2) O_+({\bf x}_3)\rangle}{\langle O({\bf x}_1)O({\bf x}_2)O({\bf x}_3)\rangle|_{\nu \rightarrow i \mu}} = \frac{1}{(\sqrt{2\mu})^3} \label{ratio3pt}
\end{equation}
Thus we see that there is an analytic continuation from overdamped to underdamped case given by \eq{anconrel} after one takes into account the ratio of normalization of $\rho$ and $\hat{\phi}$. We expect this feature to hold for any $n$ point function. Let us note that the Ward identities discussed in section \ref{WI} and Appendix \ref{Wardtest} also continue from the overdamped to the underdamped case by taking $\nu\rightarrow i \mu$ and changing the sources appropriately.

In conclusion, we see that starting from AdS one can do an analytic continuation \eq{ancnt1} to obtain the $n$ point function in overdamped case and then can do a further analytic continuation \eq{anconrel} to obtain the $n$ point function in the underdamped case.

\section{More on the Three Point Correlator}
\label{App.Hankel}
In section \ref{holo} and \ref{coherent}, we studied the holography for overdamped and underdamped fields  including interaction. Here we give more details on, $n$-point correlations function obtained in those analysis.
\subsection{Analytic Continuation of Correlation Function}
\label{analI}
Let us consider single vertex interaction in dS with three different scalar fields $\phi_1,\phi_2,\phi_3$ of masses $m_1,m_2,m_3$ respectively. Here the $m_i$'s can either be all under-damped or over-damped or mixed. For complete generality, one can consider the mixed case and obtain the $n$-point correlator as 
\begin{empheq}{multline}
	\label{wittenmom3}
	\l O_1({\bf k}_1)O_2({\bf k}_2)\cdots O_n({\bf k}_n)\r={C_n[\{a\};\{b\}]}(2\pi)^d\delta^{(d)}(\sum_{i=1}^{n} \textbf{k}_i) \\
	\times k_1^{\nu_1}k_2^{\nu_2} \cdots k_n^{\nu_n} \int_{-\infty}^{\eta_1}{d\eta'\over-\eta'}(-\eta')^{({n\over2}-1)d}\prod_{j=1}^n H_{\nu^{\ast}_j}^{(2)}(-k_j\eta')
\end{empheq}
where $\nu_a$ takes real values for overdamped fields and is purely imaginary, $\{\nu_a=i\mu_a:\mu_a\in\real\}$, for underdamped fields. 
${C_n[\{a\};\{b\}]}$ is given by
\begin{align}
	\label{cn}
	C_n[\{a\};\{b\}]= i\lambda n!\prod_{\{a\}}\left({\mathbb{C}_{\nu_a}^{\ast}\over\alpha_{\nu_a}^{\ast}}\right)\prod_{\{b\}}\left(\mathbb{C}_{\mu_b}^{\ast}\over \sqrt{2\mu}\alpha_{\mu_b}^{\ast}\right)
\end{align}   
where $\{a\}$ refers to all the overdamped fields and $\{b\}$ refers to all the underdamped fields and $\{a\}\bigcup\{b\}=\{1,2,\cdots,n\}$, with $n$ being total no.of fields. The ratio ${\mathbb{C}_\nu\over\alpha_{\nu}}$, ${\mathbb{C}_\mu\over\alpha_\mu}$ are given in \eq{defab}, \eq{defabuC} respectively.\\
Consider the integral in \eq{wittenmom3}.
\be
\label{Iexp}
I(k_1,k_2,\cdots,k_n)=\int_{-\infty}^{\eta_1}{d\eta'\over-\eta'}(-\eta')^{({n\over2}-1)d}\prod_{j=1}^n H_{\nu^{\ast}_j}^{(2)}(-k_j\eta')
\ee
The argument $-k\eta$, when written as $-i(-ik\eta)$, satisfies
\be
\label{argc}
\text{arg}(-ik\eta)={\pi\over2}
\ee     
Since,
\begin{equation}
	K_q(x)= \frac{\pi}{2}(-i)^{q+1}H^{(2)}_q(-ix),  \,  -\frac{\pi}{2}<\arg(x)<\pi
\end{equation}
we can rewrite \eq{Iexp} as,
\be
\label{IexpK1}
I(k_1,k_2,\cdots,k_n)=\left({2\over\pi}\right)^n e^{{i\pi\over2}(n+\sum_j\nu^{\ast}_j)}\int_{-\infty}^{\eta_1}{d\eta'\over-\eta'}(-\eta')^{({n\over2}-1)d}\prod_{j=1}^n K_{\nu^{\ast}_j}(-ik_j\eta')
\ee 
Substituting $\eta'=i\tau$, we get
\be
\label{IexpK2}
I(k_1,k_2,\cdots,k_n)=\left({2\over\pi}\right)^n e^{{i\pi\over2}(n+\sum_j\nu^{\ast}_j-{d(n-2)\over2})}\int_{-i\eta_1}^{i\infty}{d\tau}~\tau^{({n\over2}-1)d-1}\prod_{j=1}^n K_{\nu^{\ast}_j}(k_j\tau)
\ee 	
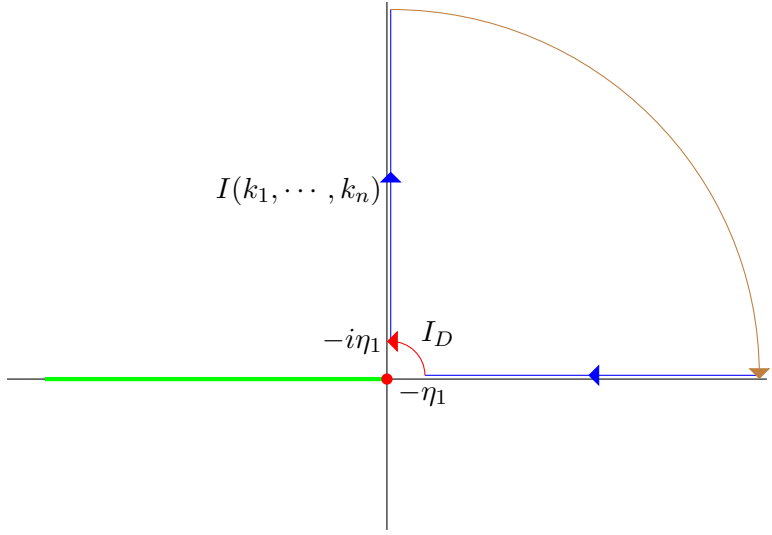
\begin{figure}[h]
	\centering
	\begin{tikzpicture}
		\draw (5,0)--(-5,0);
		\draw (0,5)--(0,-2);
		\node[text width=1.5cm] at (-0.1,0.5) {$-i\eta_1$};
		\node[text width=1.5cm] at (1.2,0.6) {$I_D$};
		\node[text width=1.5cm] at (-1.5,2.5) {$I(k_1,\cdots,k_n)$};
		\node[text width=1.5cm] at (.9,-0.2) {$-\eta_1$};
		\draw[blue] (4.9,0.05)--node{\midarrowL}(0.5,0.05);
		\draw[blue] (0.05,4.9)--node{\midarrowU}(0.05,0.5);
		\draw[brown] (4.9,0.05) node{\midarrowD} arc (0:90:4.85);
		\draw[red] (0.5,0.05) arc (0:90:0.45) node{\midarrowL};
		\draw[ultra thick, green] (0,0)--(-4.5,0);
		\node at (0,0) [circle,fill, red, inner sep=1.5pt]{};
	\end{tikzpicture}
	\caption{contour rotation for the integral ${I}(k_1,\cdots,k_n,\eta_1)$. Green line shows orientation of the branch cut on the negative real axis starting from origin. Red dot indicates pole at the origin.}
	\label{contour}
\end{figure}
The particular integral in the above equation can be represented in complex plane\footnote{$K_{q}(z)$ has pole at origin for $z\in\mathbb{C}$ and a branch cut starting from origin, see \cite{Wolfram}} by the vertical (parallel to imaginary axis) blue line in the contour drawn in Figure \ref{contour}. Since the asymptotic behaviour of the integrand is
\begin{equation}
	\label{math9}
	\tau^{({n\over2}-1)d-1} K_{\nu_1^{\ast}}(k_1 \tau)K_{\nu_2^{\ast}}(k_2 \tau)\cdots K_{\nu_n^{\ast}}(k_n \tau) \xrightarrow{\tau\rightarrow \infty}{e^{-(k_1+k_2+\cdots+k_n)\tau}\tau^{({n\over2}-1)d-({n\over2}+1)}\over\sqrt{k_1k_2\cdots k_n}}
\end{equation}
the arc (brown curve) contribution at large radius vanishes. Using \eq{Kboun}, the anti-clockwise integral contribution from the small arc (red curve) of radius $-\eta_1$ is given by
\be
\label{smallarc}
I_D=i \sum\left[(-\eta_1)^{\zeta_{m,p}}{e^{i\pi\zeta_{m,p}\over2}-1\over\zeta_{m,p}}\prod_{\{m\}}(a_{\nu_m^{\ast}}k_m^{\nu^{\ast}_m})\prod_{\{p\}}(b_{\nu_p^{\ast}}k_p^{-\nu^{\ast}_p})\right] \quad {\{m\}}\cup{\{p\}}={\{1,2,\cdots,n\}}
\ee
where $a_\nu,b_\nu$ are given in \eq{coefads}, $\sum$ denotes sum of all possible combinations of $\{m\},\{p\}$. The product $\prod_{\{m\}}k_m$, for say $\{m\}={\{1,3,4\}}$ is equal to $k_1k_3k_4$ and $\zeta_{m,p}$ is defined as   
\be
\zeta_{m,p}=\left({n\over2}-1\right)d+(\sum_{\{m\}}\nu^{\ast}_m-\sum_{\{p\}}\nu^{\ast}_p) \label{etap}
\ee 
Hence $I(k_1,\cdots,k_n)$ is equal to the real line integral (from $-\eta_1$ to $\infty$) in addition to the small arc contribution as
\be
\label{IexpK3}
I(k_1,k_2,\cdots,k_n)=\left({2\over\pi}\right)^n e^{{i\pi\over2}(n+\sum_j\nu^{\ast}_j-{d(n-2)\over2})}\left[\int_{-\eta_1}^{\infty}{d\tau}~\tau^{({n\over2}-1)d-1}\prod_{j=1}^n K_{\nu^{\ast}_j}(k_j\tau)-I_D\right]
\ee
Note that convergence of $I_D$ depends on $\nu_i$. Specifically on the power of $\eta_1$ which is given in \eq{etap}. For example for all fields being underdamped, $I_D$ goes to zero as $\eta_1\rightarrow 0$ since all $\nu$ are purely imaginary and as a result $\zeta_{m,p}$ is a positive real value. Similarly when all fields are overdamped, depending on the sign of $\zeta_{m,p}$, it will either converge or diverge. Note also that the $\tau$ integral on the RHS of \eq{IexpK3} can also diverge, depending on $\nu$, see the related discussion in section (\ref{divergence}). 

For $n=3$ the finite part of \eq{IexpK3}  can be obtained as follows: we  keep the first term within the bracket on the RHS of eq.(\ref{IexpK3}), and take the lower limit to go to $\eta_1\rightarrow 0$, this gives, eq.(\ref{IexpK}). We then evaluate the integral by analytic continuation starting from values of $\nu$ where it converges.

%

\subsection{Fourier Transform of Three Point Correlation Function}
\label{threeancnt}
In this subsection we give a detailed derivation of the position space form of three point function by Fourier transforming the momentum space form given in \eq{wittenmom3} for $n=3$. Without being too much concerned for the divergent contributions of the integral in \eq{IexpK3}, three point correlation function of the dual field $O_i({\bf k})$ in momentum space is given as follows
\begin{equation}
	\label{appeq.adso3k}
	\l O_1({\bf k}_1)O_2({\bf k}_2)O_3({\bf k}_3)\r'=A(\nu_1,\nu_2,\nu_3)k_1^{\nu_1}k_2^{\nu_2}k_3^{\nu_3}\int_{0}^{\infty}dz~z^{{d\over2}-1}K_{\nu^{\ast}_1}(k_1z)K_{\nu^{\ast}_2}(k_2z)K_{\nu^{\ast}_3}(k_3z)
\end{equation} 
where we suppress $(2\pi)^d\delta({\bf k}_1+{\bf k}_2+{\bf k}_3)$ by introducing "prime", see section \ref{holoint} for explanation. $\nu$ is used for both over-damped and under-damped cases and,
\be
\label{A}
A(\nu_1,\nu_2,\nu_3)=C_3[\{a\};\{b\}] \left({2\over\pi}\right)^3 e^{{i\pi\over2}(3+\nu^{\ast}_1+\nu^{\ast}_2+\nu^{\ast}_3-{d\over2})}
\ee
Note that in the underdamped case $\nu=i \mu$. Thus we have $K_{-i \mu} (k z)$ as one of the factors inside the integral. However since $K_{-q}  (k z) = K_q (k z)$, we can write it as $K_{i \mu} (k z)$, or equivalently $K_{\nu} (k z)$. In the overdamped case $\nu$'s are real. Hence from now on we write \eq{appeq.adso3k} as follows.
\begin{equation}
	\label{appeq.adso3k2}
	\l O_1({\bf k}_1)O_2({\bf k}_2)O_3({\bf k}_3)\r'=A(\nu_1,\nu_2,\nu_3)k_1^{\nu_1}k_2^{\nu_2}k_3^{\nu_3}\int_{0}^{\infty}dz~z^{{d\over2}-1}K_{\nu_1}(k_1z)K_{\nu_2}(k_2z)K_{\nu_3}(k_3z)
\end{equation}
Fourier transforming \eq{appeq.adso3k2} we get,
\begin{equation}
	\label{appD1.adso3x}
	\l O_1({\bf x}_1)O_2({\bf x}_2)O_3({\bf x}_3)\r=\int D^d_3{\bf k}~(2\pi)^d\delta({\bf k}_1+{\bf k}_2+{\bf k}_3)e^{i{\bf k}_j.{\bf x}_j}\l O_1({\bf k}_1)O_2({\bf k}_2)O_3({\bf k}_3)\r'
\end{equation}
where,
\begin{equation}
	D^d_n{\bf k}={1\over(2\pi)^{nd}}\prod_{i=1}^{n}d^d{\bf k}_i
\end{equation}
Using \eq{appeq.adso3k}, the position space representation of three point function comes out as 
\begin{empheq}{multline}
	\l O_1({\bf x}_1)O_2({\bf x}_2)O_3({\bf x}_3)\r=A(\nu_1, \nu_2,\nu_3){\Gamma[\Delta_1]\Gamma[\Delta_2]\Gamma[\Delta_3]\over 2^{3-\nu_1-\nu_2-\nu_3}\pi^{3d\over2}}
	\times\\
	\int_{0}^{\infty}{dz\over z^{d+1}}\int d^d{\bf x}~\prod_{i=1}^{3}\left[\left({z\over z^2+|{\bf x}-{\bf x}_i|^2}\right)^{{d\over2}+\nu_i}\right] \label{appD2.adso3x}
\end{empheq}
where we used the identity
\begin{equation}
	\label{fourierID}
	\int {d^d{\bf k}\over(2\pi)^d}e^{i{\bf k}.{\bf r}}k^\nu K_\nu(kz)={\Gamma\left[{d\over2}+\nu\right]\over2^{1-\nu}\pi^{d\over2}}z^{-{d\over2}}\left({z\over z^2+r^2}\right)^{{d\over2}+\nu}
\end{equation}
We give here a brief derivation of the above identity. One can start from RHS with $z\in\real$ as follows
\begin{align}
	&\int d^dr e^{-i{\bf k}.{\bf r}}\left(1\over z^2+r^2\right)^{\Delta}\\
	= & \int d^dr e^{-i{\bf k}.{\bf r}}{1\over\Gamma[\Delta]}\int_{0}^{\infty}ds~s^{\Delta-1}e^{-s[z^2+r^2]} \quad ({\rm using~definition~of~}\Gamma{\rm-function})\\
	=& {1\over\Gamma[\Delta]}\int_{0}^{\infty}ds~s^{\Delta-1}e^{-sz^2}\left({\pi\over s}\right)^{d\over2}e^{-{k^2\over 4s}} \quad ({\rm solving~}r{\rm-integral})\\
	=& {\pi^{d\over2}\over\Gamma[{d\over2}+\nu]}2^{1-\nu}z^{-\nu}k^\nu K_\nu(kz) \quad \left({\rm using~}\Delta={{d\over2}+\nu}\right)
\end{align}
 Rest of the derivation is analogous to \cite{freedU1}. However we include the analysis for completion. Without loss of generality, we choose ${\bf x}_3=0$ by translational invariance and change the variables as (small Latin indices refers to components while capital Latin is for the fixed positions at boundary)
\begin{equation}
	\label{vc}
	z={Z\over Z^2+X^2}\hspace{30pt} x^i={X^i\over Z^2+X^2}\hspace{30pt} x^i_A={X^i_A\over (X_A)^2}
\end{equation}
where $i=1,2,...,d$ and $A=1,2,3$. These imply the following relations \footnote{ Note that from first relation in eq.(\ref{implyvc}), ${\bf x}_3=0$ implies ${\bf X}_3\rightarrow\infty$. }
\begin{equation}
	\label{implyvc}
	{x_A^i\over (x_A)^2}=X_A^i\hspace{30pt} {z\over z^2+|{\bf x}|^2}=Z\hspace{30pt } {z\over z^2+|{\bf x}-{\bf x}_A|^2}={Z(X_A)^2\over Z^2+|{\bf X}-{\bf X}_A|^2}~(A=1,2)
\end{equation}
and the invariance of the measure
\begin{equation}
	\label{invarvol}
	{dzd^d{\bf x}\over z^{d+1}}={dZd^d{\bf X}\over Z^{d+1}}
\end{equation}
Hence the integral in eq.(\ref{appD2.adso3x}) becomes
\begin{multline}
	\label{integral1}
	\int_{0}^{\infty}{dz\over z^{d+1}}\int d^d{\bf x}~\prod_{A=1}^{3}\left[\left({z\over z^2+|{\bf x}-{\bf x}_A|^2}\right)^{{d\over2}+\nu_A}\right]\\={1\over x_1^{2\Delta_1}x_2^{2\Delta_2}}\int_{0}^{\infty}{dZ}\int d^d{\bf X}\left[{Z^{\Delta_1+\Delta_2+\Delta_3-d-1}\over \left(Z^2+|{\bf X}-{\bf X}_1|^2\right)^{\Delta_1}\left(Z^2+|{\bf X}-{\bf X}_2|^2\right)^{\Delta_2}}\right]
\end{multline}
The result of the integral in eq.(\ref{integral1}) is given in \cite{freedU1}, eq.(22). Using that, we get
\begin{multline}
	\label{AppD3.adso3x}
	\l O_1({\bf x}_1)O_2({\bf x}_2)O_3({\bf x}_3)\r=A(\nu_1, \nu_2,\nu_3){\Gamma[\Delta_1]\Gamma[\Delta_2]\Gamma[\Delta_3]\over 2^{3-\nu_1-\nu_2-\nu_3}\pi^{3d\over2}}{\cal I}(\Delta_1,\Delta_2,\Delta_3)\\
	\times{1\over |{\bf x}_1-{\bf x}_2|^{\Delta_1+\Delta_2-\Delta_3}|{\bf x}_2-{\bf x}_3|^{\Delta_2+\Delta_3-\Delta_1}|{\bf x}_3-{\bf x}_1|^{\Delta_3+\Delta_1-\Delta_2}}
\end{multline}
where we have transformed the result to old coordinates by using eq.(\ref{implyvc}) as well as restored ${\bf x}_3$ by translational invariance. ${\cal I}(\{\Delta_i\})$ is given by
\begin{equation}
	\label{I}
	{\cal I}(\Delta_1,\Delta_2,\Delta_3)={\pi^{d\over2}\over2}{\Gamma\left[{\Delta_1+\Delta_2+\Delta_3-d\over2}\right]\Gamma\left[{\Delta_1+\Delta_2-\Delta_3\over2}\right]\Gamma\left[{\Delta_1-\Delta_2+\Delta_3\over2}\right]\Gamma\left[{-\Delta_1+\Delta_2+\Delta_3\over2}\right]\over\Gamma[\Delta_1]\Gamma[\Delta_2]\Gamma[\Delta_3]}
\end{equation}
The above formula is valid in both under-damped and over-damped case where $\Delta_i$'s can be either $\Delta_i= \frac{d}{2} +\nu_i$ in over-damped case or $\Delta_i= \frac{d}{2} + i\mu_i$ in under-damped case.
\subsection{OPE limit of Three Point Function}	
\label{OPEint}

Consider the three point function obtained in \eq{AppD3.adso3x}.  This three point function should reduce to an effective two point function once two points are brought closer compared to the third. In mathematical terms one can write,
\begin{equation}
	\label{opex}
	\mathbf{x}_1\to \mathbf{x}_2\quad {\rm and} \quad |\mathbf{x}_1-\mathbf{x}_2|\ll |\mathbf{x}_1-\mathbf{x}_3|
\end{equation}
In momentum space the limits \eq{opex} correspond to taking 
\begin{align}
	\label{klimits1}
	k_1\sim k_2\gg k_3
\end{align}
In this subsection we show through an explicit calculation that the above expectation is indeed borne out for the three point function obtained in \eq{AppD3.adso3x}. We do this by analysing \eq{AppD3.adso3x} directly in the limits \eq{opex} and then through the manipulation of \eq{wittenmom3} in the limits \eq{klimits1}.\\
Consider then \eq{AppD3.adso3x}. In the limit \eq{opex}, \eq{AppD3.adso3x} reduces to
\begin{multline}
	\label{limopeo3}	
	\l O_1({\bf x}_1)O_2({\bf x}_2)O_3({\bf x}_3)\r \simeq A(\nu_1, \nu_2,\nu_3){\Gamma[\Delta_1]\Gamma[\Delta_2]\Gamma[\Delta_3]\over 2^{3-\nu_1-\nu_2-\nu_3}\pi^{3d\over2}}{\cal I}(\Delta_1,\Delta_2,\Delta_3)\\
	\times{1\over|{\bf x}_1-{\bf x}_2|^{\Delta_1+\Delta_2-\Delta_3}|{\bf x}_2-{\bf x}_3|^{2\Delta_3}}
\end{multline}
where $A(\nu_1,\nu_2,\nu_3)$ and ${\cal I}(\Delta_1,\Delta_2,\Delta_3)$ are given in \eq{A} and \eq{I} respectively. Identifying the two point function in \eq{posspa} for overdamped and \eq{rho2pt} for underdamped cases, we can write \eq{limopeo3} as 
\be
\label{scform1}
\lim_{\substack{{\bf x}_1\rightarrow{\bf x}_2 \\ |{\bf x}_1-{\bf x}_3|\gg |{\bf x}_1-{\bf x}_2|}}\l O_1({\bf x}_1)O_2({\bf x}_2)O_3({\bf x}_3)\r = {{\cal C}_{123}\over|{\bf x}_1-{\bf x}_2|^{\Delta_1+\Delta_2-\Delta_3}}\l O_3({\bf x}_2)O_3({\bf x}_3)\r
\ee 
where the structure constant ${\cal C}_{123}$ is given for $\phi_3$ being overdamped by
\be
\label{C123ov3}
{\cal C}_{123}=A(\nu_1, \nu_2,\nu_3){\Gamma[\Delta_1]\Gamma[\Delta_2]\Gamma[\nu_3]\over (2\nu_3)\times2^{3-\nu_1-\nu_2-\nu_3}\pi^{d}}{\cal I}(\Delta_1,\Delta_2,\Delta_3) e^{-{i\pi(2\nu_3-1)\over2}}
\ee
and for underdamped case by
\be
\label{C123un3}
{\cal C}_{123}=A(\nu_1, \nu_2,i\mu_3){\Gamma[\Delta_1]\Gamma[\Delta_2]\Gamma[i\mu_3]\over 2^{3-\nu_1-\nu_2-i\mu_3}\pi^{d}}{\cal I}(\Delta_1,\Delta_2,\Delta_3)\times e^{\pi\mu_3}
\ee
as discussed in section \ref{discuss}.

Now, let us understand how the singularity structure of the three point function arises in the OPE limit as in \eq{scform1} using Witten's diagram. We first consider \eq{wittenmom3}. In the limit \eq{klimits1}, we consider a region in the $\eta'$ integral of \eq{wittenmom3} where $\eta'$ takes values  so that the following conditions are met.
\begin{equation}
	\label{klimits2}
	-k_1\eta'\gg1 \quad  	-k_2\eta'\gg1 \quad 	-k_3\eta'\ll 1
\end{equation}
We will see below that the OPE limit of the three point function will arise from this region.
\begin{figure}[h]
	\centering
	\begin{tikzpicture}
		\draw (0,0)--(8,0);
		\draw (3,-1.5)--(7,0);
		\draw (3,-1.5)--(2.2,0);
		\draw (3,-1.5)--(2,0);
		\node[text width=3cm] at (2.6,-0.5) {$\mathbf{k}_1,\Delta_1$};
		\node[text width=3cm] at (4.2,-0.5) {$\mathbf{k}_2,\Delta_2$};
		\node[text width=3cm] at (7.8,-0.5) {$\mathbf{k}_3,\Delta_3$};
		\node[text width=3cm] at (9.9,0) {$\eta'=0$};
		\node[text width=3cm] at (4,-2) {$(\mathbf{x}',\eta')$};
		\node[text width=3cm] at (3.2,0.3) {$\mathbf{x}_1$};
		\node[text width=3cm] at (3.8,0.3) {$\mathbf{x}_2$};
		\node[text width=3cm] at (8.3,0.3) {$\mathbf{x}_3$};
	\end{tikzpicture}
	\caption{Witten diagram for three point correlator in OPE limit.}
	\label{Witten3mom}
\end{figure}
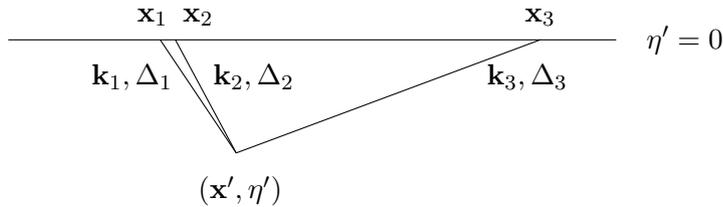
The limits can be understood diagrammatically in Figure \ref{Witten3mom} comparing with Figure \ref{Witten3} as the points ${\bf x}_1,{\bf x}_2$ come closer to each other. The dominant contribution in the bulk integral thereby comes from the bulk vertex $({\bf x}',\eta')$ close to the boundary following the condition \eq{klimits1}.\\   
Under \eq{klimits2} being met, we insert the leading behaviour of the Hankel functions of order $\nu_1,\nu_2$ at large argument as given by \eq{horH} while that of $\nu_3$ at small argument as in \eq{bounH1}. This reduces \eq{wittenmom3} to the following form
\begin{multline}
	\label{wittenmomlim}
	\l O({\bf k}_1)O({\bf k}_2)O({\bf k}_3)\r	\propto (2\pi)^d\delta^{(d)}(\mathbf{k}_1+\mathbf{k}_2+\mathbf{k}_3)\left[\tilde{\alpha}(\nu_3^{\ast}) k_1^{\nu_1+\nu_2-1} k_3^{\nu_3+\nu_3^{\ast}}\right.\\
	\times \left.\int_{-\frac{1}{k_1}}^{-\frac{1}{k_3}} d\eta' ~ (-\eta')^{\frac{d}{2}-2+\nu_3^{\ast}}e^{2 i k_2\eta'}
	+\tilde{\beta}(\nu_3^{\ast}) k_1^{\nu_1+\nu_2-1}k_3^{\nu_3-\nu_3^{\ast}} \int_{-\frac{1}{k_1}}^{-\frac{1}{k_3}} d\eta' ~ (-\eta')^{\frac{d}{2}-2-\nu_3^{\ast}}e^{2i k_2\eta'}\right]
\end{multline}
where $\tilde{\alpha}(\nu_3^{\ast})$ and $\tilde{\beta}(\nu_3^{\ast})$ are the expansion coefficients of $H^{(2)}_{\nu_3^{\ast}}(-k_3\eta')$. Now, apart from the non-local structure, it is important to show that the above  integrals over $\eta'$ are finite.
To show this, we substitute $-2 k_1\eta'=2-i z$ in both of the integrals of \eq{wittenmomlim} which yields the range of $z$ as $z\in[0,-2i(1-{k_1\over k_3})]\xrightarrow{OPE~ limit}[0,i\infty)$. So, we write
\begin{multline}
	\label{wittenmomlim2}
	\l O({\bf k}_1)O({\bf k}_2)O({\bf k}_3)\r	\propto(2\pi)^d\delta^{(d)}(\mathbf{k}_1+\mathbf{k}_2+\mathbf{k}_3)\\
	\times \left[\tilde{\alpha}(\nu_3^{\ast}) k_1^{\nu_1+\nu_2-1} k_3^{\nu_3+\nu_3^{\ast}}{i\over(2k_1)^{{d\over2}-1+\nu_3^{\ast}}}\int_{0}^{i\infty} dz ~ (2-iz)^{\frac{d}{2}-2+\nu_3^{\ast}}e^{-i(2-iz)}\right.\hspace{80pt} \\
	\left.+\tilde{\beta}(\nu_3^{\ast}) k_1^{\nu_1+\nu_2-1}k_3^{\nu_3-\nu_3^{\ast}} {i\over(2k_1)^{{d\over2}-1-\nu_3^{\ast}}}\int_{0}^{i\infty} dz ~ (2-iz)^{\frac{d}{2}-2-\nu_3^{\ast}}e^{-i(2-iz)}\right]
\end{multline}
Now, each of the above integrals over z has a branch cut originating from z=0, and a pole on the negative imaginary axis. Hence, we align the branch cut away from the first quadrant and perform a contour rotation from the positive imaginary axis to positive real axis by applying Cauchy's theorem. Since the arc contributions close to z=0 and $|z|=R$ (large) vanishes, this leads to the following result
\begin{multline}
	\label{OPEmom}
	\lim_{\substack{k_1\gg k_3\\k_2\gg k_3}}\langle {O}(\mathbf{k}_1){O}(\mathbf{k}_2){O}(\mathbf{k}_3)\rangle\propto-(2\pi)^d\delta^{(d)}(\mathbf{k}_1+\mathbf{k}_2+\mathbf{k}_3)\\
	\times\left[\left(\tilde{\alpha}(\nu_3^{\ast})\frac{\Gamma[d/2+\nu_3^{\ast}-1,2i]}{(2i)^{d/2+\nu_3^{\ast}-1}}\right)(k_1)^{-\frac{d}{2}+\nu_1+\nu_2-\nu_3^{\ast}} (k_3)^{\nu_3+\nu_3^{\ast}}\right.\hspace{135pt}\\
	\left.+\left(\tilde{\beta}(\nu_3^{\ast})\frac{\Gamma[d/2-\nu_3^{\ast}-1,2i]}{(2i)^{d/2-\nu_3^{\ast}-1}}\right)(k_1)^{-\frac{d}{2}+\nu_1+\nu_2+\nu_3^{\ast}}(k_3)^{\nu_3-\nu_3^{\ast}} \right]
\end{multline}
where in the above eq.(\ref{OPEmom}), $\Gamma[\alpha,2i]$ is the incomplete Gamma function defined as
\begin{equation}
	\Gamma[\alpha,2i]=i^{\alpha-1}\int_{0}^{\infty}dz ~e^{-i(2-iz)}(2-iz)^{\alpha-1}
\end{equation}
For $\phi_3$ being overdamped field, $\nu^{\ast}_3=\nu_3$. Hence, in the 2nd term in the square bracket of \eq{OPEmom} containing $k_3^{\nu_3-\nu_3^{\ast}}$ becomes $k_3$ independent. Now, using the following Fourier transform results
\begin{equation}
	\label{FTterm1}
	\int \prod_{j=1}^3 \frac{d^d\mathbf{k}_j}{(2\pi)^d}e^{i\mathbf{k}_a\cdot\mathbf{x}_a}(2\pi)^d\delta^{(d)}(\Sigma_{b=1}^3 \mathbf{k}_b)k_1^{\alpha}k_3^{\beta}=\frac{\mathtt{I}[d,\alpha]}{|\mathbf{x}_2-\mathbf{x}_1|^{d+\alpha}}\frac{\mathtt{I}[d,\beta]}{|\mathbf{x}_3-\mathbf{x}_2|^{d+\beta}}
\end{equation}
\begin{equation}
	\label{FTterm2}
	\int \prod_{j=1}^3 \frac{d^d\mathbf{k}_j}{(2\pi)^d}e^{i\mathbf{k}_a\cdot\mathbf{x}_a}(2\pi)^d\delta^{(d)}(\Sigma_{b=1}^3 \mathbf{k}_b)k_1^{\alpha}=\frac{\mathtt{I}[d,\alpha]}{|\mathbf{x}_3-\mathbf{x}_1 |^{d+\alpha}}\delta^{(d)}(\mathbf{x}_3-\mathbf{x}_2)
\end{equation} 
where $\mathtt{I}[d,\alpha]$ is some finite coefficient, we schematically obtain the following 
\begin{multline}
	\label{OPEopos}
	\lim_{\substack{{\bf x}_1\rightarrow{\bf x}_2 \\ |{\bf x}_1-{\bf x}_3|\gg |{\bf x}_1-{\bf x}_2|}}\langle O_1(\mathbf{x}_1)O_2(\mathbf{x}_2)O_3(\mathbf{x}_3)\rangle\simeq
	\frac{C(\nu_1,\nu_2,\nu_3)}{|\mathbf{x}_2-\mathbf{x}_1|^{\nu_1+\nu_2-\nu_3+{d\over2}}|\mathbf{x}_3-\mathbf{x}_2|^{d+2\nu_3}}
\end{multline}
where $C(\nu_1,\nu_2,\nu_3)$ is some finite coefficient.  The second term in the square bracket of \eq{OPEmom} does not contribute to the above equation since its' position space expression is related to \eq{FTterm2} which vanishes because of ${\bf x}_3\neq{\bf x}_2$.
Now, using the expression of the two point function as in \eq{posspa} and defining $\Delta_a={d\over2}+\nu_a$,  we can reduce \eq{OPEopos} to the singularity structure
same as of \eq{scform1}. This gives us the complete picture why this structure is arising.\\
Similarly, for $\phi_3$ being underdamped, $\nu^{\ast}_3=-\nu_3=-i\mu_3$, the 1st term in the square bracket of \eq{OPEmom} containing $k_3^{\nu_3+\nu_3^{\ast}}$ becomes $k_3$-independent. Using \eq{FTterm1}, \eq{FTterm2}, we get
\begin{multline}
	\label{OPEupos}
	\lim_{\substack{{\bf x}_1\rightarrow{\bf x}_2 \\ |{\bf x}_1-{\bf x}_3|\gg |{\bf x}_1-{\bf x}_2|}}\langle O_1(\mathbf{x}_1)O_2(\mathbf{x}_2)O_3(\mathbf{x}_3)\rangle\simeq
	\frac{D(\nu_1,\nu_2,\mu_3)}{|\mathbf{x}_2-\mathbf{x}_1|^{\nu_1+\nu_2-i\mu_3+{d\over2}}|\mathbf{x}_3-\mathbf{x}_1|^{d+2i\mu_3}}
\end{multline}
with $D(\nu_1,\nu_2,\mu_3)$ is some finite coefficient which is also of the same singularity structure as given in \eq{scform1}.

\subsection{Ratio of Three to Two Point Function} 
\label{3by2}
We conclude this Appendix with two comments. Firstly, the ratio of three to two point function in coherent state basis differs from that in field eigenstate basis. To see this note that in the momentum space, three point function of the dual field $O_+({\bf k})$ sourced by $\rho^{\ast}({\bf k})$ can be read off from \eq{intwvfchi} as
\be
\label{O3coh}
\l O_+({\bf k}_1)O_+({\bf k}_2)O_+({\bf k}_3)\r={3!i\lambda\over (\sqrt{2\mu}\alpha_\mu^{\ast})^3}(2\pi)^d\delta({\bf k}_1+{\bf k}_2+{\bf k}_3)k_1^{i\mu}k_2^{i\mu}k_3^{i\mu}I(k_1,k_2,k_3)
\ee
where $I(k_1,k_2,k_3)$ is given in \eq{Iexp}. Two point function of $O_+({\bf k})$ can be read off from \eq{psirho1} as
\be
\label{O2coh}
\l O_+({\bf k})O_+(-{\bf k})\r={\beta_\mu^{\ast}\over \alpha_\mu^{\ast}}k^{2i\mu}
\ee 
Hence the ratio of \eq{O3coh} to \eq{O2coh} in coherent state basis is given by
\be
\label{O3/O2coh}
{\l O_+({\bf k}_1)O_+({\bf k}_2)O_+({\bf k}_3)\r \over \l O_+({\bf k})O_+(-{\bf k})\r}=(2\pi)^d\delta({\bf k}_1+{\bf k}_2+{\bf k}_3){3!i\lambda\over (\sqrt{2\mu})^3\alpha_\mu^{*2}\beta_\mu^{\ast}}k_1^{i\mu}k_2^{i\mu}k_3^{i\mu}k^{-2i\mu}I(k_1,k_2,k_3)
\ee
Similarly in field eigenstate basis, the ratio of three to two point function of the dual field ${\hat O}_+({\bf k})$ sourced by $J_+({\bf k})$ can be derived from \eq{formco} and \eq{<O+O+>} as
\be
\label{O3/O2}
{\l {\hat O}_+({\bf k}_1){\hat O}_+({\bf k}_2){\hat O}_+({\bf k}_3)\r \over \l {\hat O}_+({\bf k}){\hat O}_+(-{\bf k})\r}=(2\pi)^d\delta({\bf k}_1+{\bf k}_2+{\bf k}_3)\left(-{3!\lambda \over d\alpha_\mu^{\ast}\beta_\mu^{\ast}}\right)k_1^{i\mu}k_2^{i\mu}k_3^{i\mu}k^{-2i\mu}I(k_1,k_2,k_3)
\ee
Clearly the results \eq{O3/O2coh} and \eq{O3/O2} differs by the factor ${-id\over\alpha_\mu^{\ast}({2\mu})^{3\over2}}$ as mentioned in section \ref{addpt}.

Finally one could ask what happens if the scalar field in anti-de Sitter (AdS) space has a different normalization than that of \cite{freedU1}. To answer this let us consider the scalar field in AdS with two different normalizations. Following \cite{freedU1} we have,
\begin{equation}
	\phi(\textbf{k},z) =\frac{z^{\frac{d}{2}} K_{\nu} (k z) }{\epsilon^{\frac{d}{2}} K_{\nu} (k \epsilon) }  \epsilon^{\frac{d}{2}-\nu} \hat{\phi} (\textbf{k}) \label{phisol1}
\end{equation}
This has the property that,
\begin{equation}
	\lim_{z \rightarrow \epsilon} \phi({\bf k},z) = \epsilon^{\frac{d}{2}-\nu} \hat{\phi} ({\bf k}) \label{rel1}
\end{equation}
Thus the relationship between the bulk field and boundary source is local. Given \eq{phisol1} one could compute the two point function as well as three point function for a $\phi^3$ interaction of strength $\lambda$ and get,
\begin{equation}
	\langle O(\textbf{k}) O(-\textbf{k})\rangle = 2\nu \frac{a_\nu}{b_\nu} k^{2 \nu} \label{2pt1}
\end{equation}
and,
\begin{equation}
	\langle O(\textbf{k}_1)O(\textbf{k}_2)O(\textbf{k}_3)\rangle = -\frac{6\lambda}{b_\nu^3} k_1^\nu k_2^\nu k_3^\nu \int_{0}^{\infty}dz~z^{{d\over2}-1}K_{\nu_1}(k_1z)K_{\nu_2}(k_2z)K_{\nu_3}(k_3z) \label{3pt1}
\end{equation}
where $a_\nu,b_\nu$ are given in \eq{coefads}.
Let us consider a different normalization for the source instead of \eq{phisol1}. 
\begin{equation}
	\phi(\textbf{k},z) = z^{\frac{d}{2}} K_{\nu} ({k} z) k^{\nu} J({\bf k}) \label{phisol2}
\end{equation}
This has the property that,
\begin{equation}
	\lim_{z \rightarrow \epsilon} \phi(\textbf{k},z)= b_\nu \epsilon^{\frac{d}{2}-\nu} J(\textbf{k}) + a_\nu \epsilon^{\frac{d}{2}+\nu} k^{2 \nu} J(\textbf{k}) \label{rel2}
\end{equation}
Thus in this case the relationship between the bulk field and boundary source is non-local unlike \eq{rel1}. Then for \eq{phisol2} the two-point function turns out to be,
\begin{equation}
	\langle \hat{O}(\textbf{k}) \hat{O}(-\textbf{k})\rangle = d a_{\nu} b_{\nu} k^{2 \nu} \label{2pt2}
\end{equation}
while the three-point function reads,
\begin{equation}
	\langle \hat{O}(\textbf{k}_1) \hat{O}(\textbf{k}_2) \hat{O}(\textbf{k}_3)\rangle = -6 \lambda k_1^\nu k_2^\nu k_3^\nu \int_{0}^{\infty}dz~z^{{d\over2}-1}K_{\nu_1}(k_1z)K_{\nu_2}(k_2z)K_{\nu_3}(k_3z) \label{3pt2}
\end{equation}
Comparing \eq{2pt1} with \eq{2pt2} and \eq{3pt1} with \eq{3pt2}, we see that the two-point and three-point coefficients for different normalization are different. Since we are allowed to choose our sources anyhow we want let us choose a different source $\hat{J}$ which is related to $J$ as follows.
\begin{equation}
	\hat{J} (\textbf{k}) = b_\nu J(\textbf{k})
\end{equation}
Then the two-point function \eq{2pt2} becomes,
\begin{equation}
	\langle \tilde{O}(\textbf{k}) \tilde{
	O}(-\textbf{k})\rangle =d \frac{a_\nu}{b_\nu} k^{2 \nu}
\end{equation}
This does not match with \eq{2pt1}. However eq,\eq{3pt2} becomes,
\begin{equation}
	\langle \tilde{O}(\textbf{k}_1) \tilde{O}(\textbf{k}_2) \tilde{O}(\textbf{k}_3)\rangle = -6 \lambda \frac{1}{b_\nu^3} k_1^\nu k_2^\nu k_3^\nu \int_{0}^{\infty}dz~z^{{d\over2}-1}K_{\nu_1}(k_1z)K_{\nu_2}(k_2z)K_{\nu_3}(k_3z)
\end{equation}
which is same as \eq{3pt1}. 

So we see that different normalizations as in \eq{phisol1} and \eq{phisol2} give rise to varying ratios of 3-point to 2-point which any suitable rescaling of boundary source can not remedy. Furthermore, from supersymmetry constraints on chiral operators we know that the ratio of two point to three point coefficient ratio is protected \cite{ShirazChiral}. Thus a different normalization as in \eq{phisol2} gives unphysical result.
\section{Transformations under Diffeomorphisms}
\label{conft}

The $SO(1,d+1$) isometry generators of $d+2$ dimensional Minkowski spacetime (signature:$(-++...+)$) with coordinates $X^A=X^0,X^1,X^2,...,X^{d+1}$ defined as
\begin{align}
	\label{Lab}
	L_{A,B}&=i\left[X_A\frac{\partial}{\partial X^B}-X_B\frac{\partial}{\partial X^A}\right]
\end{align}
satisfy the commutation relation
\begin{equation}
	\label{[Lab,Lcd]}
	[L_{AB},L_{CD}]=i\left(g_{BC}L_{AD}+g_{AD}L_{BC}-g_{BD}L_{AC}-g_{AC}L_{BD}\right)
\end{equation}
where $g_{AB}$ is the metric of $\mathbb{R}^{1,d+1}$.  
Now, $dS_{d+1}$ spacetime is an hyperboloid embedded in $\mathbb{R}^{1,d+1}$ with the constraint \eq{embhyp}. 
This $dS_{d+1}$ spacetime in Poincaré coordinate ${\bf x'}=\{\eta,x^{j}\}$ ($j=1,..,d$) can be parametrized by
\begin{align}
	\label{Xtrans}
	X^0 =\frac{R_{\rm dS}}{2}\left(\eta-\frac{1}{\eta}\right)-\frac{R_{\rm dS}}{2\eta}(x_jx^j) \hspace{20pt}
	X^i =\frac{R_{\rm dS}}{\eta}x^i \hspace{20pt}
	X^{d+1} =\frac{R_{\rm dS}}{2}\left(\eta+\frac{1}{\eta}\right)-\frac{R_{\rm dS}}{2\eta}(x_jx^j)
\end{align}
General coordinate transformations are given by 
\begin{equation}
	\label{dif}
	x'^i= x^i+v^i(\mathbf{x},\eta) \hspace{50pt}
	\eta' = \eta(1+\epsilon(\mathbf{x},\eta))
\end{equation}
The generators of the isometries in the Poincaré coordinates take the form,
\begin{align}
	P_i&=i\frac{\partial}{\partial x^i}\label{Px}\\
	M_{ij}&=-i\left(x_i\frac{\partial}{\partial x^j}-x_j\frac{\partial}{\partial x^i}\right)\label{Mx}\\
	D&=i\left(\eta\frac{\partial}{\partial \eta}+x^i\frac{\partial}{\partial x^i}\right)\label{Dx}\\
	K_{i}&=i\left[2x_i\left(\eta\frac{\partial}{\partial \eta}+x^j\frac{\partial}{\partial x^j}\right)-(x_jx^j-\eta^2)\frac{\partial}{\partial x^i}\right]\label{Kx}
\end{align}
where 
 $P_{i},M_{ij},D,K_{i}$ are related to $L_{AB}$ in the following way
\be
\label{Lconfrel}
P_{i}=L_{0,i}+L_{d+1,i}~~~~~M_{ij}=-L_{ij}~~~~~D=L_{d+1,0}~~~~~K_{i}=L_{d+1,i}-L_{0,i}
\ee
These generators give rise to infinitesimal diffeomorphisms of the form eq.(\ref{dif}) with 
\begin{align}
	\label{dif1}
	{\rm Translation}~~&:~~v^i(\mathbf{x},\eta)=\varepsilon^i\quad \quad\hspace{115pt} \epsilon(\mathbf{x},\eta)=0\\
	\label{dif2}
	{\rm Rotation}~~&:~~v^i(\mathbf{x},\eta)=\varepsilon^{i}_{~j}x^j\quad \quad\hspace{100pt} \epsilon(\mathbf{x},\eta)=0\\
	\label{dif3}
	{\rm dilatation}~~&:~~v^i(\mathbf{x},\eta)=\varepsilon x^i\quad \quad\hspace{110pt} \epsilon(\mathbf{x},\eta)=\varepsilon\\
	\label{dif4}
	{\rm SCT}~~&:~~v^i(\mathbf{x},\eta)=[2x^i(b^jx_j)-(x^jx_j-\eta^2)b^i]\quad\quad \epsilon(\mathbf{x},\eta)=2(b^jx_j)
\end{align}

Now, it turns out that there are two kinds of coordinate transformations which keep the metric as given by \eq{metadm} in the ADM gauge satisfying $N^2=(-\eta)^{-2}$ and $N^i=0$. These transformations can be thought of as residual gauge transformation as given by
\begin{align}
	\label{xrep}
	{\rm Spatial~Reparametrization}~~&:~~v^i(\mathbf{x},\eta)=v^i(\mathbf{x})\quad \quad\hspace{74pt} \epsilon(\mathbf{x},\eta)=0\\
	\label{etarep}
	{\rm Time~Reparametrization}~~&:~~v^i(\mathbf{x},\eta)=\frac{1}{2}\eta^2\partial^i\epsilon(\mathbf{x})+{\mathcal{O}(\eta^4)} \quad \quad\hspace{10pt} \epsilon(\mathbf{x},\eta)=\epsilon(\mathbf{x})
\end{align}

\subsection{Transformation of Bulk Fields}
\label{conftF}
\paragraph{Scalar field} 
Under general coordinate transformation, \eq{dif}, bulk scalar field remains invariant
\begin{align}
	\label{invphi}
	\phi'({\bf x}',\eta')=\phi({\bf x},\eta)
\end{align}
which results functional change of the field as
\begin{equation}
	\label{delphi}
	\delta\phi({\bf x},\eta)\equiv\phi'({\bf x},\eta)-\phi({\bf x},\eta)=-v^i({\bf x},\eta)\partial_i\phi({\bf x},\eta)-\epsilon({\bf x},\eta) \eta \partial_{\eta}\phi({\bf x},\eta)
\end{equation}
We can find the transformation of the bulk field under conformal transformation by putting the expressions of $v^i({\bf x},\eta)$ and $\epsilon({\bf x},\eta)$ given by \eq{dif1} to \eq{dif4}. For example, in SCT
\begin{align}
	\delta_K\phi({\bf x},\eta)=-b^i\left[2x_i\left(\eta\partial_{\eta}+x^j\partial_j\right)-(x_jx^j-\eta^2)\partial_i\right]\phi({\bf x},\eta)\label{Kphi}
\end{align}
We can get the transformation in momentum space by Fourier transform. E.g., for SCT
\begin{align}
	\delta_K\phi({\bf k},\eta)=-ib^i\left[2\eta\partial_{\eta}\dpk{i}-2d\dpk{i}-2k_j\dpk{j}\dpk{i}+k_i\dpk{j}\dpk{j}+\eta^2k_i\right]\phi({\bf k},\eta)\label{Kphik}
\end{align}

\subsection{Transformation of Sources}
\label{conftS}
\paragraph{Scalar Sources}
The source in overdamped case, $\hat{\phi}({\bf k})$, is related to the bulk field $\phi({\bf k},\eta)$ in momentum space as given in eq.(\ref{oversource}). At $\eta\rightarrow0$, the relation can be written in position space as 
\begin{equation}
	\label{ovs0}
	\phi({\bf x},\eta)=(-\eta)^{{d\over2}-\nu}\hat{\phi}({\bf x})
\end{equation} 
Using the transformation property of $\phi({\bf x},\eta)$ given in eq.(\ref{delphi}), 
we get the transformation properties of the source in position space
\begin{equation}
	\label{delphihat}
	\delta\hat{\phi}({\bf x})=-v^i({\bf x},\eta)\partial_i\hat{\phi}({\bf x})-\left(d-\Delta\right)\epsilon({\bf x},\eta)\hat{\phi}({\bf x})
\end{equation}
 One can find the transformation of the source under conformal transformations by putting the expressions of $v^i({\bf x},\eta)$ and $\epsilon({\bf x},\eta)$ given by \eq{dif1}--\eq{dif4}.\\
The sources in underdamped case, $J_\pm({\bf k})$ is related to the bulk field $\phi({\bf k},\eta)$ in momentum space at late time as given in eq.(\ref{defjpb}).
The crucial difference here from the overdamped case is that $J_+({\bf x})$ has non-local relation with $\phi({\bf x},\eta)$ given by
\begin{equation}
	\label{undJp}
	\phi({\bf x},\eta)=\alpha_\mu^{\ast}(-\eta)^{{d\over2}-i\mu}J_+({\bf x})+\beta_\mu^{\ast}(-\eta)^{{d\over2}+i\mu}{2^{2i\mu}\Gamma[{d\over2}+i\mu]\over\pi^{d\over2}\Gamma[-i\mu]}\int{d^d{\bf y}}{J_+({\bf y})\over|{\bf x}-{\bf y}|^{d+2i\mu}}
\end{equation}
The first term in  the above \eq{undJp} is local. The second term having a non-local convolution can be thought of as a response to the first term, hence the transformation of $J_{+}$ is determined only by the first term and the second term changes accordingly as a response. Therefore, the resultant transformation of $J_{+}$ is the following
\begin{equation}
	\label{delJ+}
	\delta J_+({\bf x})=-v^i({\bf x},\eta)\partial_iJ_+({\bf x})-\left(d-\Delta_+\right)\epsilon({\bf x},\eta)J_+({\bf x})
\end{equation}

\paragraph{Metric Source :}
As mentioned in \eq{metadm}, in presence of a metric perturbation $\gamma_{ij}$, the perturbed dS metric in the ADM gauge with the condition ($N^2=(-\eta)^{-2},N^i=0$) is 
\begin{align}
	\label{syngaugemet}
	ds^2&=-\frac{d\eta^2}{(-\eta)^2}+\frac{1}{(-\eta)^2}\hat{g}_{ij}dx^idx^j\\
	\label{ggamma}
	\hat{g}_{ij}&=\delta_{ij}+\gamma_{ij}
\end{align}
Now, let us consider the general  coordinate transformation given by \eq{dif}. Since $\frac{1}{\eta^2}\hat{g}_{mn}$ appears as the metric component in the full metric of \eq{syngaugemet}, it should transform like a tensor under the coordinate transformation given by \eq{dif}; which leads to 
\begin{equation}
	\label{metrictransf}
	\frac{1}{\eta'^2}\hat{g}'_{ij}(\mathbf{x'})=\frac{1}{\eta^2}\hat{g}_{mn}(\mathbf{x})\frac{\partial x^m}{\partial x'^i}\frac{\partial x^n}{\partial x'^j}+g_{\eta\eta}(\partial_i\epsilon\partial_j\epsilon)
\end{equation}
From this, we can find the  change in $\hat{g}_{ij}$ under the transformation \eq{dif} to be 
\begin{align}
	\label{delggen}
	\delta \hat{g}_{ij}=\hat{g}'_{ij}(\mathbf{x},\eta)-\hat{g}_{ij}(\mathbf{x},\eta)=-\hat{g}_{in}\partial_{j}v^n-\hat{g}_{mj}\partial_{i}v^m-v^k\partial_k\hat{g}_{ij}+2\epsilon\hat{g}_{ij}
\end{align}
to the leading order in $v^i$ and $\epsilon$. The metric perturbation $\gamma_{ij}$ defined in \eq{ggamma} serves as source for the energy momentum tensor $T^{ij}$. Using \eq{ggamma} \eq{xrep} and \eq{etarep} in \eq{delggen}, we find
\begin{align}
	\label{spatg2}
	{\rm Spatial~Rep.~:}&~~~~ \delta\gamma_{ij}(\mathbf{x})=-(\partial_i v_j(\mathbf{x})+\partial_j v_i(\mathbf{x}))+{\rm homogeneous~part} \\
	\label{delgijtime}
	{\rm Time~Rep.~:} &~~~~ \delta\gamma_{ij}(\mathbf{x})=2\delta_{ij}\epsilon(\mathbf{x})+{\rm homogeneous~part}
\end{align}
where the ``homogeneous part'' is contributed by terms proportional to $\gamma_{ij}$ which we omitted since it is not of our interest for this discussion. It is important to note that in \eq{spatg2} and \eq{delgijtime}, we lower and upper indices by $\delta_{ij}$.
In a similar way, $\delta\gamma_{ij}$ for the conformal transformations can be found by putting the changes as given by \eq{dif1} to \eq{dif4} and \eq{ggamma} in \eq{delggen}. Such as for SCT, we get
\begin{align}
	\delta_{K}\gamma_{ij}(\mathbf{x})&=-\left[2\mathcal{M}^k_{~~i}\gamma_{kj}(\mathbf{x})+2\mathcal{M}^k_{~~j}\gamma_{ki}(\mathbf{x})-\left\{\mathbf{x}^2(\mathbf{b}\cdot\partial)-2(\mathbf{b}\cdot\mathbf{x})(\mathbf{x}\cdot\partial)\right\}\gamma_{ij}(\mathbf{x})\right]\\
	\label{SCTMij}
	\mathcal{M}^i_j&=x^ib_j-b^ix_j
\end{align}
Notice that $\delta_{K}\gamma_{ij}$ in the above equation is entirely given by the homogeneous terms, the inhomogeneous contribution (terms independent of $\gamma_{ij}$) vanishes. This turns out to be true for all conformal transformations since they are a special combination of spatial and time reparametrization satisfying the relation 
\begin{equation}
	\label{confcond}
	{\rm Conformal~Transformations~:}~~~~\partial_i v_j(\mathbf{x})+\partial_j v_i(\mathbf{x})=2\delta_{ij}\epsilon(\mathbf{x})
\end{equation}
where $v^i(\mathbf{x})$ and $\epsilon(\mathbf{x})$ in the above equation are given by \eq{dif1} to \eq{dif4}.
\paragraph{Metric Determinant}
Now, let $\hat{g}$ be the determinant of the metric $\hat{g}_{ij}$ as defined in \eq{ggamma}, then
\begin{equation}
	\label{detg}
	\sqrt{\hat{g}}= 1+\frac{1}{2}\gamma_{ii}+\mathcal{O}(\gamma_{ij}^2)
\end{equation}
where index i is summed over. So the  variation gives
\be
\label{chdetg}
\delta(\sqrt{\hat{g}})=\frac{1}{2}\delta\gamma_{ii}
\ee
Therefore, using \eq{spatg2} and \eq{delgijtime} in the above equation, we get
\begin{align}
	\label{ddetgspat}
	{\rm Spatial~Reparameterisation~:}&\quad \quad \delta(\sqrt{\hat{g}(\mathbf{x})})= -\partial_i v_i(\mathbf{x})+{\rm homogeneous~part}\\
	\label{ddetgtime}
	{\rm Time~Reparameterisation~:}&\quad \quad \delta(\sqrt{\hat{g}(\mathbf{x})})= d\epsilon(\mathbf{x})+{\rm homogeneous~part}
\end{align}
For conformal transformations, $\delta\sqrt{\hat{g}}$  can be similarly found using \eq{chdetg}, which again is entirely supported by homogeneous contributions.

\subsection{Transformation of Responses}
\label{conftO}
\textbf{Change in Boundary Field Operator}\\
\\
Conformal transformations of the source produces a change in the partition function of the boundary CFT which can also be understood as due to the change in the response and hence change in the boundary field. We illustrate this statement by considering boundary partition function for free bulk theory in overdamped case
\begin{equation}
	\label{pffree}
	\log\psi[\hat{\phi}]= {1\over2}\int d^d{\bf x}d^d{\bf y}\hat{\phi}({\bf x})\langle O({\bf x})O({\bf y})\rangle\hat{\phi}({\bf y})
\end{equation}
Change in $\hat{\phi}$ produces a change in $\psi$ in 1st order as
\begin{equation}
	\label{delpsi}
	\log\psi[\hat{\phi}+\delta\hat{\phi}]=\log\Psi[\hat{\phi}]+{1\over2}\int d^d{\bf x}d^d{\bf y}(\delta\hat{\phi}({\bf x})\langle O({\bf x})O({\bf y})\rangle\hat{\phi}({\bf y})+\hat{\phi}({\bf x})\langle O({\bf x})O({\bf y})\rangle\delta\hat{\phi}({\bf y}))
\end{equation}
where $\delta\hat{\phi}$ is the transformation of the source given in eq.(\ref{delphi}) which can be put for $\delta\hat{\phi}$ in \eq{delpsi} and integrate bi-parts ignoring surface term. This gives rise an equation as
\begin{equation}
	\label{delpsio}
	\log\psi[\hat{\phi}+\delta\hat{\phi}]=\log\psi[\hat{\phi}]+\left[{1\over2}\int d^d{\bf x}d^d{\bf y}(\hat{\phi}({\bf x})\langle \delta O({\bf x})O({\bf y})\rangle\hat{\phi}({\bf y})+\hat{\phi}({\bf x})\langle O({\bf x})\delta O({\bf y})\rangle\hat{\phi}({\bf y}))\right]
\end{equation}
where the $\delta O$ can be obtained for general coordinate transformation as
\begin{equation}
	\label{delO}
	\delta O({\bf x})=[v^i\del_i+(\del_i v_i)-(d-\Delta)\epsilon]O({\bf x})
\end{equation} 

As has been noted in the paper the CFT dual to dS space is of an unconventional type, violating reflection positivity, and not admitting a continuation to Lorentzian signature in general.  For more conventional CFTs, not violating reflection positivity, primary operators are defined to transform under conformal transformations by the transformation given   in\footnote{In $d=2$ this is the definition of   quasi-primary operators.} eq.(\ref{delO}). In AdS/CFT duality bulk fields give rise to sources which couple to such primary operators. Here, for dS space, the   sources in the bulk  couple to fields, rather than operators,  in the dual theory. By analogy with the conventional case, we can define primary fields   to be those which transform as eq.(\ref{delO}) under conformal transformations. In particular, we see from the discussion above that the sources arising from bulk scalar fields couple to primary scalar fields in the boundary CFT.

Similarly in underdamped case, $J_\pm$ are associated to the boundary fields $\hat{O}_\pm$ which transforms exactly like $O$ as given by \eq{delO} with $\Delta$ replaced by $\Delta_{+}$ and $\Delta_{-}$ for $\hat{O}_{+}$ and $\hat{O}_{-}$ respectively.
\\
\\
\textbf{Change in Energy Momentum Tensor}\label{AppTijcng}\\
\\
Under the conformal transformation, invariance of the wave functional implies that  
\begin{equation}
	\Psi[\gamma_{ij}(\mathbf{x})+\delta\gamma_{ij}(\mathbf{x})]=\Psi[\gamma_{ij}(\mathbf{x})]
\end{equation}
Using \eq{ggamma} in \eq{delggen} followed by the relation satisfied by conformal transformations referred in \eq{confcond}, we find the general form of $\delta\gamma_{ij}$ for the conformal transformations (\eq{dif1} to \eq{dif4}) to be 
\begin{equation}
	\delta\gamma_{ij}\Big|_{\rm conf}=-\gamma_{im}\partial_jv^m-\gamma_{jm}\partial_iv^m-v^k\partial_k\gamma_{ij}+2\epsilon\gamma_{ij}
\end{equation}
In a similar way, as we found the change in operator O from the change in field source, we can find the change in $T^{ij}$ as a response to the change in source $\gamma_{ij}$ by performing integration by parts. 
The change under conformal transformations is of the form
\begin{equation}
	\label{Tch}
	\delta T^{ij}=-T^{im}\partial_m v^j-T^{mj}\partial_m v^j+(\partial_k v^k)T^{ij}+v^k\partial_k T^{ij}+2\epsilon T^{ij}
\end{equation}
where $v^i$ and $\epsilon$ are given by \eq{dif1} to \eq{dif4}.

\section{Ward Identities}
\label{Wardtest}

Ward identities were discussed in section \ref{WI} above. Here we will provide some more details. Our discussion below will mostly focus on the Ward identities in the coherent state representation for underdamped fields. 
\subsection{U(1) Gauge Field}
\label{U(1)ward}
The Ward identities for a $U(1)$ symmetry were discussed in section \ref{WIU1}. 
We follow the same conventions below and will explicitly check that the Ward identity eq.(\ref{condaab}) involving a quadratic term in the scalar, and a cubic term with one  gauge boson  and two scalars is valid. 

One of the reasons for the very explicit check below is that similar Ward identities, involving gauge fields, played an important role during the development of the AdS/CFT correspondence for elucidating the connection  between  bulk fields and their corresponding boundary sources \cite{freedU1}.   Here we have proposed a somewhat novel way of identifying sources for under damped fields by working in the coherent state representation and an explicit check is  worth while to do.

We work to leading order in ${R_{ds}^{d-1}\over G_N}$ where the wave function is given in terms of   the on-shell action, obtained by solving the equations of motion with appropriate boundary conditions.  
The coefficient function, $\l O_+({\bf x})O_+({\bf y})\r$, in the case of a real bulk scalar was obtained in eq.(\ref{rho2pt}). 
From this we get, noting the relations,  eq.(\ref{defcra}), eq.(\ref{defcrb}),  that for a complex bulk scalar,
\be
\label{twoptre}
\l O_+^\dagger({\bf x}) O_+({\bf y})\r=2{i\Gamma[1-i\mu]\Gamma[{d\over2}+i\mu]\over\pi^{1+{d\over2}}(1+\coth(\pi\mu))}{1\over|{\bf x}-{\bf y}|^{d+2i\mu}}
\ee
Next we will   calculate the coefficient function for the cubic term $\l J_+({\bf z})O_+^\dagger({\bf x})O_+({\bf y})\r$, eq.(\ref{expwv}). This will allow us then to check the Ward identity. 
The calculation of this term will be carried out after setting ${\hat g}_{ij}=\delta_{ij}$. i.e.  the boundary value of the metric to be flat.

In the approximation we are working with, it was argued in section \ref{coherent}, see discussion after eq.(\ref{twpaf}), that this cubic term can be obtained by expressing the boundary values for the scalar  $\phi({\bf x},\eta_1)$ in terms of $J_+$ and then using the relation, eq.(\ref{Jrelchi}), to express $J_+$ in terms of  $\rho^{\ast}$. 
To be more precise, the complex field $\phi$ is defined in terms of the  real fields $\phi_1,\phi_2$ as given in eq.(\ref{dec}).
For $\phi_1$ a source $J_{+1}$ can be defined as given in \eq{defjpb} and it is then related to the coherent state eigenvalue $\rho_1^{\ast}$ as given by, eq.(\ref{Jrelchi}),
\be
\label{reljpr}
\rho_1^{\ast}({\bf k})=\sqrt{2\mu} \alpha_\mu^{\ast} J_{+1}(-{\bf k})
\ee
Similarly for the second real field $\phi_2$. 
For the complex fields $\phi,\phi^\dagger$, related to the real fields as given in eq.(\ref{dec}), eq.(\ref{decb}) we can define the sources to be 
\begin{align}
\phi({\bf k},\eta)&=f_\mu^{\ast}(k,\eta)k^{i\mu}J_{+}({\bf k}) \label{J+iden}\\
\phi^\dagger({\bf k},\eta)&=f_\mu^{\ast}(k,\eta)k^{i\mu}J^\dagger_{+}({\bf k}) \label{J+siden}
\end{align}
where 
\begin{eqnarray}
\label{jpal}
J_+({\bf k}) & \equiv  &  {1\over \sqrt{2}} [J_{+1}({\bf k})+i J_{+2}({\bf k})]\\
J_+^\dagger({\bf k}) & \equiv  & {1\over \sqrt{2}} [J_{+1}({\bf k})-i J_{+2}({\bf k})]
\end{eqnarray}

The coherent state eigenvalue for the complex field $\phi$ is given in eq.(\ref{defcra}), we see using our definitions above that 
it is related to  the source $J_+$ by 
\be
\label{sdef}
\rho^{\ast}({\bf k})=\sqrt{2\mu}\alpha_\mu^{\ast}\left[{J_{+1}(-{\bf k})+i J_{+2}(-{\bf k})\over \sqrt{2}}\right]=\sqrt{2\mu}\alpha_\mu^{\ast} J_+(-{{\bf k}})
\ee
Similarly from eq.(\ref{defcrb}) we get that 
\be
\label{defcc}
(\rho^{\ast})^\dagger({\bf k})=\sqrt{2\mu} \alpha_\mu^{\ast} \left[{J_{+1}(-{\bf k})-iJ_{+2}(-{\bf k})\over \sqrt{2}}\right]=\sqrt{2\mu}\alpha_\mu^{\ast}J_+^\dagger(-{\bf k})
\ee

The cubic term can be then  calculate by working it out in terms of the boundary values of $\phi({\bf x},\eta_1), \phi({\bf x},\eta_1)^\dagger$, expressing this result in terms of $J_+,J_+^\dagger$, eq.(\ref{J+iden}), eq.(\ref{J+siden}) and then using eq.(\ref{sdef}), eq.(\ref{defcc}) to obtain it in terms of $\rho^{\ast}, (\rho^{\ast})^\dagger$.


We will   accordingly first calculate the cubic term working in the field eigenstate basis $\phi({\bf x},\eta)$.
In fact for checking the Ward identity of interest  it is enough to calculate the cubic coefficient function taking the  gauge boson to be purely longitudinal, i.e.  with $A_i({\bf k})\propto {\bf k}_i$.
Since the wave function is gauge invariant\footnote{The action eq.(\ref{actsca}) is of course manifestly gauge invariant; this ensures the gauge invariance of the wave function which is given by the value of the on-shell action.}  we also note that 
\be
\label{invwf}
\Psi[A_i({\bf x})=\partial_i\chi({\bf x}), \phi({\bf x},\eta_1), \phi^\dagger({\bf x},\eta_1)]=\Psi[A_i=0, \phi({\bf x},\eta_1)+\delta \phi({\bf x},\eta_1), \phi^\dagger({\bf x},\eta_1)+\delta \phi^\dagger({\bf x},\eta_1)]
\ee
where on the RHS we have carried out a gauge transformation to set the longitudinal mode of $A_i$ to vanish resulting in the changes $\delta \phi, \delta\phi^\dagger$  to the scalars given by, eq.(\ref{condga}),eq.(\ref{condgb}),
\be
\label{chngpa}
\delta \phi({\bf x},\eta_1)=-i e \chi ({\bf x}) \phi({\bf x},\eta_1), \ \ \delta \phi^\dagger({\bf x},\eta_1)=+i e \chi({\bf x}) \phi({\bf x},\eta_1)^\dagger
\ee
The required cubic term can now be conveniently calculated from the wave function on the RHS of eq.(\ref{invwf})  where $A_i$  vanishes. 

Next, we note that  $\Psi[A_i=0, \phi+\delta \phi, \phi^\dagger+\delta \phi^\dagger]$ can be obtained, as was discussed above,  from the on-shell action, for the scalar fields which solve the free equation of motion subject to the boundary condition that they take the values $\phi+\delta \phi, \phi^\dagger+\delta \phi^\dagger$ at the boundary  $\eta=\eta_1$. 
Starting from the action, \eq{actsca}, and using the fact that we are computing it for an on-shell solution we obtain that the change in the action due to changing the boundary values of the scalar from $(\phi,\phi^\dagger)$ to $(\phi+\delta \phi, \phi^\dagger+\delta \phi^\dagger)$ is given by 
\be
\label{delSphib}
\delta{S}={ie \over (-\eta_1)^{d-1}}\int{d^d{\bf x}}\chi(\phi^\dagger(\eta,\mathbf{x})\del_\eta\phi(\eta,\mathbf{x})-\phi\del_\eta(\eta,\mathbf{x})\phi^\dagger(\eta,\mathbf{x}))\big|_{\eta=\eta_1}
\ee 
or in k-space by, 
\begin{empheq}{multline}
\label{delSphib1x}
\delta{S}={ie \over (-\eta_1)^{d-1}}\int \prod_{i=1}^{3} {d^d{\bf k}_i\over(2\pi)^d}(2\pi)^d\delta({\bf k}_1+{\bf k}_2+{\bf k}_3)\\
\times\chi({\bf k}_3)[\phi^\dagger({\bf k}_1,\eta)\del_\eta\phi({\bf k}_2,\eta)-\phi({\bf k}_2,\eta)\del_\eta\phi^\dagger({\bf k}_1,\eta)]\big|_{\eta=\eta_1}
\end{empheq}

For underdamped fields , we have the identifications eq.(\ref{J+iden}), eq.(\ref{J+siden}).
Putting these relations  in \eq{delSphib1x}, and then expanding the RHS of \eq{delSphib1x},  using \eq{deffu}, we get
\be
\label{delS2ku}
\delta S=-{e}\int_k\chi({\bf k}_3)J^\dagger_{+}({\bf k}_1)J_{+}({\bf k}_2)2\mu\alpha_\mu^{*2}\left[{\beta_\mu^{\ast}\over\alpha_\mu^{\ast}}k_1^{2i\mu}-{\beta_\mu^{\ast}\over\alpha_\mu^{\ast}}k_2^{2i\mu}\right]
\ee
where
$$\int_k\equiv \int \prod_{i=1}^{3} {d^d{\bf k}_i\over(2\pi)^d}(2\pi)^d\delta({\bf k}_1+{\bf k}_2+{\bf k}_3)$$
Next, substituting for $J_+, J_+^\dagger$ in terms of $\rho^{\ast}, (\rho^{\ast})^\dagger$, eq.(\ref{sdef}),  eq.(\ref{defcc}) and going to position space we get 
\begin{empheq}{multline}
	\label{poscha}
	\delta{S}=-{e}\int d^d{\bf x}d^d{\bf y}d^d{\bf z}~ \chi({\bf z})(\rho^{\ast})^\dagger({\bf x})\rho^{\ast}({\bf y})\\{i\Gamma[1-i\mu]\Gamma[{d\over2}+i\mu]\over\pi^{1+{d\over2}}(1+\coth(\pi\mu))}\left[\delta({\bf z}-{\bf y}){1\over|{\bf x}-{\bf z}|^{d+2i\mu}}-\delta({\bf z}-{\bf x}){1\over|{\bf z}-{\bf y}|^{d+2i\mu}}\right]
\end{empheq}

From the relation between $\Psi$ and $S$, \eq{wfe} 
 we therefore get that the required cubic term in the wave function is 
\begin{empheq}{multline}
\label{retw}
\delta \log \Psi= -{ie}\int d^d{\bf x}d^d{\bf y}d^d{\bf z}~ \chi({\bf z})(\rho^{\ast})^\dagger({\bf x})\rho^{\ast}({\bf y})\\{i\Gamma[1-i\mu]\Gamma[{d\over2}+i\mu]\over\pi^{1+{d\over2}}(1+\coth(\pi\mu))}\left[\delta({\bf z}-{\bf y}){1\over|{\bf x}-{\bf z}|^{d+2i\mu}}-\delta({\bf z}-{\bf x}){1\over|{\bf z}-{\bf y}|^{d+2i\mu}}\right]
\end{empheq}

We are now in a position to check the Ward identity. 
The cubic term in eq.(\ref{expwv}) after substituting for $A_i({\bf x})=\partial_i \chi({\bf x})$  becomes 
\be
\label{cubict}
{e\over 2} \int {d^d{\bf x} d^d{\bf y} d^d{\bf z}}~ \partial_{i}^{\bf z}\chi({\bf z})\rho^{\ast}({\bf x})(\rho^{\ast})^\dagger({\bf y})\l J_i({\bf z}) O_+({\bf x})O_+^\dagger({\bf y})\r
\ee
Carrying out an integration by parts and comparing with eq.(\ref{retw}) then gives 
 \begin{empheq}{multline}
 \label{exppart}
 \partial_i^{\bf z}\l J_i({\bf z}) O_+^\dagger({\bf x}) O_+({\bf y})\r ={2i\Gamma[1-i\mu]\Gamma[{d\over2}+i\mu]\over\pi^{1+{d\over2}}(1+\coth(\pi\mu))}\left[i\delta^{(d)}({\bf z}-{\bf y}){1\over|{\bf x}-{\bf z}|^{d+2i\mu}}\right.\\
 \left.-i\delta^{(d)}({\bf z}-{\bf x}){1\over|{\bf z}-{\bf y}|^{d+2i\mu}}\right]  
 \end{empheq}
 Finally noting   from eq.(\ref{twoptre}) the value of the  two point function $\l O_+^\dagger({\bf x}) O_+({\bf y})\r$ we  see that the Ward identity we had to verify, eq.(\ref{condaab}),  is indeed valid. 

\subsection{Ward Identities for Spatial Reparametrisations} \label{appSrep}
These Ward identities were discussed in section \ref{WISR}. 
Here we provide some more details. 

As argued in section \ref{WISR} the wave function is invariant under spatial reparametrisations of the form, eq.(\ref{sptr}) with the metric and the coherent state 
eigenvalue $\rho^{\ast}$ transforming as given in eq.(\ref{chngrhospat}), eq.(\ref{mettransspat}). In the classical limit where the wave function is obtained from the on-shell action the invariance of the wave function follows from the invariance  of the action under spatial reparametrisations. The resulting condition due to this invariance  is given in eq.(\ref{condinaa}). 

Expanding the wave function as given in eq.(\ref{coeffiexp}), starting with a flat metric, ${\hat g}_{ij}=\delta_{ij}$  and imposing the invariance condition 
eq.(\ref{condinaa}) leads to 
\be
\label{condinab}
\Psi[\rho-v^i\partial_i\rho, \delta_{ij}-\partial_iv_j-\partial_jv_i,\eta_1]=\Psi[\rho,\delta_{ij},\eta_1]
\ee
Expanding the wave function with  the trilinear term given in eq.(\ref{coeffiexp}), up to linear order in $v_i$,  leads to the condition 
\begin{empheq}{multline}
\label{eondaa}
{1\over 2} \int d^d{\bf x}d^d{\bf y}[\partial_i^{\bf x} (v_i ({\bf x})\rho^{\ast}({\bf x})) \rho^{\ast}({\bf y})+ \rho^{\ast}({\bf x}) \partial_i^{\bf y}(v_i({\bf y})\rho^{\ast}({\bf y}))] \l O_+({\bf y})O_+({\bf x})\r \\
+ {1\over 2}\int d^d{\bf x} d^d{\bf y}d^d{\bf z} ~\rho^{\ast}({\bf x}) \rho^{\ast}({\bf y})\partial_i^{\bf z}v_j({\bf z}) \l T_{ij}({\bf z})O_+({\bf x})O_+({\bf y})\r=0
\end{empheq}
Varying with respect to $v_i({\bf z})$ then gives rise to the Ward identity in eq.(\ref{condtwothree}).

\subsection{Ward Identities for Time Reparametrisations}
\label{twi}
These Ward identities were discussed in section (\ref{witra}). 
Under the coordinate transformation which asymptotically takes the form 
$\eta\rightarrow \eta(1+\epsilon({\bf x}))$, $x^i \rightarrow x^i$, with the metric and $\rho^{\ast}$ fields transforming as 
eq.(\ref{metrcompt}), eq.(\ref{tanrh}), the wave function is invariant, once we also account for the change in the cut-off with $\eta_1\rightarrow \eta_1(1+\epsilon({\bf x}))$.
This gives rise to the condition, eq.(\ref{condwca}) on the wave function.

Now consider the wave function obtained by expanding up to the trilinear term as given in eq.(\ref{coeffiexp}).  We will only be interested in the cut-off independent terms 
in the coefficient functions.  Imposing the invariance condition eq.(\ref{condwca}) then leads to 
\begin{empheq}{multline}
\label{condtri}
\int d^d{\bf x} d^d{\bf y} [\epsilon({\bf x}) +\epsilon({\bf y})] (d-\Delta_-)\rho^{\ast}({\bf x})\rho^{\ast}({\bf y})\l O_+({\bf x})O_+({\bf y})\r \\
+ \int d^d{\bf z}d^d{\bf x}d^d{\bf y} \epsilon({\bf z})\rho^{\ast}({\bf x})\rho^{\ast}({\bf y}) \l T^i_i({\bf z})O_+({\bf x})O_+({\bf y})\r =0
\end{empheq}
where $\Delta_-$ is defined in \eq{dimdelm}. Varying with respect to $\epsilon({\bf x})$ gives the Ward identity eq.(\ref{condconf}).  

For good measure let us verify that the  action for a free scalar is indeed invariant under time reparametrisations. 
Denoting the bulk coordinates by $x^\mu=(\eta,x^i)$, consider a general coordinate transformation 
$x^\mu\rightarrow x^\mu+v^\mu$, under which the metric and  scalar transform as 
$\phi\rightarrow \phi-v^\mu\partial_\mu \phi, g_{\mu\nu}\rightarrow g_{\mu\nu}-\nabla_\mu v_\nu -\nabla_\nu v_\mu$.
The action changes under this transformation due the alteration in the metric, the scalar field and the cut-off, $\eta_1$.
The sum of these three must cancel. 
The change in the action due to the alteration in the metric is given by 
\begin{eqnarray}
\label{changemet}
\delta_g S & = &\frac{1}{2} \int d^{d+1}x \sqrt{-g} \delta g_{\mu\nu} T^{\mu\nu}\\
& = &  -\int d^{d+1}x \sqrt{-g} \nabla_\mu v_\nu T^{\mu\nu}\\
&=& -\int d^d{\bf x} {1\over (-\eta_1)^{d+2}} \epsilon({\bf x}) T^{\eta\eta}
\end{eqnarray}
Here we have used the conservation of the stress tensor $\nabla_\mu T^{\mu\nu}=0$ and the fact that asymptotically the only non-zero component of $v_\mu$ is $v_\eta=-{\epsilon \over \eta}$. 

The change due to the alteration in the scalar field is given by 
\be
\label{changescalar}
\delta_\phi S= \int {d^d{\bf x}}{\epsilon({\bf x})\over (-\eta)^d}(\eta\del_{\eta}\phi)^2
\ee
where we have use the fact that the scalar field is on-shell and satisfies the equation of motion \eq{Eq.m}. 

Finally the change in the cut-off is given by $\eta_1\rightarrow \eta_1 +\epsilon({\bf x}) \eta_1$. The resulting change in the action is given by 
\be
\label{changesc}
\delta_{co}S=\int d^d{\bf x}\sqrt{-g} \epsilon({\bf x}) \eta_1 {\cal L}
\ee
where ${\cal L}$ is the scalar Lagrangian given in \eq{action} and the integral is to be evaluated on the surface $\eta=\eta_1$. 

Adding the three terms in eq.(\ref{changemet}), eq.(\ref{changescalar}) eq.(\ref{changesc}), it is easy to see gives a vanishing result.

\section{More Details on the Stress Tensor}
\label{stressapp}
\subsection{Stress Tensor Two Point Function}
\label{App.2ptTT}
Einstein-Hilbert action in dS$_{d+1}$ is given in \cite{klemm} as \footnote{We have taken different sign convention in defining $K_{ab}$, see \eq{Kabdef}}
\be
\label{EHactG}
S={R_{dS}^{d-1}\over16\pi G}\int{d^{d+1}x\sqrt{-g}}[R-d(d-1)]-{R_{dS}^{d-1}\over8\pi G}\int{d^{d}{\bf x}\sqrt{\gamma}}[K-(d-1)]
\ee
where $R_{dS}$ is the radius of dS. Setting $8\pi G=1$, \footnote{Note that in \cite{ViswaTTads}, the authors have set $16 \pi G=1$ while we have set $8 \pi G=1$. This results in an extra half factor in front of the action \eq{EHact} compared to eq.(10) of their paper.}.
\be
\label{EHact}
S=\frac{R_{dS}^{d-1}}{2}\int{d^{d+1}x\sqrt{-g}}[R-d(d-1)]-R_{dS}^{d-1} \int{d^{d}{\bf x}\sqrt{\gamma}}[K-(d-1)]
\ee
where $R$ and $K$ are the Ricci scalar in bulk and extrinsic curvature at boundary $\eta=0$ respectively, $\gamma$ is induced metric determinant on the boundary. $R$ can be expressed by generalising eq.(3.43) of \cite{poisson2004relativist} for $d$ dimensional Ricci scalar $R_{\gamma}$ as
\be
\label{Rdecom}
R=R_{\gamma}+\varepsilon(K^2-K_{ab}K^{ab})+2\varepsilon\nabla_\alpha(n^\beta\nabla_{\beta}n^{\alpha}-n^{\alpha}\nabla_\beta n^\beta)
\ee
where $n^{\alpha}$ is the unit outwards normal to the boundary given by 
\be
\label{n}
n_{\alpha}={1\over\eta}\delta_\alpha^0
\ee
and $\varepsilon=n^\alpha n_{\alpha}$. To note, Greek letters $\alpha,\beta,\cdots$ denotes bulk coordinates, Latin letters $a,b,\cdots$ denotes boundary coordinates. According to eq.(3.3) in \cite{poisson2004relativist}, $\varepsilon=-1$ for the spacelike boundary $\eta=0$. Using eq.(3.36) in \cite{poisson2004relativist} and eq.(\ref{Rdecom}), we derive from eq.(\ref{EHact})
\be
\label{EHactred}
S= \frac{R_{dS}^{d-1}}{2}\int{d^{d+1}x\sqrt{-g}}[R_{\gamma}+K^{ab}K_{ab}-K^2-d(d-1)+2(d-1)K]
\ee
which is similar to the action eq.(11) in \cite{ViswaTTads}.
Now, we introduce metric perturbation $\zeta_{ab}$ on the background dS metric as
\be
\label{habdef}
\gamma_{ab}={1\over\eta^2}[\delta_{ab}+\zeta_{ab}]
\ee
and obtain extrinsic curvature $K_{ab}$ as 
\begin{align}
K_{ab}&={1\over2}(\nabla_bn_a + \nabla_a n_b)\label{Kabdef}\\
&={1\over\eta^2}[\delta_{ab}+\zeta_{ab}]-{1\over2\eta}\del_0\zeta_{ab} \label{Kab}
\end{align}
where we use \eq{n} to derive 2nd equality and $\del_0$ denotes partial derivative w.r.t. $\eta$.
Using the definition $K=\gamma^{ab}K_{ab}$, the action eq.(\ref{EHactred}) can be expanded in the 2nd order of $\zeta_{ab}$ as
\be
\label{dSact1}
S = -\frac{R_{dS}^{d-1}}{8} \int \frac{d^{d+1} x}{(-\eta)^{d-1}}  \left(\zeta^{k}_{j,i} \zeta^{j,i}_k -\zeta_{,i} \zeta^{,i} -2 \zeta^{k}_{j,i} \zeta^{i,j}_k + 2 \zeta_{,i} \zeta^{i,j}_j- \zeta^{i}_{j,0} \zeta^{j}_{i,0} + \zeta_{,0} \zeta_{,0}\right)
\ee
where $\zeta$ is the trace of $\zeta_{ab}$. Notice that the above action can be obtained from eq.(22) of \cite{ViswaTTads}using analytic continuation \footnote{In \cite{ViswaTTads}, the AdS Radius $L$ is set to one. One first needs to restore the $L$ factors in eq.(22) before performing the analytic continuation.},
\begin{equation}
	z \rightarrow i (-\eta), L \rightarrow i R_{dS}
\end{equation}
 Using the transverse and traceless condition we can write \eq{dSact1} as follows.
\begin{equation}
	S = -\frac{R_{dS}^{d-1}}{8} \int \frac{1}{(-\eta)^{d-1}} d^{d+1} x \left(\tilde{\zeta}^{k}_{j,i} \tilde{\zeta}^{j,i}_k -\tilde{\zeta}^{i}_{j,0} \tilde{\zeta}^{j}_{i,0} \right) \label{dSact2}
\end{equation}
where $\tilde{\zeta}^{i}_j$ is transverse traceless part of $\zeta^i_j$. From now on we will specialize to $d$ odd for simplicity.\\ 
Given the action \eq{dSact2} the equation of motion turns out to be
\begin{equation}
	\partial_0^2 \tilde{\zeta}^i_j - \nabla^2 \tilde{\zeta}^i_j + \frac{1-d}{\eta} \partial_0 \tilde{\zeta}^i_j=0 
\end{equation}
whose solution reads
\begin{equation}
	\tilde{\zeta}^i_j({\bf k},\eta) = \frac{(-\eta)^{\frac{d}{2}} H^2_{\frac{d}{2}} (-k \eta)  }{(-\eta_1)^{\frac{d}{2}} H^2_{\frac{d}{2}} (-k \eta_1)} \hat{\zeta^i_j} ({\bf k}) \label{soldS}
\end{equation}
Then from \eq{dSact2} we get
\begin{equation}
	S = \frac{R_{dS}^{d-1}}{8} \int \frac{d^d {\bf x}}{(-\eta)^{d-1}}  \tilde{\zeta}^i_j \partial_0 \tilde{\zeta}^j_i |_{\eta=\eta_1} \label{dSact3}
\end{equation}
Using \eq{soldS} this can be reduced to
\begin{equation}
	S =-\frac{R_{dS}^{d-1}}{8} \int  d^d {\bf k} \hat{\zeta}^i_j ({\bf k})  \hat{\zeta}^j_i (-{\bf k})  d \frac{\beta^{\ast}_{\frac{d}{2}}}{\alpha^{\ast}_{\frac{d}{2}}}k^d \label{dSact4}
\end{equation}
where $\alpha_{\nu}, \beta_\nu$ are given in \eq{valtab}.
Let,
\begin{equation}
	\lim_{\eta \rightarrow \eta_1} \zeta^i_j = h^i_j
\end{equation}
and $h$ is the trace of $h^i_j$. Then in terms of the full graviton field $h^i_j$
\begin{equation}
	\hat{\zeta}^i_j = h^i_j - \frac{k_j k^l}{k^2} h^i_l - \frac{k_l k^i}{k^2} h^l_j + \frac{k^i k_j k^k k_l}{k^4} h^l_k - \frac{1}{d-1} \left(\delta^i_j - \frac{k_j k^i}{k^2} \right) \left(h - \frac{k_k k^l}{k^2} h^k_l\right) \label{TTreldS}
\end{equation}
The above relation obeys the constraints
\begin{align}
	&	k_i \hat{\zeta}^i_j = k_i h^i_j - \frac{k_i k_j k^l}{k^2} h^i_l - k_l h^l_j + \frac{k_j k^k k_l}{k^2} h^l_k = 0 \nonumber \\
	& k^j \hat{\zeta}^i_j= k^j h^i_j- k^l h^i_l - \frac{k^j k_l k^i}{k^2} h^l_j + \frac{k^i k^k k_l}{k^2} h^l_k = 0 \nonumber \\
	& \hat{\zeta}^i_i = h- \frac{k_i k^l}{k^2} h^i_l - \left(h - \frac{k_k k^l}{k^2} h^k_l\right) =0 
\end{align}
The relation \eq{TTreldS} can be rewritten as follows
\begin{equation}
	\hat{\zeta}^i_j=  \left(P^i_k P^l_j - \frac{1}{d-1} P^{i}_j P^l_k\right) h^k_l \label{TTreldS2}
\end{equation} 
where,
\begin{equation}
    P^{i}_j = \delta^i_j - \frac{k_j k^i}{k^2}
\end{equation}
Now,
\begin{align}
	\hat{\zeta}^i_j \hat{\zeta}^j_i &= \left(P^i_k P^l_j - \frac{1}{d-1} P^{i}_j P^l_k\right)  \left(P^j_m P^n_i - \frac{1}{d-1} P^{j}_i P^n_m\right) h^k_l h^m_n \nonumber \\
	&= \left(P^n_k P^l_m - \frac{1}{d-1} P^{l}_k P^n_m\right) h^k_l h^m_n
\end{align}
This can be further simplified to
\begin{align}
	\hat{\zeta}^i_j \hat{\zeta}^j_i = h^m_i h^l_j \left( \frac{1}{2}P^i_l P^j_m + \frac{1}{2} P^{ij} P_{lm} - \frac{1}{d-1} P^j_l P^i_m\right)
\end{align}
Using the above result \eq{dSact4} can be written as
\begin{equation}
	S =-\frac{R_{dS}^{d-1}}{16} \int d^d {\bf k}  k^d  \frac{d \beta^{\ast}_{\frac{d}{2}}}{\alpha^{\ast}_{\frac{d}{2}}} h^m_i({\bf k}) h^l_j(-{\bf k}) \left( P^i_l P^j_m + P^{ij} P_{lm} - \frac{2}{d-1} P^j_l P^i_m\right)
\end{equation}
Then the stress tensor two point function for $d$ being odd, takes the form 
\begin{equation}
	\langle T^i_m (\textbf{k}) T^j_l (-\textbf{k})\rangle = -i\frac{R_{dS}^{d-1}}{8}  k^d  \frac{d \beta^{\ast}_{\frac{d}{2}}}{\alpha^{\ast}_{\frac{d}{2}}} \left( P^i_l P^j_m + P^{ij} P_{lm} - \frac{2}{d-1} P^j_l P^i_m\right) \label{doddTT}
\end{equation}
For $d$ being even, the stress tensor two point function takes the form
\begin{equation}
	\langle T^i_m (\textbf{k}) T^j_l (-\textbf{k})\rangle = -i\frac{R_{dS}^{d-1}}{8}  k^d \log(k)  \frac{d \bar{\beta}^{\ast}_{\frac{d}{2}}}{\bar{\alpha}^{\ast}_{\frac{d}{2}}} \left( P^i_l P^j_m + P^{ij} P_{lm} - \frac{2}{d-1} P^j_l P^i_m\right) \label{devenTT}
\end{equation}
where $\bar{\alpha}^{\ast}_\nu,\bar{\beta}^{\ast}_\nu$ are given in \eq{appalp}, \eq{appbet}.
We can convert \eq{doddTT} and \eq{devenTT} to position space using the relations,
\begin{align}
	&\int {d^d{\bf k}\over(2\pi)^d}k^de^{i{\bf k}.r}={A\over r^{2d}}\label{kdfou}\\
	&\int {d^d{\bf k}\over(2\pi)^d}k^{d}{k_ik_j\over k^2} e^{i{\bf k}.r}=-\del_i \del_j {B\over r^{2d-2}}\label{kd-2fou}\\
	&\int {d^d{\bf k}\over(2\pi)^d}k^{d}{k^ik^jk^kk^l\over k^4} e^{i{\bf k}.r}=\del_i \del_j\del_k\del_l {C\over r^{2d-4}}\label{kd-4fou}
\end{align}
where, $\textbf{r}= \textbf{x}- \textbf{y}, r= \abs{\textbf{r}}$ and the derivatives are with respect to $\textbf{r}$ and,
\begin{equation}
	A={2^d\over\pi^{d\over2}}{\Gamma[d]\over\Gamma[-{d\over2}]}~~~~~B={2^{d-2}\over\pi^{d\over2}}{\Gamma[d-1]\over\Gamma[-{d\over2}+1]}~~~~~C={2^{d-4}\over\pi^{d\over2}}{\Gamma[d-2]\over\Gamma[-{d\over2}+2]} \label{abc}
\end{equation}
Using \eq{abc} and putting the expressions for $\alpha_{d\over2},\beta_{d\over2}$ from \eq{valtab}
the stress tensor two point function takes the following form in the position space,
\begin{equation}
	\langle T^i_m (\textbf{x}) T^j_l (\textbf{y})\rangle =  \frac{e^{i \frac{\pi}{2} (d-1)}R_{dS}^{d-1}\Gamma[d+2]}{8(d-1)\pi^{\frac{d}{2}} \Gamma[\frac{d}{2}] r^{2d}} \left(J^i_l (\textbf{r}) J^j_m (\textbf{r}) + J^{i j} (\textbf{r}) J_{l m} (\textbf{r}) - \frac{2}{d} \delta^i_m \delta^j_l \right) \label{TTtwopoint}
\end{equation}
where,
\begin{equation}
J^i_j (\textbf{x}) = \delta^i_j - \frac{2 x^i x_j}{x^2}
\end{equation}
Note that the above equation \eq{TTtwopoint} agrees with \eq{stresspos} taking $R_{dS}=1$. Note also that \eq{TTtwopoint} can easily be obtained from eq.(43) of \cite{ViswaTTads} by doing the analytic continuation,
\begin{equation}
	L \rightarrow i R_{dS}.
\end{equation}
as mentioned in section \ref{2dcft}.
\subsubsection{Discussion}
Let us examine the reflection positivity for the two point function given in \eq{TTtwopoint}. 
\begin{itemize}
	\item For $d=2n, n\in\mathbb{Z}, n>1$
	\begin{equation}
		e^{i \frac{\pi}{2} (d-1)} = i(-1)^{n+1}
	\end{equation}
	and as a result we see from \eq{TTtwopoint} that reflection positivity is violated. For $n=1$, we are in dS$_3$ and the dual theory is CFT$_2$. In Appendix \ref{App.Central} we calculate the central charge $c$ and show that it is imaginary. On general grounds $c$ is related to the stress tensor two point correlator as
	\begin{equation}
		c\sim \langle TT \rangle	
	\end{equation}
	showing that reflection positivity will be violated in this case too. 
	
	\item Let us now move on to $d=2n+1$. Then we get,
	\begin{equation}
		e^{i \frac{\pi}{2} (d-1)} = (-1)^n
	\end{equation}
	Let us consider the situation when $n$ is odd. This is the case for our universe, $d=3$. We choose the stress tensor to be only dependent on $x_1$ and set all the other $d-1$ coordinates to zero. Then from \eq{TTtwopoint}, setting all indices $i,j,l,m$ to $2$ for simplicity, we have, 
	\begin{equation}
		\langle T^2_2 (x_1) T^2_2 (-x_1)\rangle = - \frac{R_{dS}^{d-1}}{4} \frac{\Gamma[d+2]}{\pi^{\frac{d}{2}} d\Gamma[\frac{d}{2}] r^{2d}} < 0 \label{TT2ptsp}
	\end{equation}
	So we see that the reflection positivity is violated.
	\item Finally we consider the case when $d=2n+1$ and $n=2m, m\in\mathbb{Z}$. Analogous to \eq{TT2ptsp}, we get in this case,
	\begin{equation}
		\langle T^2_2 (x_1) T^2_2 (-x_1)\rangle = \frac{R_{dS}^{d-1}}{4} \frac{\Gamma[d+2]}{\pi^{\frac{d}{2}} d\Gamma[\frac{d}{2}] r^{2d}} > 0 \label{TT2ptsp2}
	\end{equation}
	Thus for $d=4 m+1$, we see that the two point function is reflection positive.
\end{itemize}

\subsection{Central Charge in dS$_3$}
\label{App.Central}
The central charge in the dual CFT was calculated in \cite{klemm}.
This calculation added local counter terms to remove divergences in the stress tensor. 
As emphasised above,  unlike in the AdS case, the divergences in dS space are physical and should not be removed by adding counter terms. 
Here  we do not include local counter terms in the calculation. It turns out that   the finite part of the stress tensor trace does not change.  The    value of the central charge  agrees with \cite{klemm} and is imaginary for the dual CFT of dS$_3$, as mentioned above. 

We work  in the global coordinates \footnote{The metric eq.\eqref{globalmet} becomes eq.\eqref{dsmet} with the help of the transformation $r= R_{dS} \cosh(\theta)$}. 
\begin{equation}
	ds^2 = - \frac{dr^2}{\frac{r^2}{R_{dS}^2}-1} + r^2 (d\theta_1^2 + \sin[2](\theta_1) d\phi^2) \label{globalmet}
\end{equation}

 The Brown York stress tensor in the case of dS can be  found to be,
\begin{equation}
	T^{ab}  =i\frac{2}{\sqrt{h}}\frac{\delta S}{\delta h_{ab}}= i r^4 \frac{1}{8 \pi G}  \left(K^{ab} - K \gamma^{ab}\right) \label{Teq2}
\end{equation}
The above stress tensor is obtained by varying the on-shell action with respect to the boundary source  $h_{ab}$ identified as
\begin{equation}
   h_{ab} =\lim_{r\rightarrow\infty} \frac{\gamma_{ab}}{r^2}
\end{equation}
Calculating the various components of the stress tensor one gets,
\begin{equation}
	T^{\theta_1 \theta_1} = - \frac{i r^2}{8 \pi G R} + \frac{i R_{dS}}{16 G \pi}, T^{\phi \phi} = T^{\theta_1 \theta_1} \csc[2](\theta_1)
\end{equation}
Considering only the finite part of the stress tensor one gets for the trace,
\begin{equation}
	T_{finite} = \frac{i R_{dS}}{8 G \pi}
\end{equation}
Comparing it with the conformal anomaly 
\begin{equation}
	\label{cftanom}
	\frac{c \textbf{R}}{12\pi}
\end{equation}
with $\textbf{R}=2$, one gets the central charge to be,
\begin{equation}
	c = \frac{3 i R_{dS}}{2 G}
\end{equation}
as mentioned in \eq{cenc}. 
This result can also be obtained through analytic continuation of the central charge in ${\rm AdS_3/CFT_2}$ correspondence by $L\rightarrow iR_{ds}$.

\section{Alternate Source Identifications}
\label{app.alts}
\subsection{Alternate Quantisation in AdS/CFT}
\label{altQads}
Scalars in AdS with a mass lying in the range 
\be
\label{Madsaq}
-{d^2\over4}\leq M^2 \leq -{d^2\over4}+1
\ee
can be quantised in two different ways\cite{freedU1,wittklev}.
Asymptotically both  solutions to the free scalar equation in this mass range fall -off towards the boundary going, in terms of the  Poincaré $z$ coordinate, as 
$z^{{d\over 2}\pm \nu}$, where $\nu=\sqrt{{d^2\over4}+M^2}$.
The  different quantisations correspond to taking either of the two fall-offs as the source. 
Choosing the source to correspond to the fall off $z^{{d\over2}-\nu}$ the dual operator has dimension ${d\over2}+\nu$ with a two point function given in \eq{oxoyads}.

The two-point function if the second choice, namely of  taking $z^{{d\over2}+\nu}$ as the source, is made can be obtained by carrying out a Legendre transform. 
Denoting the sources corresponding to the two choices to be $\phi_b({\bf k})$ and $J_b({\bf k})$ respectively we have 
\be
\label{Wactads}
W_{AdS}[J_b]=\int D{\phi_b}Z_{\rm AdS}[{\phi_b}]\exp[\int{d^d{\bf k}\over(2\pi)^d}J_b({\bf k})\phi_b({\bf k})]
=\exp[\frac{1}{2}\int \frac{d^d\mathbf{k}}{(2\pi)^d} \left\{ - k^{-2\nu} \frac{b_{\nu}}{2 \nu a_{\nu}}  \right\} J_b({\bf k}) J_b (-{\bf k})]
\ee
 In position space,
with $J_b$ as the source,  the resulting two-point function is given by 
\be
\label{OxOyW}
\l O({\bf x})O({\bf y})\r={\Gamma[{d\over2}-\nu]\Gamma[\nu]\over 2\pi^{{d\over2}+1}}{\sin(\pi\nu)\over|{\bf x}-{\bf y}|^{d-2\nu}}
\ee
(where using the identity $\Gamma[z]\Gamma[1-z]={\pi\over\sin(\pi z)}$).

The range eq.(\ref{Madsaq}) corresponds to 
\be
\label{rangenu}
0\le \nu\le 1
\ee
For $d\ge 2$ we see from eq.(\ref{OxOyW}) that the coefficient in the two point function in the  range, eq.(\ref{rangenu}), 
  is always positive, as is needed for the two point correlator to satisfy reflection positivity.

\subsection{Overdamped Scalar in dS space}
\label{app.altov}
\paragraph{Momentum Basis :} The identification of the source in the momentum basis for overdamped scalars in dS space  has been discussed in section \ref{over}, here we give a few more details. 
There are two  essential difference with the AdS case discussed in the previous few paragraphs. First, we need to keep the cut-off dependent terms in dS space. 
Second, we carry out a Fourier transformation to go to the momentum space basis rather than a Legendre transformation. The wave function after the Fourier transformation will continue to be normalisable.

%
Starting with the wave function for the free case given in eq.(\ref{psifa})  and carrying out the Fourier transformation from $\phi({\bf k}), \eta)$ to its momentum conjugate 
$\pi({\bf k},\eta)$ we get 
\begin{align}
	{\cal W}[{\pi},\eta]&=\int D{\phi}~\psi[{\phi},\eta]\exp\left[-i\int{d^d{\bf k}\over(2\pi)^d}{\pi}({\bf k},\eta){\phi}({\bf k},\eta)\right]\\
	&=\exp\left[i\int \frac{d^d\mathbf{k}}{(2\pi)^d} ~\pi(\mathbf{k},\eta)\left(\frac{(-\eta)^d(\mathcal{F}_\nu(k,\eta))^\ast}{2\eta \partial_\eta(\mathcal{F}_\nu(k,\eta))^\ast}\right)\pi(-\mathbf{k},\eta)\right]  \label{altovw}
\end{align}
where we use the saddle point relation
\be
\pi({\bf k},\eta)=-\left(\frac{\partial_\eta(\mathcal{F}_\nu(k,\eta))^\ast}{(-\eta)^{d-1}(\mathcal{F}_\nu(k,\eta))^\ast}\right)\phi(-{\bf k},\eta)
\ee
Near boundary, defining the source ${\hat \pi}$ as
\be
\pi({\bf k},\eta)=(-\eta)^{-{d\over2}-\nu}{\hat \pi}({\bf k})
\ee
we can express \eq{altovw} in terms of ${\hat \pi}$ using the boundary limit of ${\cal F}_\nu(k,\eta)$, \eq{altf} as
\be
\label{W3}
{\cal W}[{\hat \pi},\eta_1]=\exp\left[\frac{1}{2}\int \frac{d^d\mathbf{k}}{(2\pi)^d} ~\hat{\pi}(\mathbf{k})\left[{i(-\eta_1)^{-2\nu}\over\left(\frac{d}{2}-\nu\right)}-\frac{\beta^\ast_\nu}{\alpha^\ast_\nu}{(2i\nu)\over\left(\frac{d}{2}-\nu\right)^2} k^{2\nu}\right]\hat{\pi}(-\mathbf{k})\right]
\ee 
which agrees with \eq{momwf}.
We see, as was discussed in section \ref{altrepov}, that the dual operator which couples to ${\hat \pi}$ also has dimension $\Delta$, \eq{overdim}. 
In contrast in the AdS case, after the Legendre transformation one gets the two point function for an operator of dimension ${d\over2}-\nu$, see discussion above. 
The essential reason for this difference is that we have retained the cut-off terms, which are physical when describing the wave function in the dS case, for carrying out the Fourier transformation. 

\paragraph{ Coherent State Basis }In section \ref{over}, we have studied the holography of overdamped field using  the source identification \eq{oversource}. Here we discuss another way of identification similar to the study for underdamped field in section \ref{coherent}. 
Starting with the creation and annihilation operators defined in section \ref{freewvf}, we define two operators $\hat{a}$ and $\hat{b}$ in terms of $a_{\bf k}$ and $a^{\dagger}_{-{\bf k}}$ as
\be
\label{abhat}
\hat{a}({\bf k})=k^{-\nu}(\alpha_\nu a_{\bf k}+\alpha_\nu^{\ast} a^\dagger_{-{\bf k}})~~~~~~\hat{b}({\bf k})=k^{\nu}(\beta_\nu a_{\bf k}+\beta_\nu^{\ast} a^\dagger_{-{\bf k}})
\ee
where ${\hat{a}},{\hat{b}}$ are Hermitian and satisfies the commutation relation
\be
\label{abcom}
[\hat{a}({\bf k}),\hat{b}({\bf k}')]=-{i\over2\nu}(2\pi)^d\delta({\bf k}+{\bf k}')\left({k'\over k}\right)^{\nu}
\ee
Expanding $\Phi({\bf k},\eta)$ and $\Pi({\bf k},\eta)$ given by \eq{Phik} and \eq{Pik} near boundary using \eq{altf} for overdamped case, we get
\begin{align}
	\label{Phigather}
	\Phi({\bf k},\eta)&=\hat{a}({\bf k})(-\eta)^{{d\over2}-\nu}+\hat{b}({\bf k})(-\eta)^{{d\over2}+\nu}\\
	\Pi({\bf k},\eta)&=-\left[\hat{a}({\bf k})\left({d\over2}-\nu\right)(-\eta)^{-{d\over2}-\nu}+\hat{b}({\bf k})\left({d\over2}+\nu\right)(-\eta)^{-{d\over2}+\nu}\right]\label{Pigather}
\end{align}   
In terms of field operators, $\hat{a}$ and $\hat{b}$ can be written as 
\begin{align}
	\label{ahatf}
	\hat{a}({\bf k})&={{d\over2}+\nu\over2\nu(-\eta)^{{d\over2}-\nu}}\Phi({\bf k},\eta)+{(-\eta)^{{d\over2}+\nu}\over2\nu}\Pi({\bf k},\eta)\\
	\hat{b}({\bf k})&=-{{d\over2}-\nu\over2\nu(-\eta)^{{d\over2}+\nu}}\Phi({\bf k},\eta)-{(-\eta)^{{d\over2}-\nu}\over2\nu}\Pi({\bf k},\eta) \label{bhatf}
\end{align}
Note that unlike underdamped case, $\hat{a}$ and $\hat{b}$ are Hermitian in position space. The eigenbasis of $\hat{a}$ is defined as
\be
\label{eighata}
\hat{a}({\bf k})\ket{{\rho}_a}={\rho}_a({\bf k})\ket{{\rho}_a}
\ee
Consequently, ${\rho}_a$, the eigenvalue of $\hat{a}$, is real in position space. Proceeding in similar way as shown in section \ref{coherent}, we get the wave function expressed in ${\rho}_a({\bf k})$ as
\be
\label{ovwvfrho}
\psi[{\rho}_a,\eta_1]=\exp\left[{1\over2}\int{d^d{\bf k}\over(2\pi)^d}\rho_a({\bf k})\rho_a(-{\bf k})\left\{-2i\nu(-\eta_1)^{-2\nu}-2i\nu{\beta_\nu^{\ast}\over\alpha_\nu^{\ast}}k^{2\nu}\right\}\right]
\ee
where we use the saddle point relation between $\phi$ and $\rho_a$ given as
\be
\label{phirhoov}
\phi({\bf k},\eta)={f_\nu^{\ast}\over\alpha_\nu^{\ast}}k^\nu{\rho}_a({\bf k})
\ee
where $\phi$ is the eigenvalue of $\Phi$, \eq{eigphia}. Comparing the coefficient function with that we obtain in $\psi[\hat{\phi},\eta_1]$, \eq{bounwvpexp}, we see that although the coefficient of the cut-off dependent term is different, the finite term which we identified as correlator comes out to be the same.


\subsection{Underdamped Scalar}
\label{app.altun}
\paragraph{$J_-$ as the Source :} Here we consider an underdamped scalar. In section \ref{under}, we have identified $J_+$ as the source through \eq{subst1}. We also noted that there can be an alternate identification as in \eq{subst2} with $J_-$ being the source. The wave function, in the later identification, reduces to
\begin{empheq}{multline}
\label{wvfJ-}
\psi[J_-,\eta_1]=-\int {d^d{\bf k}\over (2\pi)^d}\left( \frac{i}{2}\right)\left[(\beta_\mu^\ast)^2\left(\frac{d}{2}+i\mu\right)(-\eta_1)^{2i\mu}\right.\\\left.+(\alpha_\mu^\ast)^2\left(\frac{d}{2}-i\mu\right)(-\eta_1)^{-2i\mu}k^{-4i\mu}+\alpha_\mu^\ast\beta_\mu^\ast d~ k^{-2i\mu}\right]J_-(\mathbf{k})J_-(-\mathbf{k})
\end{empheq}
Similar to section \ref{under}, we identify the cut-off independent term as two point correlator which signifies the dimension of the dual field as $\Delta_-={d\over 2}-i\mu$. There are as well two cut-off dependent terms --- first term is local in position space but the second term is non-local. As discussed in section \ref{WI}, the finite part of the coefficient function in this identification satisfy the standard  Ward identities of a CFT. 

\section{Understanding Mixed Boundary Condition in Harmonic Oscillator}
\label{Harmo}
In \cite{Isono} a mixed boundary condition was discussed, along with an extra boundary term in the action, as was also discussed in section \ref{coherent} above, where we mentioned that this leads to the wave function for the Buch Davies vacum  in the coherent state basis. 
Here we consider a simple  example of the harmonic oscillator  and show that similar mixed  boundary conditions, with an   additional  boundary term in the action, leads to the ground state wave function being obtained in the coherent state basis.

We take the action for the harmonic oscillator to be 
\begin{equation}
	\label{harmoS}
	S_H=\int dt \left[{1\over2}\dot{x}^2-{1\over2}\omega^2x^2\right]
\end{equation}
The ground state wave function (upto a normalisation factor) is given by 
\be
\label{gstwf}
\l x|0 \r=\exp[-{1\over2}\omega x^2]
\ee

The creation and annihilation operators are 
\begin{eqnarray}
	a & = & {1\over\sqrt{2}}[\sqrt{\omega}\hat{x}+i{\hat{p}\over\sqrt{\omega}}]\\
	a^\dagger & = & {1\over\sqrt{2}}[\sqrt{\omega}\hat{x}-i{\hat{p}\over\sqrt{\omega}}]
\end{eqnarray}

A coherent state $\ket{\alpha}$ satisfies the condition
\begin{equation}
	\label{harmocoh}
	a\ket{\alpha}=\alpha\ket{\alpha}
\end{equation} 
 and is an eigenstate of $a$ with eigenvalue $\alpha$. 
Its wave function is given,  upto an overall normalisation, by  
\begin{equation}
	\label{xalpha}
	\braket{x}{\alpha}= \exp[-{1\over2}\omega x^2+\alpha\sqrt{2\omega}x]
\end{equation} 
The ground state wave function in the $\ket{\alpha}$ basis then takes the form 
\be
\label{gscb}
\l\alpha|0\r=\int dx \l\alpha|x\r\l x|0\r
\ee
From eq.(\ref{xalpha}) and eq.(\ref{gscb}),   this leads to,
\begin{equation}
\label{costappng}
\l\alpha|0\r=\exp[{\alpha^{*2}\over2}]
\end{equation}

We will now show that the same exponential form of the wave function can be obtained by doing the path integral, with appropriate boundary conditions, etc,  in the semiclassical approximation. 
Analogous to the discussion above for the Bunch Davies vacuum we  carry out the path integral along a contour where the   time coordinate, given by $t(1-i\epsilon)$,  has a small imaginary part, The range of  $t$ is  $[-\infty,t_f]$ and we impose the boundary condition that 
\be
\label{condappg}
\phi\rightarrow 0, \ \  {\rm  when} \ t\rightarrow -\infty.
\ee
As a warm up we first check that one gets the correct position space wave function for the ground state by carrying out the path integral with the usual harmonic oscillator action,
\be
\label{hsoa}
\l x|0\r= \int^{x,t_f} D\phi~ e^{iS_H}
\ee
The classical saddle point  solution meeting the condition eq.(\ref{condappg}) is given by, 
\be
\label{coddya}
x(t)=A e^{i\omega t}
\ee
 (we are suppressing the small imaginary part in $t$),
with $x(t)$  meeting the condition at $t_f$,
\be
\label{condappa}
x(t_f)=A e^{i\omega t_f}=x.
\ee
This condition determines $A$ in terms of $x$ and $t_f$.
The saddle point value is then given by $iS_H=-{\omega \over 2} x^2$, which leads to eq.(\ref{gstwf}). 

Next we change the boundary condition and impose that at $t_f$, the condition to be met is 
\be
\label{condbappn}
{\dot x}(t_f) + i \omega x(t_f)=i\sqrt{2\omega} \alpha^*
\ee
To obtain a well defined variational principle leading to the equation of motion
\be
\label{eom}
{\ddot x}+\omega^2 x=0
\ee
we now need to add an extra boundary term to the action at $t=t_f$ of the form,
\be
\label{sbaappn}
S_B=i\omega {x(t_f)^2\over 2}-i\sqrt{2\omega}\alpha^*x(t_f)
\ee
The saddle point solution meeting eq.(\ref{condappg}) still takes the form eq.(\ref{coddya}) but now the boundary condition eq.(\ref{condbappn}) implies that the constant $A$ is determined in terms of $\alpha^*$ by the relation
\be
\label{addcapp}
A e^{i\omega t_f}={1\over \sqrt{2 \omega}} \alpha^*
\ee
and the saddle point value of the action $S_H+S_B$ is then given by $-i {\alpha^{*2}\over 2}$ leading to eq.(\ref{costappng}).

\newpage
\bibliographystyle{JHEP}
\bibliography{dscft}

\end{document}